\documentclass[aps,prb,twocolumn,amsmath,amssymb,reprint,superscriptaddress,10pt]{revtex4-1}
\pdfoutput=1

\usepackage{float}
\usepackage{graphicx}  
\usepackage{xcolor}
\usepackage{dcolumn}   
\usepackage{amssymb}   
\usepackage{epstopdf}
\usepackage{MnSymbol,wasysym}

\usepackage{lipsum,adjustbox}

\usepackage{amsmath}
\usepackage{amsfonts}
\definecolor{darkblue}{rgb}{0.1,0.2,0.6}
\definecolor{darkred}{rgb}{0.8,0.1,0.2}
\usepackage[colorlinks,citecolor=darkblue,linkcolor=black,urlcolor=darkblue]{hyperref}
\usepackage{tikz}
\usepackage{enumitem}

\usepackage{url}  

\numberwithin{equation}{section}
\renewcommand\theequation{\arabic{section}.\arabic{equation}}

\newcommand{\beginsupplement}{%
        \setcounter{table}{0}
        \renewcommand{\thetable}{S\arabic{table}}%
        \setcounter{figure}{0}
        \renewcommand{\thefigure}{S\arabic{figure}}%
     }

\makeatletter
\def\@fnsymbol#1{\ensuremath{\ifcase#1\or\dagger\or \ddagger\or
   \mathsection\or \mathparagraph\or \|\or **\or \dagger\dagger
   \or \ddagger\ddagger \else\@ctrerr\fi}}
\makeatother

\begin{document}

\title{Fock space localization of polaritons in the Jaynes-Cummings dimer model}
\author{Hassan Shapourian}
\affiliation{Department of Physics, University of Illinois at Urbana-Champaign, Urbana Illinois 61801, USA}
\author{Darius Sadri}
\thanks{Deceased.}
\affiliation{Department of Electrical Engineering, Princeton University, Princeton New Jersey 08544, USA}

\date{\today}

\begin{abstract}
We present a new method to study the semiclassical dynamics of the Jaynes-Cummings dimer model, describing two coupled cavities each containing a two-level system (qubit).
We develop a Fock space WKB approach in the polariton basis where each site is treated exactly while the intersite polariton hopping is treated semiclassically.  We show that the self-trapped states can be viewed as Fock space localized states.
We find that this picture yields the correct critical value of interaction strength at which the delocalization-localization transition occurs.
Moreover, the validity of our WKB approach is supported by showing that the quantum spectrum can be derived from a set of Bohr-Sommerfeld quantization conditions and by confirming that the quantum eigenstates are consistent with the classical orbital motion in the polariton band picture.
The underlying idea of our method is quite general and can be applied to other interacting spin-boson  models.
\end{abstract}

\maketitle


\section{Introduction}

Interacting light-matter systems have opened new possibilities in simulating various types of strongly correlated phases ranging from the superfluid and Mott insulating states~\cite{Greentree2006,Hartmann2006,Hartmann2008,Tomadin2010,Houck2012,Schmidt2013,PhysRevLett.106.153601,Schiro2012,
PhysRevLett.110.163605,PhysRevLett.111.113001,PhysRevLett.110.233601,PhysRevLett.108.233603,RevModPhys.85.299,LeHur_Marco} to the quantum Hall fluids~\cite{PhysRevA.82.043811,PhysRevA.86.053804,PhysRevLett.101.246809,Hafezi2013,PhysRevA.84.043804,Hafezi_FQH}.
The property that makes the coupled light-matter systems unique compared to conventional condensed matter systems is that they are equipped with optical probes to manipulate arbitrary states, beyond the ground state, as well as to monitor non-equilibrium dynamics.
The strongly correlated light-matter physics have been experimentally realized in various implementations of cavity QED
with atoms~\cite{Haroche1996,Kimble2005}, excitons~\cite{Reithmaier2004,Yoshie2004,Bloch2005}, or superconducting qubits~\cite{Wallraff2004,Blais2004}.
The building block of all these systems is the celebrated Jaynes-Cummings (JC) model,
\begin{align} \label{eq:JCH}
H_{JC} &= \nu_q \frac{\sigma^z}{2} + \nu_c a^\dagger a + g\ (\sigma^+ a+ a^\dagger \sigma^-)
\end{align}
which describes a two-level system (qubit) interacting with a single mode of the electromagnetic field inside a cavity. The qubit is characterized by the resonance frequency $\nu_q$, and the pseudo-spin operators $\sigma^z$, and $\sigma^{\pm}=(\sigma^x\pm i \sigma^y)/2$.
 The cavity field with frequency $\nu_c$ is described by the annihilation and creation operators $a$ and $a^\dag$ which are linearly coupled to the qubit via the coupling constant $g$. Throughout this paper, we focus on the resonant case where $\nu_q=\nu_c=\nu$. The cavity-qubit coupling induces an anharmonicity in the spectrum of the JC Hamiltonian that can be  qualitatively regarded as an effective interaction between photons. A tunnel-coupled array of JC sites would then realize an ideal setup for strongly correlated photons.

The Jaynes-Cummings dimer model (JCDM) is the simplest, yet non-trivial, system to study interacting photons~\cite{Tureci}. The model Hamiltonian
\begin{align} \label{eq:JCDM}
H= H_{JC, R}+ H_{JC, L} - J (a_R^\dagger a_L + a_L^\dagger a_R),
\end{align}
describes two identical JC sites coupled through a tunneling (kinetic) term in the photonic channel $a$ and $a^\dag$ where the subscript $L(R)$ specifies the left(right) sites. As the cavity-qubit interaction is increased, the JCDM displays a transition from a Josephson oscillating (delocalized) regime where photons coherently tunnel between cavities to a self-trapped (localized) regime where photons are frozen inside one cavity.
A similar transition caused by  the Hubbard interaction has been studied in a Bose-Einstein condensate double-well (BECDW) system~\cite{Milburn,smerzi_prl,BEC_exp1,BEC_exp2,BEC_exp3,Bloch2013}.
A theoretical study of this transition in JCDM was originally done in the pioneering work of Schmidt, {\it et~al.}~\cite{Tureci} where a semiclassical picture as well as numerical exact solutions were provided. They also proposed a superconducting circuit implementation of the JCDM.
This proposal was subsequently realized experimentally and the transition was successfully observed~\cite{James}.

In this article, we revisit the localization-delocalization transition from a different point of view beyond just numerics.
We note that the JCDM conserves the total polariton number $N=n_L+n_R$ where $n_s= \sigma^+_s\sigma^-_s+a_s^\dagger a_s$ for $s=L/R$, left/right JC islands; i.e.~$[N,H]=0$.
We use this property to develop a new approach in the polariton basis where each JC site is 
treated exactly while the hopping of polaritons between sites is treated semiclassically.  
To this end, we introduce a Fock space picture in which the tunneling terms act like hopping operators and map the Hamiltonian onto a one-dimensional tight-binding model. The Schr\"odinger equation in this basis obeys a discrete form and hence can be approximately solved by a discrete version of the Wentzel-Kramers-Brillouin (WKB) approach~\cite{Braun_RMP}.
Using our Fock space representation, we show that the self-trapping transition in the JCDM can be viewed as a localization transition in Fock space~\cite{Altshuler1997}.
 This phenomenon has also been discussed in the context of BECDW~\cite{PhysRevA.71.043603,JuliaDiaz}.
In fact, we find that the WKB approximation maps the JCDM into a four-band one-dimensional tight-binding Hamiltonian.  
Each band looks like a BECDW with a more complicated interaction.  These bands are associated with the four possible polariton states of two JC sites: upper-upper, upper-lower, lower-upper, and lower-lower polaritons.  As we will see the upper-upper and lower-lower polariton bands are equivalent to BECDW with attractive and repulsive interactions, respectively.

Our article is organized as follows. In Sec.~\ref{sec:fock}, we introduce the Fock space picture and find the wave-functions in the WKB limit.
In Sec.~\ref{sec:quant}, we derive a set of Bohr-Sommerfeld quantization conditions in the form of $S(E_n)=\oint p\cdot dq = 2\pi h (n+1/2)$ which yields a quantized energy $E_n$ in each polariton band. The main results of the paper are in Sec.~\ref{sec:local_trans} where we study the delocalization-localization transition using the presented WKB picture.
Finally, we discuss the conclusions and possible applications of our methods in Sec.~\ref{sec:conc}.


\section{\label{sec:fock} Fock space representation and WKB analysis}
In this section, we write the JCDM in the Fock space of photons and show that the Hamiltonian is tridiagonal. Thus, it can interpreted as a one-dimensional tight-binding model with nonuniform nearest neighbor hoppings.
This representation of the Hamiltonian is going to be the basis for the exact diagonalization and the WKB analysis in the upcoming sections.

\begin{figure}
\includegraphics[scale=0.063]{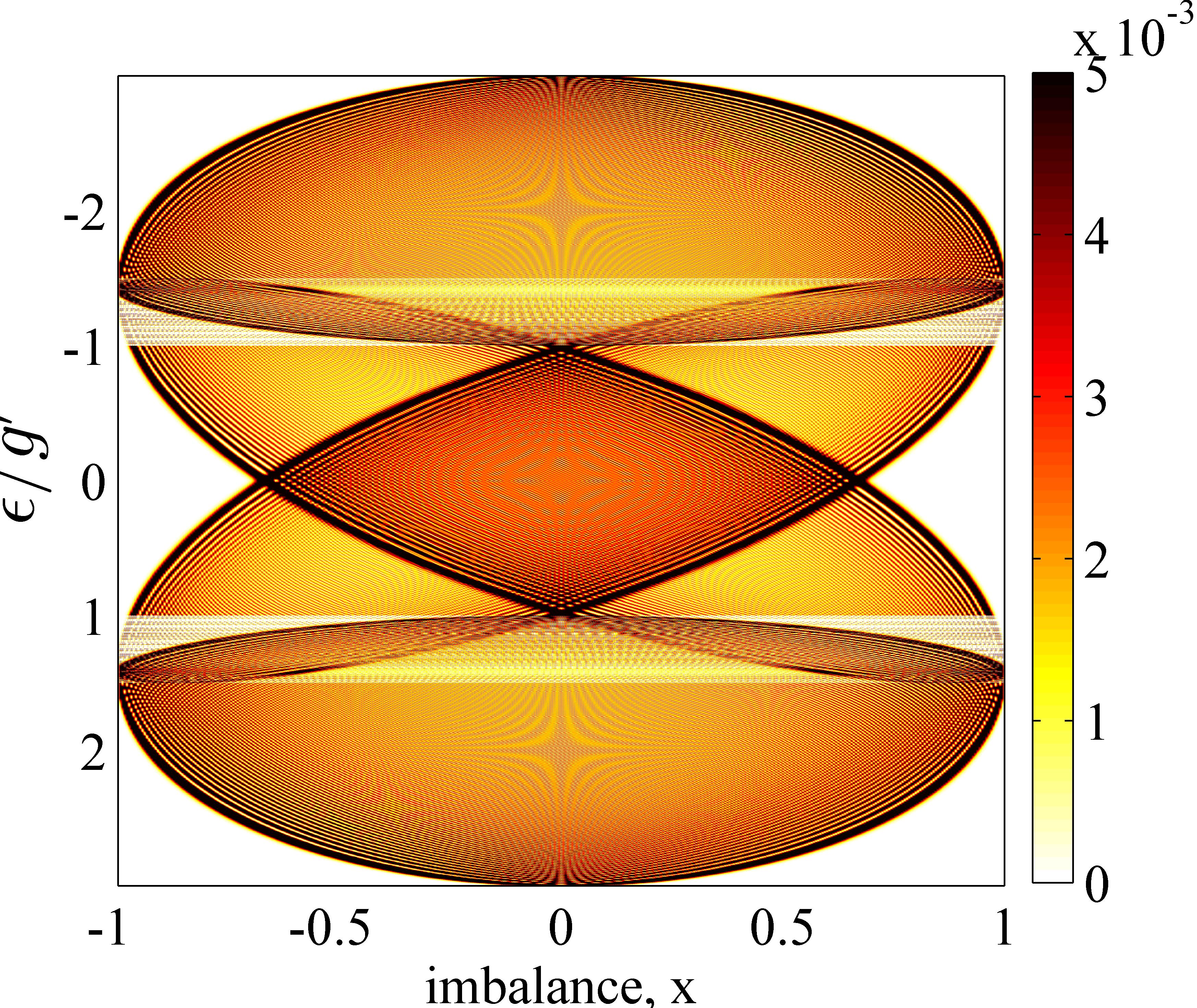}
\includegraphics[scale=0.063]{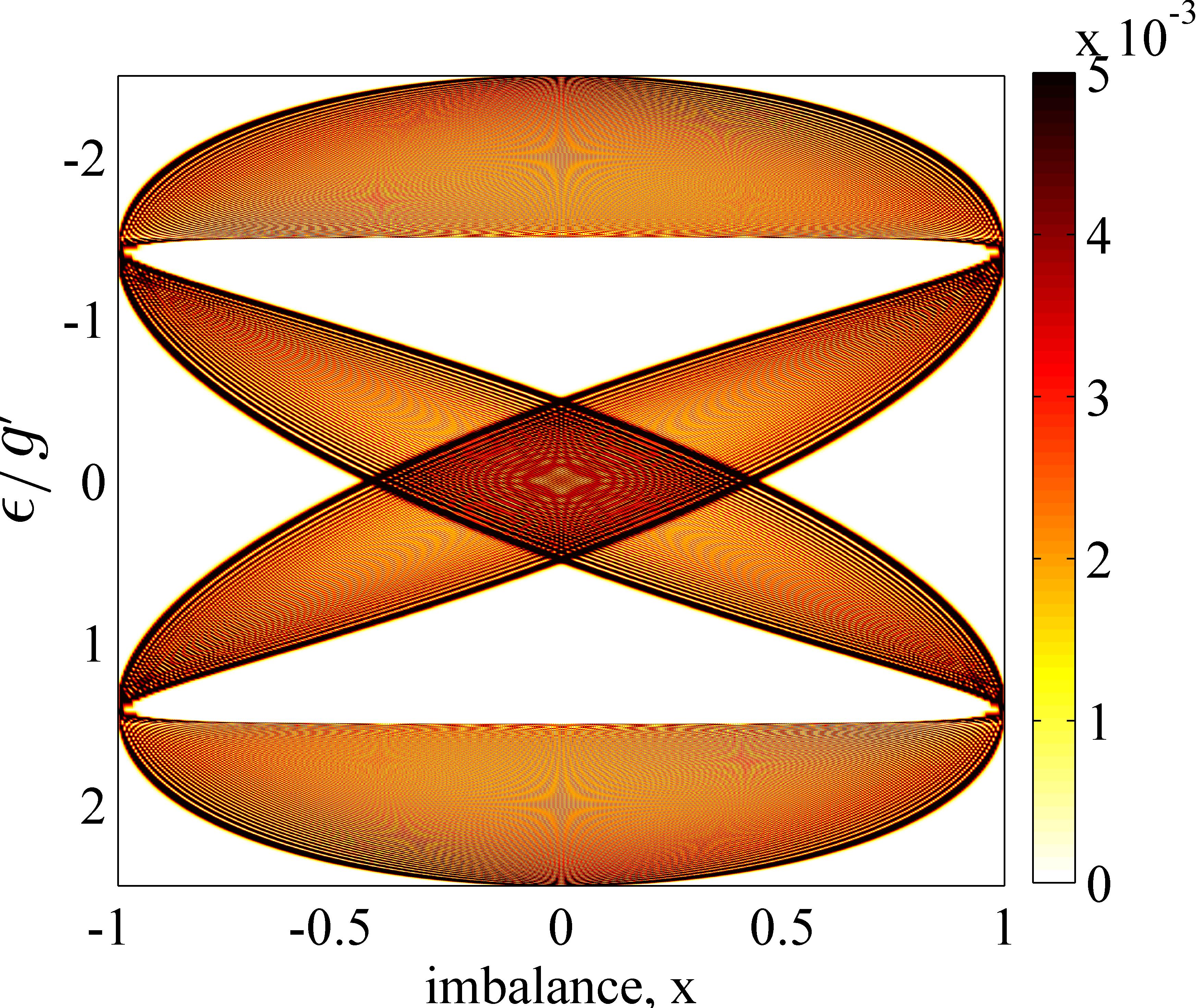}
\caption{\label{fig:nodal_structure} (Color online) Eigenstates and spectrum on the JC dimer model.  Color represents the amplitude squared of the eigenstates in $Z$-space defined by Eq.~(\ref{eq:amplitude}).  The horizontal axis is normalized imbalance $x=Z/N$, while the vertical axis is a scaled eigenenergy.  Here $g/J\sqrt{2N}=1.0$ (top) and $2.0$ (bottom), and $N=400$.}
\end{figure}

As mentioned earlier, the total polariton number operator $N=n_L+n_R$ commutes with the Hamiltonian and we can study this model in a subspace with a fixed $N$.
The simplest basis to span this subspace is a set of $4N$ orthonormal states denoted by $|n_{c,L},m_L\rangle \otimes |n_{c,R},m_R\rangle$ which refers to a state with $n_{c,L(R)}=a^\dagger_{L(R)}a_{L(R)}$ photons and $m_{L(R)}=\uparrow/\downarrow$, up/down qubit on the left (right) site. The only constraint is that the total polariton number must be N.
 Consequently, any generic state can be decomposed into a superposition of these states
\begin{align*}
|\Psi\rangle = \sum A^{n_{c,L} m_L n_{c,R} m_R}_\Psi |n_{c,L}, m_L\rangle \otimes | n_{c,R}, m_R\rangle\ .
\end{align*}
The polariton imbalance of a state $|\Psi\rangle$ is defined as the expectation value of
\begin{align} \label{eq:imbalance}
Z= \langle \Psi | (a_L^\dagger a_L+\sigma^+_L\sigma^-_L)- (a_R^\dagger a_R+\sigma^+_R\sigma^-_R)|\Psi \rangle
\end{align}
and the normalized imbalance is defined by $x=Z/N$.  Note that the integer $Z$ ranges from $-N$ to $+N$ in increments of 2.
In our WKB analysis the imbalance $Z$ will play the role of the position.
We chose our basis in terms of the eigenstates of imbalance operator $|Z,m_L,m_R\rangle$, where
\begin{align}
|Z,\downarrow,\downarrow\rangle&= |(N+Z)/2, \downarrow\rangle \otimes | (N-Z)/2, \downarrow\rangle \nonumber \\
|Z,\downarrow,  \uparrow\rangle&=|(N+Z)/2, \downarrow\rangle \otimes | (N-Z)/2-1, \uparrow\rangle  \nonumber  \\
|Z,  \uparrow,\downarrow\rangle&=|(N+Z)/2-1,  \uparrow\rangle \otimes | (N-Z)/2, \downarrow\rangle  \nonumber  \\
\label{eq:basis}
|Z,  \uparrow,  \uparrow\rangle&=|(N+Z)/2-1,  \uparrow\rangle \otimes | (N-Z)/2-1,  \uparrow\rangle
\end{align}
provided that $|Z| \less N$. Note that for $Z=\pm N$ there are only two states.
In the absence of cavity-qubit coupling $g$, the states $|Z,m_L,m_R\rangle$ are fourfold degenerate. This degeneracy is lifted by $g$ through the following term
\begin{align} \label{eq:ons}
H_Z=& g \sum_{m_R}\sqrt{\frac{N+Z}{2}}(|Z,\uparrow,m_R\rangle\langle Z,\downarrow,m_R|+\text{H.c.}) \nonumber \\
 &+ g \sum_{m_L}\sqrt{\frac{N-Z}{2}}(|Z,m_L,\uparrow\rangle\langle Z,m_L,\downarrow|+\text{H.c.})~.
\end{align}
Therefore, for any fixed $N$ the Hamiltonian in this basis can be written as
{\small
\begin{align} \label{eq:TBJCD}
H =& - \frac{J}{2} \sum_{Z,m_L,m_R}  {\Big(} T^{m_L m_R}_{Z,Z+2} |Z,m_L,m_R\rangle \langle Z+2,m_L,m_R| + \text{h.c.} {\Big)}  \nonumber \\
&+ \sum_Z H_Z
\end{align}
}
where the hopping terms are
\begin{align*}
T^{\downarrow\downarrow}_{Z,Z+2} = &\ [{(N+Z+2)(N-Z)}]^{1/2}, \nonumber \\
T^{\downarrow\uparrow}_{Z,Z+2} =&\ [{(N+Z+2)(N-Z-2)}]^{1/2},\nonumber \\
T^{\uparrow\downarrow}_{Z,Z+2} =&\ [{(N+Z)(N-Z)}]^{1/2},\nonumber \\
T^{\uparrow\uparrow}_{Z,Z+2} =&\ [{(N+Z)(N-Z-2)}]^{1/2} ,
\end{align*}
and the first two terms in Eq.~(\ref{eq:JCH}) are dropped as their sum is constant in the resonant case $\nu_c=\nu_q$.
We represent wave-functions in terms of a four-component ``position''-dependent vector $C(Z)= [C_{\downarrow \downarrow},C_{\downarrow\uparrow},C_{\uparrow\downarrow},C_{\uparrow\uparrow}]^T$,
\begin{align} \label{eq:amplitude}
|\Psi\rangle = \sum_{Z,m_L,m_R} C_{ _{m_L m_R}}(Z)\  |Z,m_L,m_R\rangle
\end{align}
and diagonalize the Hamiltonian in this basis. The spectral map is then constructed by tracing out the spin degrees of freedom as illustrated in Fig.~\ref{fig:nodal_structure}; i.e.~we compute $|\Psi(Z)|^2 =  \sum_i |C_{i}(Z)|^2$ for every eigenstate and show it as a false color map. The white regions correspond to the classically forbidden regime where the eigenstates have exponentially small probability.  The localization occurs when the semiclassical description corresponds to a particle in a double well with the two wells separated by a barrier that the particle has to tunnel through.
The statement of localization is that the tunneling time from one half to the other is exponentially large.
For instance, for a small ratio of couplings $g/J$ as in Fig.~\ref{fig:nodal_structure}(top) there is no barrier and the localization does not occur, whereas for larger couplings as in Fig.~\ref{fig:nodal_structure}(bottom) a barrier (white regions) emerges over some energy ranges, and localization is possible.  The ``two wells'' correspond to the photons being mostly on one of the two JC sites.  When the barrier is present, the tunneling time to the other site is exponentially large in $N$.
The spectral map consists of four bands and each band is associated with a certain polariton configuration.
We shall explain this in more detail after deriving the WKB solution.
The delocalization-localization transition of the BECDW problem has also been studied in the Fock space representation~\cite{PhysRevA.71.043603,JuliaDiaz,FuLiu_dynamics}.  The important difference here is that the BECDW Hamiltonian can be formulated fully in terms of the imbalance and the relative phase between two condensates whereas our JCDM has two qubits as extra degrees of freedom that ultimately lead to four bands in the Fock space representation.

An important remark is that the above Hamiltonian, Eq.~(\ref{eq:TBJCD}), possesses a $\mathbb{Z}_2$ (left-right) parity symmetry whose operator $\cal P$ is defined as ${\cal P} |Z\rangle = |-Z\rangle$; thus, the eigenstates must respect this symmetry, too.
This means that strictly speaking there is no localized eigenstate and the localization is purely a dynamical effect in the following sense: If the system is prepared with some initial non-zero imbalance the dynamics would still preserve this non-zero imbalance for a long time. The time scale at which the imbalance changes sign is divergent as a function of system size.


Let us now work out a semiclassical framework to study the Hamiltonian in Eq.~(\ref{eq:TBJCD}) in the large-N limit. To this end, we divide the Schr\"odinger equation by $N$
\begin{align*}
\frac{i}{N}\frac{\partial |\Psi\rangle}{\partial t} = H |\Psi \rangle,
\end{align*}
and bring the $1/N$ factor into the definition of $H$. Using the basis introduced in Eq.~(\ref{eq:amplitude}), we define the effective Planck's constant $h=1/N$ and the position (normalized imbalance) $x=Z/N$ and rewrite the Schr\"{o}dinger equation as
\begin{align} \label{eq:schrodinger}
i h \frac{\partial}{\partial t} \Psi(x,t)=& W(x) \Psi(x) -  B(x+h) \Psi(x+2h)\nonumber \\
&-  B(x-h) \Psi(x-2h).
\end{align}
where $\Psi(x,t)$ is the continuum version of the four component vector $C(Z)$ (Eq.~(\ref{eq:amplitude})) and we introduce matrices
\begin{align}
B(x) = \frac{J}{2}  \text{diag}{\Big(}&[{(1+x+h)(1-x+h)}]^{1/2},\nonumber \\
& [{(1+x+h)(1-x-h)}]^{1/2},\nonumber \\
& [{(1+x-h)(1-x+h)}]^{1/2},\nonumber \\
& [{(1+x-h)(1-x-h)}]^{1/2} {\Big)},\nonumber  \\
 \label{eq:B_mat}
W(x)= g' \sqrt{1+ x}& \ \sigma^x \otimes \mathbb{I}_2 + g' \sqrt{1-x} \ \mathbb{I}_2 \otimes \sigma^x .
\end{align}
where $\mathbb{I}_2$ is a $2\times 2$ identity matrix and the rescaled coupling is $g'=\frac{g}{\sqrt{2N}}$.
Let us introduce the conjugate operator to position $\hat{p}$ such that $[\hat{p},x]=-ih$. So, the Schr\"{o}dinger equation can be recast into its continuum version as
\begin{align*}
i h \frac{\partial}{\partial t} \Psi(x,t)=& W(x) \Psi(x,t) - e^{i\hat{p}} B(x) e^{i\hat{p}} \Psi(x,t)\nonumber \\
&-  e^{-i\hat{p}} B(x) e^{-i\hat{p}} \Psi(x,t).
\end{align*}
The next step is to find a semiclassical ansatz for the wave-function. A comprehensive discussion of solving the discrete Schr\"odinger equation in the semiclassical WKB limit can be found in~\cite{Braun_RMP}. A similar many-body WKB approach has been applied to the BECDW problem~\cite{PhysRevA.76.032116,PhysRevA.78.023611,Keeling}. A brief review of this procedure is provided in Appendix~\ref{sec:BEC}. 
Before we apply the WKB approach to the JCDM, it is essential to write down the matrices in the polariton basis given by the modes
\begin{align*}
|\Psi_1\rangle&= |N(1+x)/2,+\rangle \otimes |N(1-x)/2,+\rangle, \\
|\Psi_2\rangle&= |N(1+x)/2,+\rangle \otimes |N(1-x)/2,-\rangle, \\
|\Psi_3\rangle&= |N(1+x)/2,-\rangle \otimes |N(1-x)/2,+\rangle, \\
|\Psi_4\rangle&= |N(1+x)/2,-\rangle \otimes |N(1-x)/2,-\rangle,
\end{align*}
where $|n,\pm\rangle=( |n,\downarrow\rangle \pm |n-1,\uparrow\rangle)/\sqrt{2}$ are called upper
\begin{figure}
\centering
\hspace{-1cm}
\includegraphics[scale=0.8]{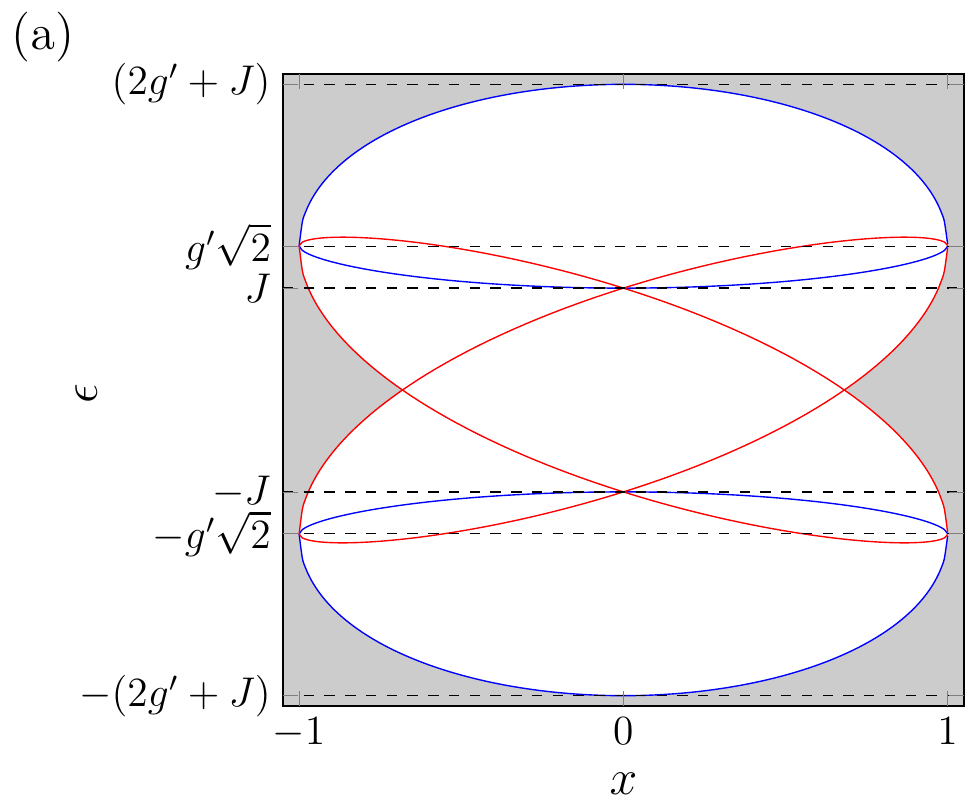}
\vspace{0.2cm}

\hspace{-1.14cm}
\includegraphics[scale=0.8]{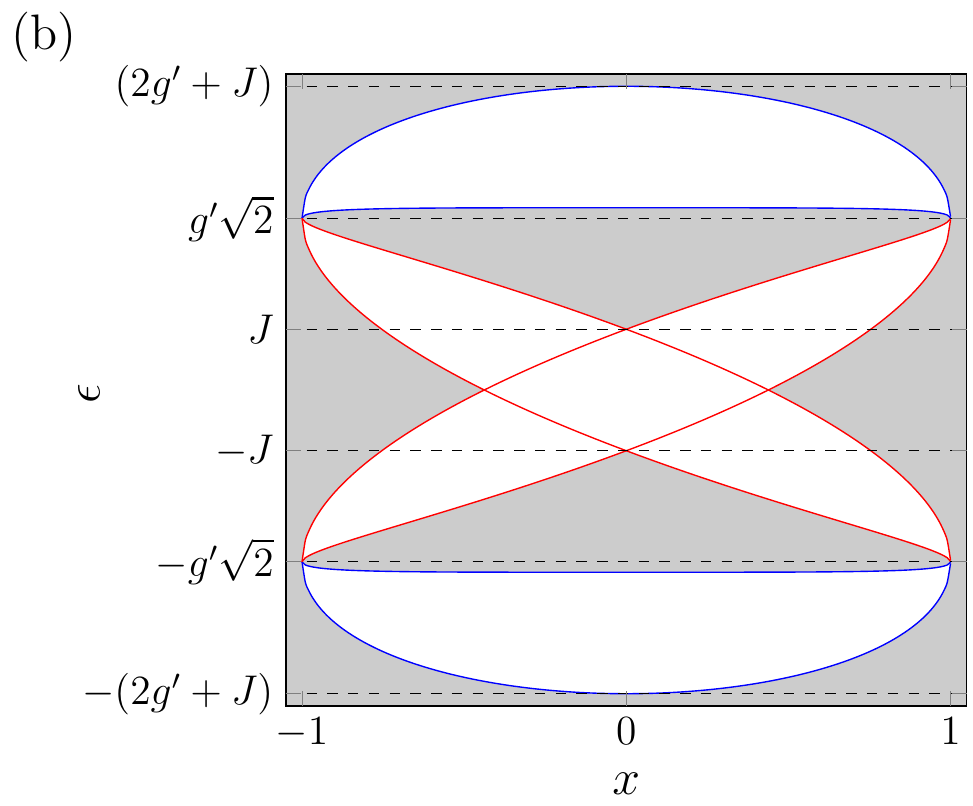}
\caption{\label{fig:spec_WKB} (Color online) Classically allowed regions (white area) of different polariton components for $J/g'=1$ (a) and $1/2$ (b). The classical potentials, Eq.~(\ref{eq:V_pm}), of the first and fourth polariton bands are shown in blue and those of the second and third bands are shown in red.  }
\end{figure}
 (lower) polariton states which are eigenstates of the JC Hamiltonian in Eq.~(\ref{eq:JCH}),
\begin{align}
H_{JC} |n,\pm \rangle = (\nu\ n \pm g \sqrt{n}) |n,\pm \rangle\ .
\end{align}
In the rest of this paper, we refer to the states $|\Psi_1\rangle, \dots, |\Psi_4\rangle$  by calling them upper-upper (first), upper-lower (second), lower-upper (third), and lower-lower (fourth) polariton states or by noting their respective polariton index shown inside parentheses.
Using this basis, the JC interaction $W(x)$ is diagonal
\begin{align*}
W(x)=g'\ \text{diag} {\Big[}A_p+A_m, A_p-A_m, -A_p+A_m, -A_p-A_m{\Big]}
\end{align*}
where $A_m=\sqrt{1-x}$ and  $A_p=\sqrt{1+x}$.
A perturbative expansion of $B(x)$ in powers of $h$ is
\begin{align*}
B(x)= B^{(0)}(x) + h  B^{(1)}(x)  +\dots
\end{align*}
where
\begin{align}
B^{(0)}=&  \frac{J}{2} \sqrt{1-x^2}\ \mathbb{I}_4
\end{align}
and
\begin{align} \label{eq:1st_coupling}
B^{(1)}(x)=\frac{J}{2\sqrt{1-x^2}}\left(
\begin{array}{cccc}
 0 & \frac{x+1}{2} & \frac{1-x}{2} & 0 \\
 \frac{x+1}{2} & 0 & 0 & \frac{1-x}{2} \\
 \frac{1-x}{2} & 0 & 0 & \frac{x+1}{2} \\
 0 & \frac{1-x}{2} & \frac{x+1}{2} & 0 \\
\end{array}
\right).
\end{align}
Note that in contrast to the BECDW problem, we have four polariton modes. The wave-function takes the form
\begin{align*}
\Psi(x,t) = \left( \begin{array}{c}
\alpha_1(x,t) e^{\frac{i}{h}S_1(x,t)}\\
\alpha_2(x,t) e^{\frac{i}{h}S_2(x,t)}\\
\alpha_3(x,t) e^{\frac{i}{h}S_3(x,t)}\\
\alpha_4(x,t) e^{\frac{i}{h}S_4(x,t)}
\end{array}
\right)
\end{align*}
for the Schr\"{o}dinger equation $H\Psi= i h \frac{\partial}{\partial t} \Psi$.
The real and imaginary parts of the Schr\"odinger equation are straightforward to compute. Since $W(x)$ and $B^{(0)}(x)$ are diagonal, we get four decoupled equations at zeroth order in $h$,
\begin{align*}
-\partial_t S_i=& -2B_{ii}^{(0)} \cos\left(2\partial_x S_i\right)+ W_{ii}(x).
\end{align*}
From the zeroth-order equations, we can define classical Hamiltonians associated with each polariton mode
\begin{align*}
H_1 &= g' \left( \sqrt{1-x}+ \sqrt{1+x}\right)- J \sqrt{1-x^2} \cos 2\varphi_1, \nonumber \\
H_2 &= g' \left(- \sqrt{1-x}+ \sqrt{1+x}\right)- J \sqrt{1-x^2} \cos2\varphi_2, \nonumber \\
H_3 &= g' \left( \sqrt{1-x}- \sqrt{1+x}\right)- J \sqrt{1-x^2} \cos2\varphi_3, \nonumber \\
H_4 &= g' \left( -\sqrt{1-x}- \sqrt{1+x}\right)- J \sqrt{1-x^2} \cos2\varphi_4,
\end{align*}
where at the classical level in each mode  $\varphi_i=\frac{\partial S_i^{(0)}}{\partial x}$ and $x$ are canonical variables such that their Poisson bracket is $\{x,\varphi\}=1$. The four phase variables introduced above emerge from the original angle variables associated with qubit degrees of freedom; however, the angles defining qubit states (say in the Bloch representation) are coupled to each other and have enormous quantum fluctuations, while the emergent angle variables in the present polariton basis are decoupled from each other in the classical limit and have exponentially small (in system size) fluctuations in the strong coupling (large $g$) limit.
We introduce the classical velocity,
\begin{align} \label{eq:cl_velocity}
v_i= \frac{\partial H_i}{\partial \varphi_i}= 4B^{(0)}_{ii}(x) \sin2\varphi_i \ .
\end{align}
For each Hamiltonian, one can define the allowed region for classical solution bound by two turning points where the velocity vanishes at $2\varphi_i=0$ and $\pi$  with $V_i^{l}$ and $V_i^{h}$ :
\begin{align} \label{eq:V_pm}
V_1^{l(h)}(x) &= g' \left( \sqrt{1-x}+ \sqrt{1+x}\right)\pm J \sqrt{1-x^2} \nonumber \\
V_2^{l(h)}(x) &= g' \left(- \sqrt{1-x}+ \sqrt{1+x}\right)\pm J \sqrt{1-x^2} \nonumber \\
V_3^{l(h)}(x) &= g' \left( \sqrt{1-x}- \sqrt{1+x}\right)\pm J \sqrt{1-x^2}\nonumber \\
V_4^{l(h)}(x) &= g' \left( -\sqrt{1-x}- \sqrt{1+x}\right)\pm J \sqrt{1-x^2}
\end{align}
and the classical motion is constrained to be within the hard barriers defined by the above expressions.
Figure~\ref{fig:spec_WKB} illustrates the allowed region for each polariton mode. It is helpful to view the JCDM in this basis as a quantum particle confined inside the classically allowed region. In fact, the bounded regions in Fig.~\ref{fig:nodal_structure} coincide with the classical potentials in Fig.~\ref{fig:spec_WKB} (see also an illustration of wave-functions in Fig.~\ref{fig:wf_ll}). Note that  in the weakly interacting limit (Fig.~\ref{fig:spec_WKB}(a)), all bands overlap and there is no gap (forbidden region) in the middle (i.e.~no localization) while a forbidden region appears in the strong coupling (Fig.~\ref{fig:spec_WKB}(b)).

We look for the stationary solution in form of $\psi_{i}=  \alpha_i(x)  e^{-\frac{i}{h} \epsilon t}e^{\frac{i}{h} S_i(x)}$ where $\psi_i$ is the $i$-th polariton component of $\Psi(x,t)$. Note that $\epsilon=E/N$ is the normalized energy where the original  eigenvalues of the Hamiltonian is denoted by $E$. In the parameter regime $J/g' \leq 2-\sqrt{2}$, bands do not overlap and the classical solutions in terms of energy $\epsilon$ can be found easily (see Fig.~\ref{fig:spec_WKB}(b)):
\begin{enumerate}[label= (\roman*)]
\item $g'\sqrt{2} <\epsilon< (2g'+J)$

Only the first polariton component is allowed.
\item $-g'\sqrt{2} <\epsilon<-J$ and $J<\epsilon<g'\sqrt{2}$

Both second and third polariton  components are allowed. However, only one component is allowed for each value of $x$. In other words, they are not simultaneously non-zero over the same range of $x$.
\item $|\epsilon|<J$

Both second and third polariton  components are allowed. There exists a common range of $x$ in which both components are non-zero.
\item $- (2g'+J) <\epsilon<-g'\sqrt{2}$

 Only the fourth polariton component is allowed.
\end{enumerate}
The WKB solution for the $i-$th polariton mode in the classically allowed regions is
\begin{align}\label{eq:allowed}
\varphi_i = \partial_x S_i^{(0)}=\frac{1}{2}  \text{Arccos}\left(\frac{\epsilon-W_{ii}(x)}{-2B_{ii}^{(0)}(x)} \right),
\end{align}
and in the forbidden region is
\begin{align} \label{eq:forbid}
\phi_i = \partial_x Q_i^{(0)} &= \frac{1}{2} \text{cosh}^{-1}\left( \frac{\epsilon-W_{ii}(x)}{-2B_{ii}^{(0)}(x)} \right), \end{align}
where we define $ S_i(x)=i Q_i(x)$ to get an exponentially growing/decaying wave-function.
We note that the solution is oscillating in a classically allowed region whereas it is exponentially decaying in a forbidden region.

The first-order correction mixes different modes. For a generic case, some modes are forbidden and some modes are allowed, so one can write
\begin{align} \label{eq:WKB_1st}
\partial_x S_i^{(1)} &= \frac{ \sum_{j} B^{(1)}_{ij}(x) \alpha_j(x) e^{\frac{i}{h}(S_j-S_i)} \text{cos}\left(2\partial_x S_j^{(0)}(x)\right)}{{2B^{(0)}_{ii}(x)}\alpha_i(x)\ \text{sin}\left(2\partial_x S_i^{(0)}(x)\right)}
\end{align}
It is important to keep in mind that for the classically forbidden solutions $S^{(0)}$ is pure imaginary and sinusoidal functions must be replaced by the hyperbolic functions.


\section{\label{sec:quant} Boundary matching and quantization rules}
\begin{figure}
\includegraphics[scale=0.78]{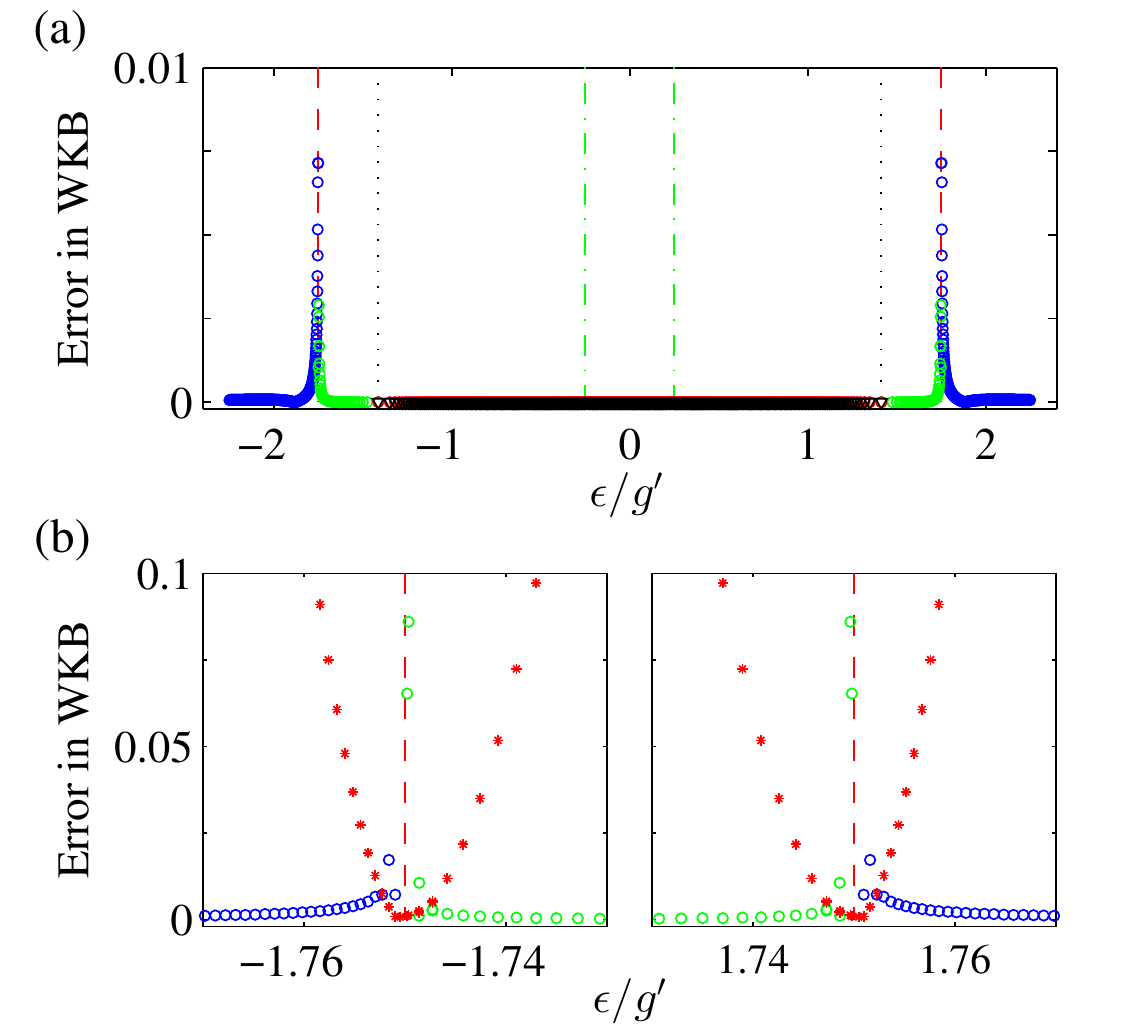}
\caption{\label{fig:BS_quant} (Color online) (a) Error in the Bohr-Sommerfeld quantization rules over the entire quantum spectrum. Different symbols represent different bands: blue circles are Eq.~(\ref{eq:main_BS1_ll}) and its analog for the upper-upper band, green circles is Eq.~(\ref{eq:main_BS2_ll}) and its analog for the upper-upper band, and red and black triangles are middle bands. The vertical lines represent the following energy scales: red dashed, $\epsilon=\pm (2g'-J)$, black dotted, $\epsilon=\pm g'\sqrt{2}$, and green dash-dotted, $\epsilon=\pm J$. (b) Zoomed figure around  critical energies $\epsilon_{c,4}=-2g'+J$ (left) and $\epsilon_{c,1}=2g'-J$ (right). The red stars are the modified quantization rules at the critical level given by Eq.~(\ref{eq:main_BS_cr1}) and its analog for the upper-upper band, the other symbols are the same as in (a). Here $J/g'=1/4$ and $N=400$.}
\end{figure}

In this section, we obtain quantization rules for the entire energy spectrum in the strong coupling regime. This is to illustrate the usefulness of the semiclassical polariton band picture to describe the quantum dynamics in terms of classical orbits. For any given energy, one component is large and others are exponentially small; except for the middle band where both second and third components are non-zero. As Fig.~\ref{fig:BS_quant} shows, the quantization rules agree with the quantum spectrum in all regions as long as bands do not overlap or $J/g'< 2-\sqrt{2}$. A detailed discussion of the validity of the WKB approximation is provided in Appendix~\ref{sec:valid}. In short, the classical motion in polariton bands is valid for a wide range of photon numbers up to such small numbers as $N=6$.

In order to derive the quantization rules, we write a standard connection formula at each boundary between classically allowed and forbidden regions; that is to match the WKB wave-functions with the exact solutions in the neighborhood of each boundary (derived in Appendix~\ref{sec:exact_bc}).
For the remainder of this section, we write down WKB wave-functions for each band and show that they completely match with the exact eigenstates as a result of diagonalizing the Hamiltonian. We derive the quantization rules as a function of energy and check that they are consistent with the quantum spectrum.
We shall only show the resulting quantization rules for the lower-lower polariton mode in the following, the derivation details can be found in Appendix~\ref{sec:der_ll}. The same process can be carried out for other modes which leads to quite similar quantization conditions (see Appendix~\ref{sec:der_others}).

\begin{figure}
\centering
\hspace{-1.1cm}
\includegraphics[scale=0.8]{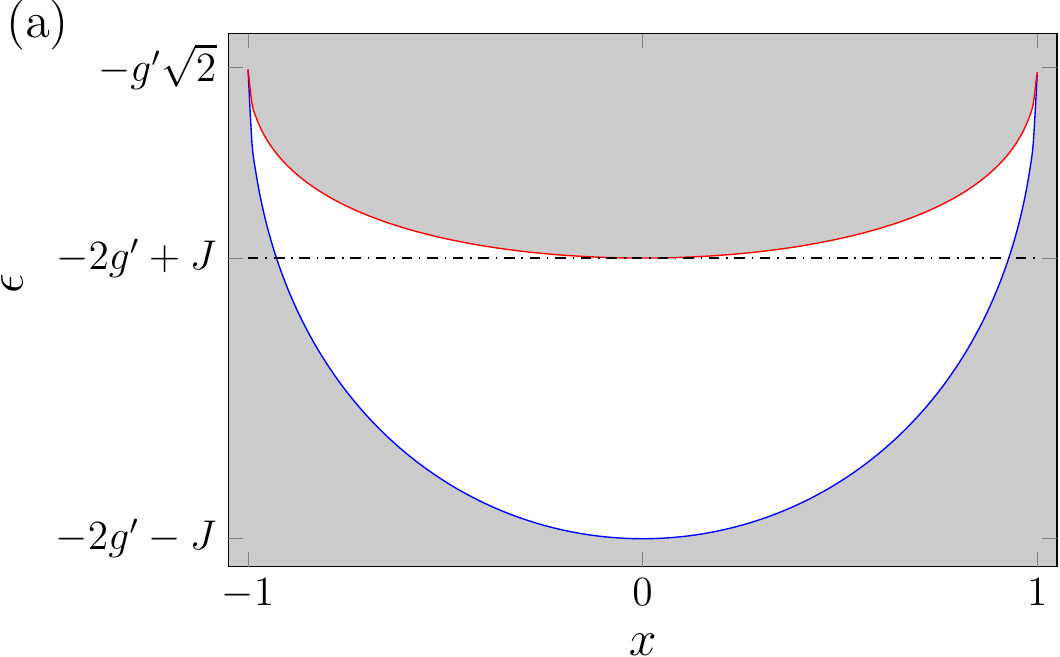}
\includegraphics[scale=0.81]{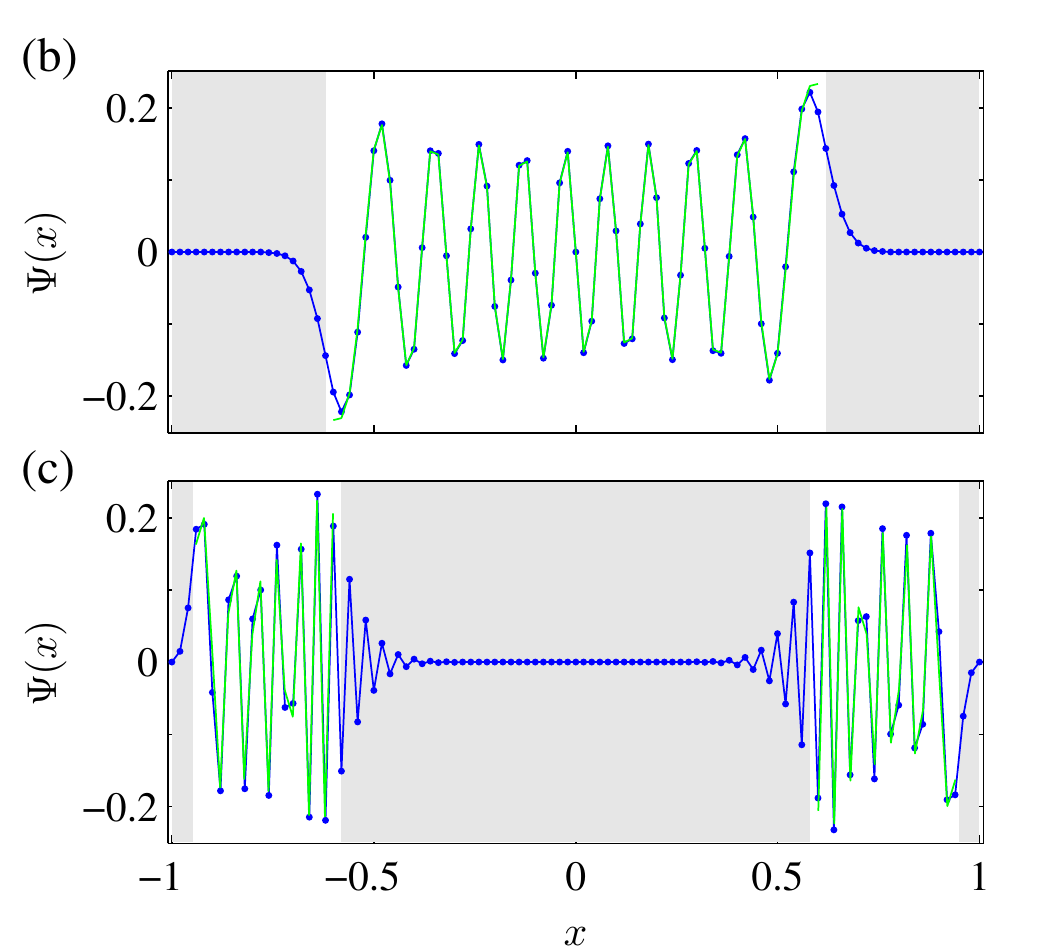}
\caption{\label{fig:wf_ll} (Color online) (a) Classically allowed region for the lower-lower polariton band. It is divided into the localized and delocalized regions. The critical energy $\epsilon_{c,4}=-2g'+J$ (dash-dotted line) separates these two regions. (b) Delocalized state $\epsilon < \epsilon_{c,4}$, and (c) localized state $\epsilon > \epsilon_{c,4}$. The blue is exact diagonalization results and green is the WKB wave-function. Here $J/g'=1/4$ and $N=100$.}
\end{figure}

The lower-lower polariton band is defined for energies $-J-2g'\leq \epsilon\leq -g'\sqrt{2}$.  This band is similar to BECDW with repulsive interaction.
As shown in Fig.~\ref{fig:wf_ll}(a), there are three regions in this band: (i) $\epsilon<J-2g'$, where the delocalized wave-functions center around $x=0$ (Fig.~\ref{fig:wf_ll}(b)); (ii) $\epsilon>J-2g'$, where the localized wave-functions form, in which the probability density is maximum near the boundary points $x=\pm 1$ separated by a barrier (classically forbidden region) in the middle (Fig.~\ref{fig:wf_ll}(c)); (iii) the critical region close to the bifurcation point $\epsilon_{c,4}=J-2g'$ where the Schr\"{o}dinger equation will not be linear and as a result the quantization rule will be different from the usual Bohr-Sommerfeld formulas. Let us now discuss the quantization condition in these three regions one by one.

(i) {\it Delocalized states} ($\epsilon< J-2g'$):
The particle is confined in a two-sided potential and the probability density is large close to the center ($x=0$).
Given that the classically allowed region is $|x| <z_l$ where $\pm z_l$ are turning points $\epsilon=V_4^l(x=\pm z_l)$,
the quantization condition in this case is derived to be
\begin{align} \label{eq:main_BS1_ll}
\frac{1}{h} \Delta S_4(\epsilon) = \left( n+\frac{1}{2}\right) \pi
\end{align}
where $\Delta S_4= \int_{-z_l}^{z_l} \varphi_4(x)$ is the classical WKB phase and $\varphi_4(x)$ is the canonical momentum in the fourth band given by Eq.~(\ref{eq:allowed}).
As expected, this result is similar to the familiar Bohr-Sommerfeld quantization for a particle confined in a potential well~\cite{Ballentine}.
Note that other modes contribute through Eq.~(\ref{eq:forbid}) which are exponentially small and are hence neglected.

(ii) {\it Localized states} ($J-2g'<\epsilon<-g'\sqrt{2}$):
The potential profile consists of two isolated classically allowed regions close to $x=\pm1$. Hence, there are two sets of turning points $z_l$ and $z_h$:
One is bouncing off the $\varphi_4=0$ potential, which is similar to the previous case $\epsilon=V_4^l(x=z_l)$,
and the other is bouncing off the $2\varphi_4=\pi$ potential, which is the solution of $\epsilon=V_4^h(x=z_h)$.
Provided that the classically allowed region is $z_h <|x| <z_l$, the quantization rule is found to be
\begin{align}  \label{eq:main_BS2_ll}
\frac{1}{h} \Delta S_4(\epsilon) =  n \pi \pm \frac{1}{4} e^{-\frac{1}{h} \Delta Q_4(\epsilon)}\ .
\end{align}
where $\Delta S_4$ is to be integrated over the classically allowed region and $\Delta Q_4$ is the WKB tunneling amplitude (see Appendix~\ref{sec:der_ll} for explicit expressions). Note that this condition is similar to the usual Bohr-Sommerfeld except for the tunneling term.
The important message here is that the tunneling rate is related to the energy splitting in the energy spectrum through
\begin{align}
\delta \epsilon_{\text{(split)}} = \frac{1}{2} e^{-\frac{1}{h} \Delta Q_4(\epsilon)} \left|\frac{d\Delta S_4}{d\epsilon} \right|^{-1}.
\end{align}
which is consistent with the quantum spectrum; i.e.~for energies in the range $|\epsilon|<2g'-J$, there are pairs of eigenvalues that come exponentially close together as the system size $N=1/h$ is increased. In the classical limit $N\to \infty$, tunneling is suppressed and we get pairs of  fully localized degenerate states. Figure~\ref{fig:wf_ll}  illustrates a few typical examples that WKB solutions match well with the exact wave-functions.

(iii) {\it Critical region} ($\epsilon\sim J-2g'$):
Consider a small deviation $\lambda \ll 1$ from the critical energy $\epsilon=J-2g'+\lambda/N$. The quantization formula is derived to be
\begin{align} \label{eq:main_BS_cr1}
\text{arg} \left[e^{-\frac{2i}{h}\Delta S_4} \frac{\sqrt{2\pi}}{\Gamma(1/2-i\chi)}- e^{\pi\chi/2} \right]= n\pi + \pi/2
\end{align}
where $\Delta S_4$ is calculated over the classical region, $\Gamma(x)$ denotes the Gamma function and $\chi=\lambda/\sqrt{2J(g'-2J)}$. Notice that the quantization condition in this case is more complicated than the standard Bohr-Sommerfeld quantization (Eq.~(\ref{eq:main_BS1_ll})) and as we show later, this leads to smaller level spacing and a larger density of states near the critical levels compared to regular regions in the spectrum. This new form of the quantization rule close to energy levels dividing localized and delocalized states was originally found for the BECDW problem in Ref.~\cite{Keeling}. As shown therein, an immediate implication of this formula is that the quantum break times, at which the classical and quantum solutions start to differ, near critical regions grow logarithmically with $N$ instead of algebraically. This conclusion also applies to our case at two critical regions of the spectrum for upper-upper and lower-lower bands.


Figure~\ref{fig:BS_quant} summarizes the main results of this section. We plug in the quantum eigenvalues of the Hamltonian to the  quantization condition for the appropriate polariton band determined by the location of the eigenvalue and subtract the integer multiple of $\pi$ from it. The difference is shown as the defect of WKB quantization. It is evident in Fig.~\ref{fig:BS_quant}(a) that the WKB error is quite small except for the critical energy levels $\epsilon_{c}=\pm (2g'-J)$. Close to critical levels, we must use the modified Bohr-Sommerfeld quantization rules,  Eq.~(\ref{eq:main_BS_cr1}), as shown in Fig.~\ref{fig:BS_quant}(b).
In Appendix~\ref{sec:valid}, we investigate the validity of the polariton band picture by introducing a quantum-to-classical correspondence based on the Husimi distribution. Using the phase space distribution to extract the classical aspects of the Hamiltonian eigenstates, we show that the properties associated with polariton bands remain valid up to such small total polariton numbers as $N=6$. We also note that the quantization conditions start to fail as we go to a weak coupling regime where the polariton bands start to overlap and higher order corrections become important. Nonetheless, the quantization conditions stay valid away from the overlapping regions in the spectrum (see Fig.~\ref{fig:quant_dev} in Appendix~\ref{sec:valid}).


\section{\label{sec:local_trans} Localization transition}

Before we discuss the localization transition in the polariton band picture, let us make few remarks about the transition in the classical limit of the JCDM. In Appendix~\ref{sec:cl}, using a coherent-state path integral formalism in the large photon number and large spin limit, we show that the classical dynamics of the JCDM is described in an eight dimensional phase space, two for photons and two for spins per JC site.
The resulting equations of motion are found to be the same as the factorization of Heisenberg equations. The critical value of coupling $g_c$, above which localization occurs, is found to be dependent on initial spin configurations and its minimum value is analytically derived to be $g_c=2 J \sqrt{N}$. This is the same as the quantum mechanical value for $g_c$ found numerically~\cite{Tureci,James} and analytically (as shown below). This result is in contrast with the previous numerical analysis of~\cite{Tureci}, $g_c\approx 2.8 J\sqrt{N}$, where  dynamics in a reduced four dimensional phase space was only studied. In the same reference it was argued that the difference between classical and quantum values of $g_c$ is related to large quantum fluctuations. However, we see that the difference is an artifact of only considering part of the classical phase space, and taking  into account  the full phase space resolves this issue.

\begin{figure}
\hspace{-0.5cm}
\includegraphics[scale=0.8]{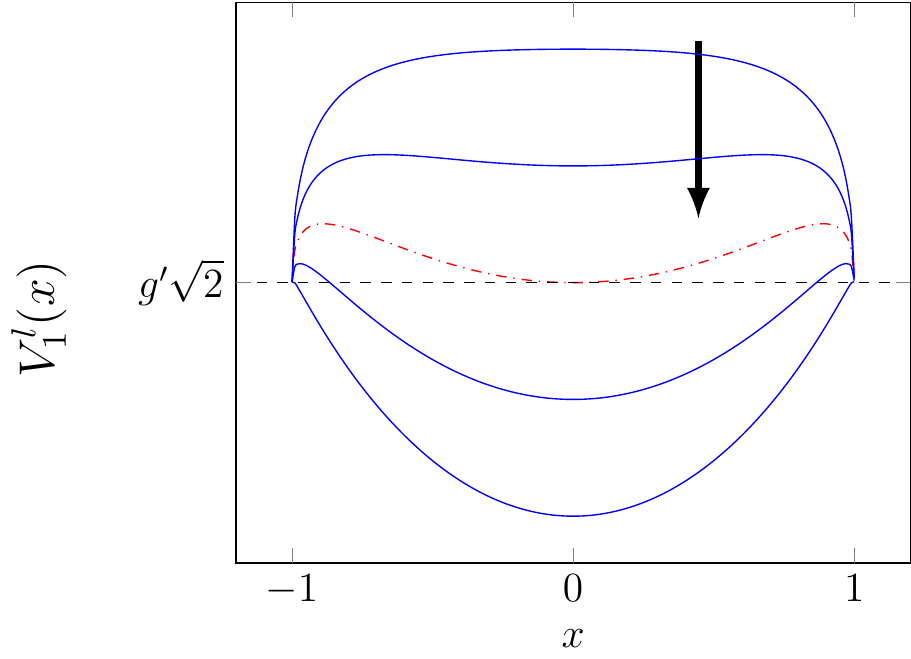}
\caption{\label{fig:V_l1_ev} (Color online) Evolution of the minimum potential curve $V^l_1(x)$ of upper-upper band, Eq.~(\ref{eq:V_pm}), as $J/g'$ is tuned up from top to bottom as shown by the arrow. The maximum potential curve $V^h_4(x)$ of the lower-lower band is just the mirror image of these curves with respect to the horizontal line $\epsilon=0$. The red (dash-dotted) curve for $J/g'= 2-\sqrt{2}$ is the critical point at which the minimum moves from $x=0$ to $x=\pm 1$.}
\end{figure}

\begin{figure}
\includegraphics[scale=0.07]{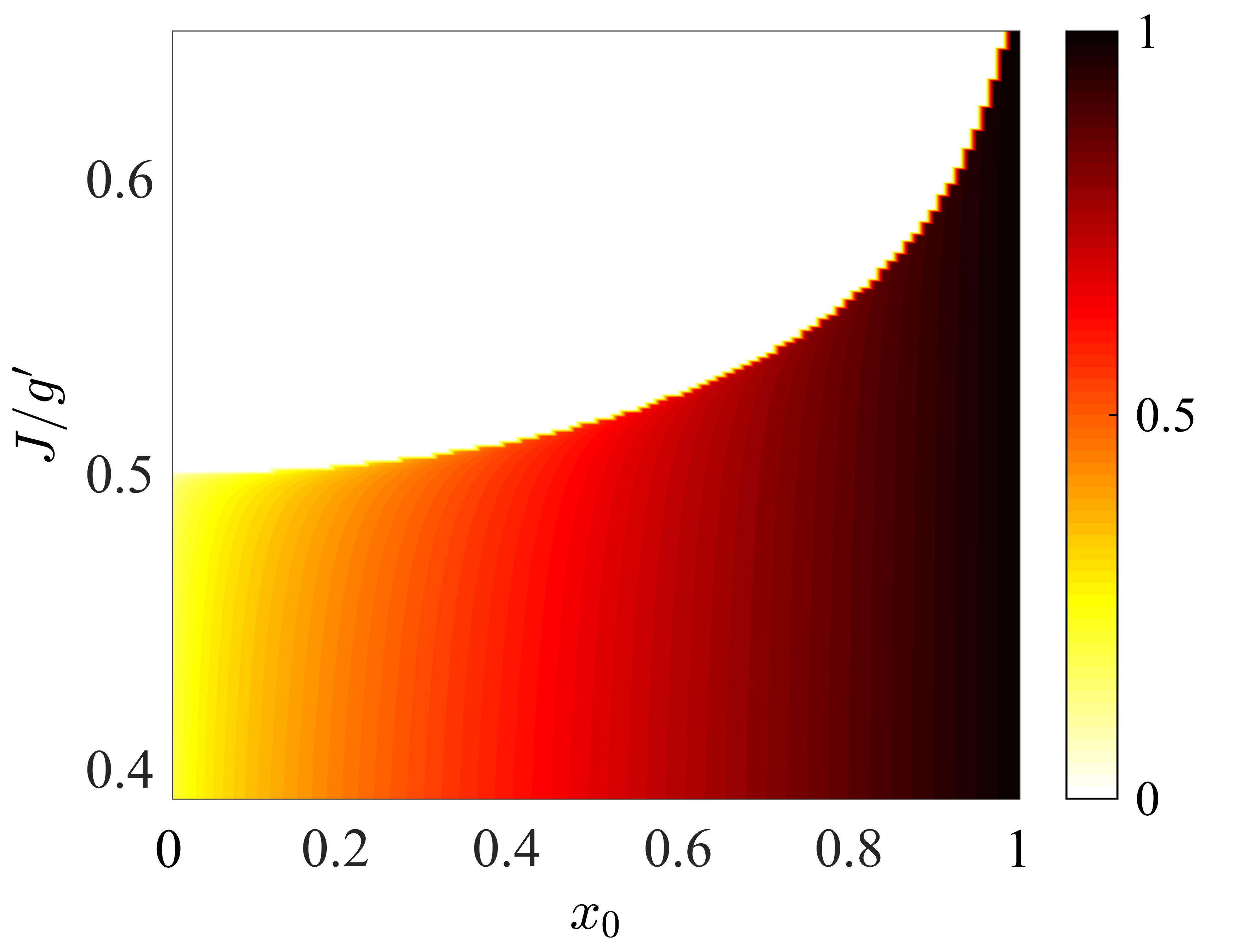}
\includegraphics[scale=0.78]{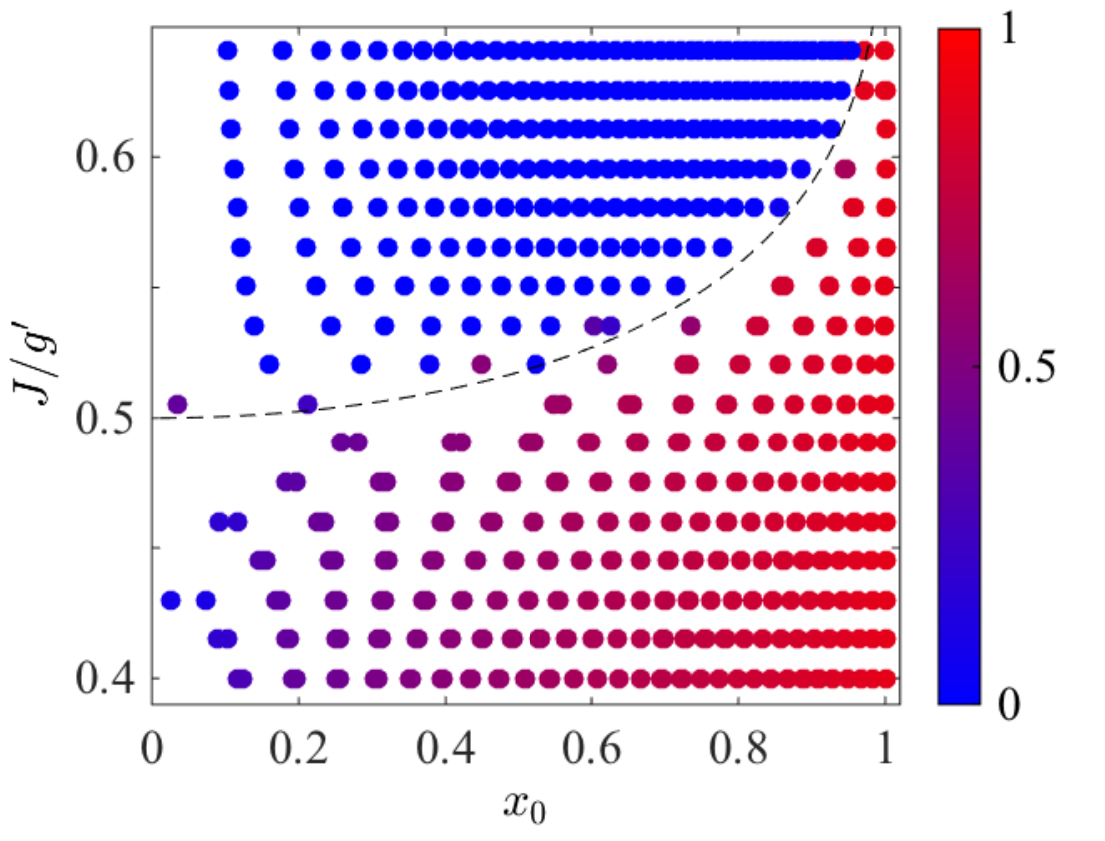}
\caption{\label{fig:loc_comp} (Color online) Top: the classical average imbalance (color code) as a function of $J/g'$ for various initial conditions ($\epsilon=V_1^{l}(x_0)$). Bottom: the expectation value of the imbalance (Eq.~(\ref{eq:imbalance})) as the color code for the eigenstates of the Hamiltonian close to the touching point between upper-upper and middle bands. The dashed line is the classical phase boundary given by Eq.~(\ref{eq:cl_phase_boundary}).}
\end{figure}

Now, let us turn to the polariton band picture. As we have seen in the previous section, the localized states appear at energies close to where the upper-upper or the lower-lower polariton bands meet the middle bands. As Fig.~(\ref{fig:V_l1_ev}) suggests, the localization can be understood by studying the changes in the curvature of the $V_1^{l}(x)$ or $V_4^{h}(x)$ functions. Here, we investigate the localization transition for the upper-upper polariton band in this section. The result is exactly the same for the lower-lower band.
For the purely classical solution, where the particle is bound to be in the allowed regions, one can easily obtain the period of motion for a given trajectory $T=\oint dx/v_1$ and the average imbalance $\langle x\rangle=(1/T)\oint  xdx/v_1$  where the velocity $v_1$ is given by Eq.~(\ref{eq:cl_velocity}) in terms of the initial position $x_0$ where  $\epsilon=V_1^{l}(x_0)$.
Figure~\ref{fig:loc_comp}(top) shows the average imbalance for various values of the $J/g'$. The phase boundary is given by
\begin{align} \label{eq:cl_phase_boundary}
\frac{g}{J \sqrt{2N}}= \left(\frac{1-\sqrt{1-x_0^2}}{2x_0^2}\right)^{-1/2}
\end{align}
Interestingly, the critical value for $x_0 \to 1$ becomes $g_c/J= 2\sqrt{N}$ precisely the same as the exact quantum simulations in~\cite{Tureci,James}.
The phase boundary in Eq.~(\ref{eq:cl_phase_boundary}) above is derived as follows: As we see in Fig.~\ref{fig:V_l1_ev}, for coupling strengths in the range $1/2<J/g'<1/\sqrt{2}$, there always exists a local maximum at  $x_m(J/g')=[(4(J/g')^2-1)/4(J/g')^4]^{1/2}$. This gives rise to a barrier for the bound states confined within $x_m\leq x\leq 1$. Therefore, the initial condition determines whether the particle is localized or not by being greater or smaller than $x_m$; hence, $x_0=x_m(J/g')$ gives the expression for the boundary between localized and delocalized states in Eq.~(\ref{eq:cl_phase_boundary}). Moreover, when $J/g' \leq 1/2$ the maximum occurs at $x=0$ and any initial condition with energies $g'\sqrt{2}<\epsilon<-J+2g'$ is classically localized.

\begin{figure}
\includegraphics[scale=0.7]{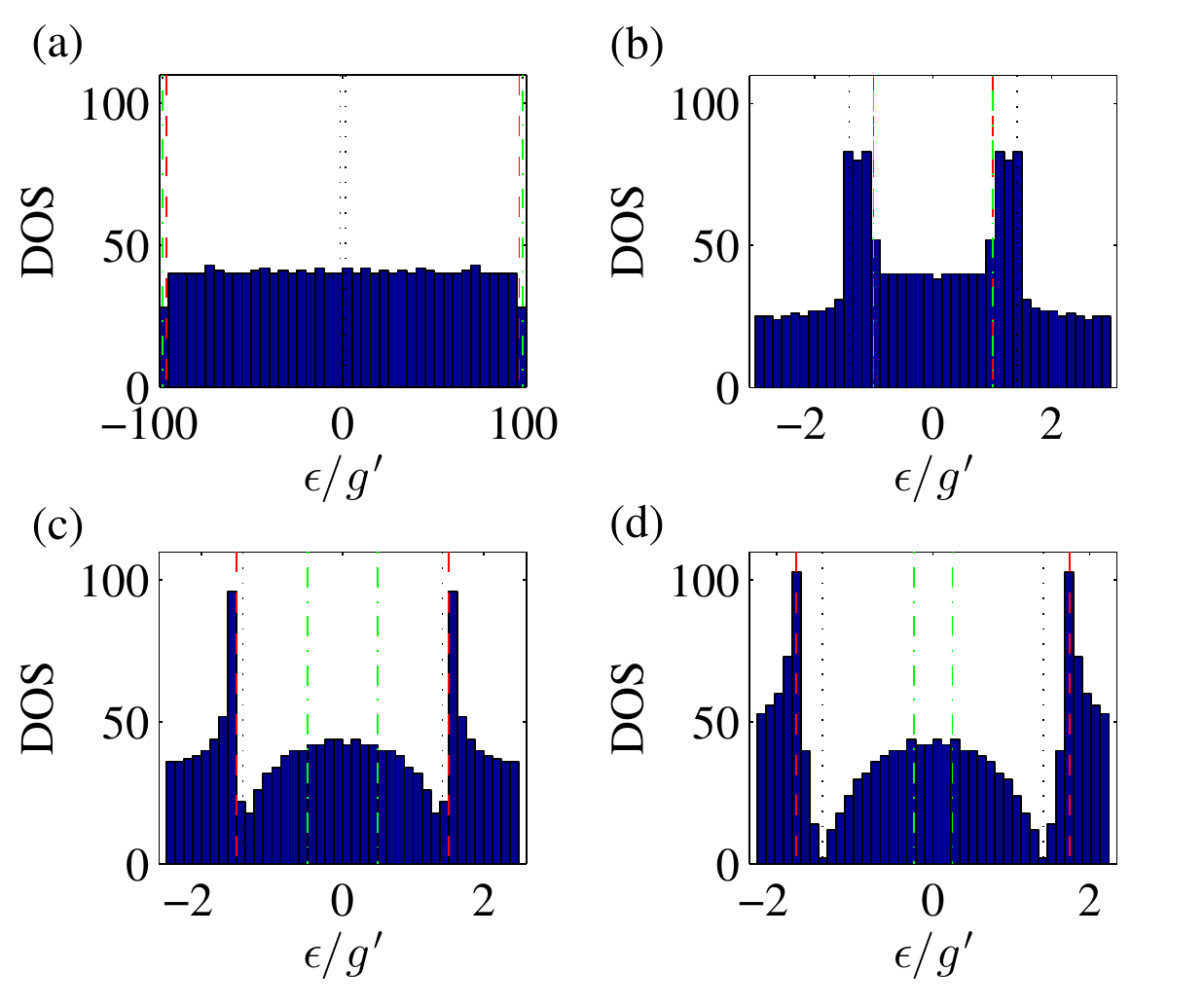}
\caption{\label{fig:DOS} (Color online) Density of states for different values of coupling $g$. From (a) to (d) $J/g'=100,\ 1,\ 0.5\ ,\text{and}\ 0.25$. The vertical lines are explained in the caption of Fig.~\ref{fig:BS_quant}}
\end{figure}

Let us compare the above result with the quantum spectrum. At the quantum level, we can study the localization of eigenstates after adding an infinitesimal symmetry breaking term
\begin{align*}
H_{\text{imb}}= \varepsilon (n_L-n_R)
\end{align*}
where $n_{L(R)}$ is polariton number on the left (right) site.
In order to compare the quantum results with their classical counterparts, we define an equivalent classical trajectory for a given eigenstate of energy $\epsilon_n$ with initial position $\epsilon_n=V_1^l(x_{n0})$. A comparison with the classical phase boundary is illustrated in Fig.~\ref{fig:loc_comp}(bottom). For any value of $J/g'$, each eigenstate is represented by a point whose horizontal position is $x_{n0}$ and whose color reflects the quantum expectation value of imbalance defined in Eq.~(\ref{eq:imbalance}).
 In this figure, we draw the classical phase boundary using the criterion whether $x_{n0}$ is greater or smaller than $x_m$.
The agreement between the quantum results and classical picture in polariton bands is quite remarkable.

A signature of the delocalization-localization transition may also be illustrated as changes in the density of states (DOS), (see Fig.~\ref{fig:DOS}). In the weakly interacting limit $g\ll J$, the system is linear and DOS is completely uniform (Fig.~\ref{fig:DOS}(a)). As the coupling $g$ is increased, anharmonicities emerge in the DOS  and eigenvalues start to accumulate near two critical levels  $\epsilon_{c,1}=2g'-J$ and $\epsilon_{c,4}=-2g'+J$ (Fig.~\ref{fig:DOS}(b)). As we continue increasing the ratio $g/J$ (Fig.~\ref{fig:DOS}(c)), more eigenvalues are depleted at two regions near touching points of the first and fourth bands with the middle bands, where $\epsilon=\pm g'\sqrt{2}$, and eventually a gap opens in the DOS (Fig.~\ref{fig:DOS}(d)).
Using our WKB quantization conditions near the critical levels~Eq.~(\ref{eq:main_BS_cr1}), it is easy to see that the DOS in this region is indeed logarithmically diverging in the system size (total polariton number).
The level spacing can be generically computed from the quantization condition through
\begin{align*}
\delta \epsilon = \epsilon_{n}-\epsilon_{n-1}\simeq \frac{\pi}{N} \left(\frac{d \Delta S}{d\epsilon}\right)^{-1}\ .
\end{align*}
Away from the critical levels, the spectrum is harmonic and the density of states is uniform $D(\epsilon)\sim 1/\delta \epsilon \sim N$. Near the critical points $\epsilon=\epsilon_{c,i}+\lambda/N$, we have ${d \Delta S}/{d\epsilon} \sim S^{(1)} \sim \log N$ (see Appendix~\ref{sec:der_ll} for details) and the DOS scales as $D(\epsilon)\sim N \log N$ that is greater than the regular DOS by a factor of $\log N$.
The fact that $\delta \epsilon \sim 1/N \log N$ also implies that the quantum break times, the time scale at which the quantum and classical dynamics start to differ~\cite{Keeling}, scale logarithmically with $N$.
This result is also consistent with the $N$-th order degenerate perturbation theory for the energy splitting, discussed in~\cite{Tureci}, which gives a long-time frequency of $\Delta \sim J (J/g)^{N-1}$ for quantum oscillations, the characteristic time of which scales logarithmically with $N$.

The anharmoncity observed in the DOS is translated into a chaotic behavior at the classical level (see Fig.~\ref{fig:classical_chaos} in Appendix~\ref{sec:chaos}). At the quantum level, the signatures of quantum chaos can be characterized further in terms of level statistics and the Brody parameter~\cite{Brody_orig, Manfredi}. This quantum-to-classical correspondence of the JCDM will be presented elsewhere. A recent work~\cite{Brody_trimer} studies this behavior for a BEC system where one must consider at least three sites (trimer) to make the system non-integrable at the classical level. Remarkably for JC systems, a two-site (dimer) model would be sufficient for classical non-integrability due to extra degrees of freedom added by qubits.


\section{\label{sec:conc} Discussion}

In summary, we have developed a new technique to study the delocalization-localization transition in the Jaynes-Cummings dimer model.
In this method, we introduced an effective one-dimensional tight-binding model (polariton basis) for this system and treat each site  exactly while we treated the intersite polariton hopping semiclassically. In order to construct the polariton basis, we wrote the Fock space representation for the JCDM where the Schr\"odinger equation is mapped onto a discrete equation. The continuum approximation can be made for large system sizes (total polariton numbers) and the JCDM can be viewed as a moving particle confined in a potential well details of which are determined by the cavity-qubit interactions.
The localization in this picture is equivalent to the existence of localized wave-functions near the band edges.
We also presented another view for the localization transition in terms of a gap opening in the density of states. We have found that the critical coupling for the transition is the same in both quantum mechanical and fully classical calculations, which resolves the issue raised by Ref.~\cite{Tureci}.
Furthermore, we employed a WKB approximation in this picture to derive a set of Bohr-Sommerfeld quantization conditions for the entire energy spectrum in the strong coupling regime. We showed that the quantization rules match with the eigenvalues of the Hamiltonian.

The presented method is quite general and can be applied to a variety of models including interactions between spins and bosonic fields. The usual classical limit of these models in the large spin and large photon number limit could involve some difficulties if one wants to include higher order fluctuations due to nonlinear terms for a spin path integral (see Appendix~\ref{sec:cl}). Remarkably, our method gets around this difficulty by expanding around a different classical limit, namely polariton bands. The polariton band picture remains valid even at small photon numbers.
In particular, this method can be generalized to other interesting systems where the exact treatment of spin-$1/2$ is essential, such as the Rabi model, the driven-dissipative JC model, the multi-mode JC model, the Dicke model and so on.
Moreover, the equations of motion can be used to study quench dynamics of these models. A recent work~\cite{Biroli} has developed a similar semiclassical approach to study the dynamical transitions due to quench dynamics in various models including the Dicke model.

Our focus in this article has been the dynamical properties of the JCDM as a closed system.
It is worth noting that the experimental setup~\cite{James} is an open system where photons can escape from the cavities and the qubits may relax and decohere over time. The experimental finding for the critical coupling is however different from the theoretical treatments for the closed system. This difference still remains an open question.
In order to accommodate the dissipation due to environment, it is straightforward to generalize the classical limit of the JCDM for open systems using various non-equilibrium path-integral formalisms originally developed by Kadanoff and Baym~\cite{Kadanoffbook} or Keldysh~\cite{Kamenevbook}. Along these lines, Ref.~\cite{Mandt} has recently studied the dissipative JCDM by mapping the dynamics onto a set of Fokker-Planck equations in terms of photon and spin coherent-states in the positive-$P$ representation. They have shown that this method can qualitatively reproduce the experimental measurements. They also considered a driven-dissipative JCDM where it was argued that the suppression of the steady-state tunneling current is a manifestation of the localization transition.
It is important to note that the calculations of Ref.~\cite{Mandt} have been done for spin coherent states, which can only be justified for large spins. As we have seen here, treating qubits as spin-1/2 is quite crucial and only this way can one obtain the full picture. Hence, one interesting future direction is how to generalize the polariton bands to capture the dynamics of an open JCDM.

\acknowledgments
This work was initiated by Darius Sadri and has benefited immensely from his ideas. Darius tragically passed away before the work was completed.
The authors would like to acknowledge insightful discussions with A. A. Houck, Y. Liu, S. Mandt, J. Raftery, W. E. Shanks, S. J. Srinivasan, H. E. T\"ureci, and  D. L. Underwood.
In particular, the authors would like to thank D. A. Huse and M. Schir\'o for valuable discussions and helpful comments on the manuscript.

\begin{appendix}
\beginsupplement

\renewcommand\theequation{A\arabic{equation}}
\section{\label{sec:cl} Classical dynamics and the localization transition}

In this appendix, we use the standard coherent-state path integral formulation of quantum mechanics~\cite{fradkinbook,Altlandbook} and write down a classical action for the JCDM. We derive the classical equations of motion and obtain the localization transition after making an analogy with a driven pendulum. In Appendix~\ref{sec:chaos}, we visualize the time-evolution in the phase space using the Poincar\'e sections and show that the nonlinear dynamics governed by the classical  equations leads to chaos.
The semiclassical picture derived in~\cite{Tureci} is based on fully factorizing expectation values of products of the photon and the qubit operators in Heisenberg equations of motion. Although this type of factorization is quite customary in the study of JC-based models, there is no physical intuition to what the underlying assumptions are and it is not clear how one can systematically improve this process by including new terms.

\begin{figure}
\centering
\includegraphics[scale=0.55]{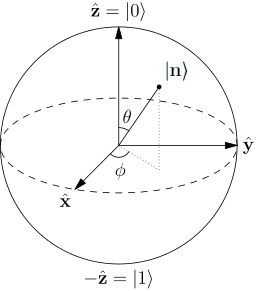}
\caption{\label{fig:topSpin} The Bloch representation of spin coherent-state.}
\end{figure}

The dynamics of the JCDM with spin-$S$ can be described in terms of the following action in the real-time coherent-state path integral formalism
\begin{align*}
{\cal S}=\sum_{s=L,R} {\cal S}^s_{JC}[\boldsymbol{n}_s,\bar{\psi}_s,\psi_s] + {\cal S}_t[\psi,\bar{\psi}]
\end{align*}
where the tunneling term becomes
\begin{align} \label{eq:tunnel_action}
{\cal S}_t[\psi_{L/R},\bar{\psi}_{L/R}]=J \int dt \left(\bar{\psi}_R \psi_L + \bar{\psi}_L \psi_R\right)
\end{align}
and the JC action is
\begin{align} \label{eq:JC_action}
{\cal S}_{JC} [\textbf{n} ,\bar{\psi},\psi] =& {\cal S}_B[\textbf{n}, \bar{\psi},\psi] - \int dt  \left( S \textbf{n} \cdot\textbf{B} + \nu_c \bar{\psi}\psi  \right),
\end{align}
in which
\begin{align} \label{eq:Bfield}
\textbf{B}(t)= \nu_q \hat{\textbf{z}} + i g  (\psi- \bar{\psi} ) \hat{\textbf{y}} + g (\psi+ \bar{\psi} ) \hat{\textbf{x}} ,
\end{align}
here $\bar{\psi}$ and $\psi$ are complex fields representing the photon fields and the spin coherent-state is defined in terms of the Bloch state $|\textbf{n}\rangle$ defined in the coordinate system $(\hat{\textbf{x}},\hat{\textbf{y}},\hat{\textbf{z}})$ such that
\begin{align*}
\langle \textbf{n} | \hat{\boldsymbol{S}} | \textbf{n} \rangle = S \textbf{n}
\end{align*}
where $\textbf{n}$ is a unit vector (see Fig.~\ref{fig:topSpin}) spanning the Bloch sphere and $\hat{\boldsymbol{S}}$ on the LHS  is the spin-$S$ operator while $S$ on the RHS  denotes the number $S$. For qubits, we have $S=1/2$. The first term in the action, Eq.~(\ref{eq:JC_action}), is the Berry phase contribution
\begin{align*}
{\cal S}_B[\textbf{n}, \bar{\psi},\psi] = \int dt {\Big[} \bar{\psi}\partial_t \psi  +  S \langle \textbf{n} | \partial_t | \textbf{n} \rangle {\Big]}.
\end{align*}
It is important to note that the spin part of the above action is always imaginary even in the imaginary-time formalism and this signals the fact that this term is topological. Indeed, this term is equal to the area on the Bloch sphere enclosed by the path traveled by the Bloch vector and can be written as
\begin{align} \label{eq:spin_berry}
 \langle \textbf{n} | \partial_t | \textbf{n}\rangle = \int dt \int_0^1 d\tau\ \textbf{n}(t,\tau)\cdot (\partial_t\textbf{n}(t,\tau) \times\partial_\tau \textbf{n}(t,\tau) )
\end{align}
at which $\tau$ is a parameter to define the area on the Bloch sphere such that
\begin{align*}
\textbf{n}(t,0) = \textbf{n}(t)  \ \ \ \ \ \ \ \ \ \  \textbf{n}(t,1) = \textbf{n}_0
\end{align*}
 here $\textbf{n}_0=\hat{\textbf{z}}$ is the spin quantization direction.
In the limit $N\to \infty$ and $S \to \infty$, dynamics is described by the stationary solutions of the action,
\begin{align*}
\partial_t\textbf{n}_s &= \textbf{B}_s(t)\times\textbf{n}_s
\end{align*}
for the qubit and,
\begin{align*}
i \partial_t \psi_s = \nu_c \psi_s + g S [n_{s,x} - i n_{s,y}] - J  \psi_{\bar{s}}
\end{align*}
for the photon field.
Note that the Heisenberg equations of motion for JCDM Hamiltonian in Eq.~(\ref{eq:JCDM}) are
\begin{align*}
\partial_t \hat{S}_{s,k} &= i [ H, \hat{S}_{s,k}] = (\textbf{B}_s\times \hat{\boldsymbol{S}}_s)_{k} \\
i \partial_t \hat{a}_s &= i [ H, \hat{a}_s]= \nu_c \hat{a}_s + g [\hat{S}_x - i \hat{S}_y]
\end{align*}
where $\textbf{B}$ is defined in Eq.~(\ref{eq:Bfield}) and the additional subscript $k$ denotes the $k$-th component of a vector. The classical equations of motion are equivalent to factorizing approximation of the expectation values of quantum operators; i.e.
\begin{align*}
\partial_t \langle \hat{\boldsymbol{S}} \rangle &\approx \langle \textbf{B} \rangle \times \langle \hat{\boldsymbol{S}} \rangle
\end{align*}
where $\langle ... \rangle$ denotes the expectation value over the initial state and we make the following identifications: $\langle a \rangle = \psi$ and  $\langle \hat{\boldsymbol{S}} \rangle =  S\textbf{n}$.

It is more convenient to represent spin in the spherical coordinate system $\textbf{n}=(\sin\theta \cos\phi,\sin\theta\sin\phi,-\cos\theta)$, as in Fig.~\ref{fig:topSpin}, and to split the cavity fields into their real and imaginary parts $\psi_s=R_s + i I_s$. Using this parametrization, the equations of motion become
\begin{subequations}
 \label{eq:cl_full_dyn}
\begin{align}
\dot{\phi}_s &=  -\nu_q  - 2g (R_s \cos\phi_s - I_s \sin\phi_s) \text{cot}\theta_s \\
\dot{\theta}_s &= 2g (R_s \sin\phi_s + I_s \cos\phi_s)
\end{align}
for the spin and
\begin{align}
 \label{eq:JCDdyn1}\dot{ R}_s &= \nu_c I_s -gS \sin\theta_s\sin\phi_s - J  I_{\bar{s}} \\
\label{eq:JCDdyn2} \dot{I}_s  &=-\nu_c R_s - gS \sin\theta_s\cos\phi_s + J R_{\bar{s}}
\end{align}
\end{subequations}
for the cavity fields, where $s=L,R$ and its opposite $\bar{s}=R, L$.
Here, our focus is the resonant case where $\nu_q=\nu_c$ and the free dynamics of the spin and photon field can be removed in the rotating frame.

\begin{figure}
\includegraphics[scale=0.8]{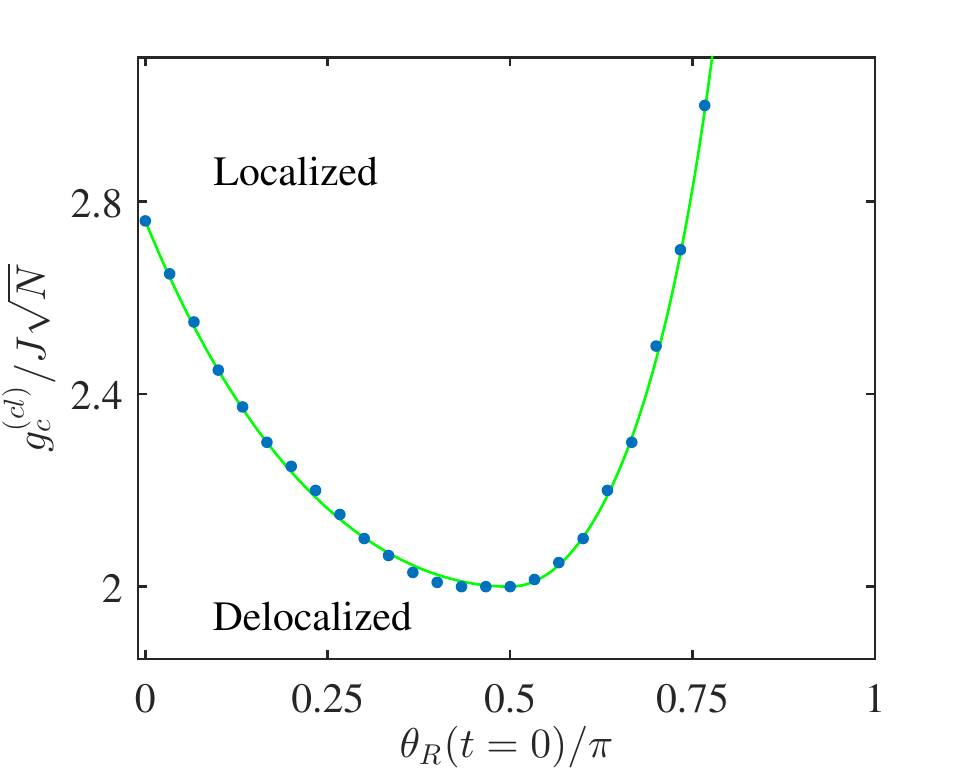}
\caption{\label{fig:cl_phase} (Color online) The classical critical ratio $g/J$ as a function of initial configuration for the spin. Here, the dynamics is given by Eqs.~(\ref{eq:cl_full_dyn}a-d) and the other spin angles are initially fixed at $\theta_L(0)=\phi_L(0)=0$. The green curve is given by Eq.~(\ref{eq:cl_cr}), based on analogy to the pendulum, and the blue circles is the critical boundary obtained numerically after calculating the long-time averaged imbalance. We put $S=1/2$.}
\end{figure}

The classical dynamics of the JCDM takes place in an eight dimensional phase space.
In order to find a lower bound on the critical value of $g/J$ where the delocalization-localization transition occurs, we approach the critical point from the localized side. In the localized phase, we start with all polaritons on the left cavity and we can assume $\psi_L(t)\approx \sqrt{N}$ and $\psi_R(t)\approx 0$ in all times. This is a plausible assumption as long as $N\gg S$.
So, the dynamics in the rotating frame simplifies into
\begin{subequations}
\begin{align}
\label{eq:cl_loc1} \dot{ R}_L &=  -gS \sin\theta_L \sin\phi_L,  \\
\dot{ R}_R &=  -gS \sin\theta_R \sin\phi_R,  \\
\dot{I}_L &= - gS \sin\theta_L\cos\phi_L,  \\
\label{eq:cl_loc} \dot{I}_R &= - gS \sin\theta_R\cos\phi_R + J \sqrt{N},
\end{align}
\end{subequations}
where only the fourth equation couples the two JC islands.
This equation describes a motion subject to an external drive $F=J \sqrt{N}$.
Depending on the ratio $\eta=gS/J\sqrt{N} $, the sign of the RHS may be allowed to change over time or not. These two modes of motion correspond to delocalized and localized phases.
If $\eta < 1$, the RHS of Eq.~(\ref{eq:cl_loc}) always remains positive and the right (empty) cavity absorbs more and more energy. Therefore, excitations can be fully transferred to the empty cavity and this process continues repeatedly. However, if there exists a turning point at which the RHS of Eq.~(\ref{eq:cl_loc}) vanishes (i.e.~$\eta \sin\theta_0\cos\phi_0 =1$),  there exists initial conditions such that the RHS oscillates between positive and negative values. In other words, the motion goes in and out of phase with the driving force and the averaged energy absorbed by the empty cavity becomes zero. In this case, the time-averaged transferred excitations is zero and the system is in the localized phase. Thus, the critical value is defined as the boundary between these two limits, $\eta_c=1$.
So, the lowest value of the coupling ratio $g/J$ to get a localized phase is given by
\begin{align} \label{eq:cl_critical}
\frac{g^{(cl)}_c}{J}= \frac{1}{S} \sqrt{N}\ .
\end{align}
After plugging $S=1/2$, we see that this result is the same as the quantum simulations~\cite{Tureci,James} as well as the WKB picture studied in this article.
However, this is different from the previous numerical analysis on the semiclassical factorized Heisenberg equations of motion~\cite{Tureci} where the critical ratio is found to be $g/J\approx 2 \sqrt{2N}$ that is off by a factor of $\sqrt{2}$ from the exact quantum dynamics. This contrast is because Ref.~\cite{Tureci} only considered the initial condition with both qubits down and ignored the fact that the critical ratio $g/J$ actually depends on the initial states of qubits.
Figure~\ref{fig:cl_phase} illustrates this dependence explicitly. As we see in this plot, the phase boundary in terms of $g/J\sqrt{N}$ starts around $2\sqrt{2}\simeq 2.8$, the same as Ref.~\cite{Tureci}, but reaches a minimum at $2$ for $\theta_R=\pi/2$. Hence, the classical dynamics (i.e. the dynamics  in the limit $N\gg S\gg 1$) implies that the minimum value of $g$ required to obtain self-trapping is given by Eq.~(\ref{eq:cl_critical}).

Few remarks regarding Fig.~\ref{fig:cl_phase} are in order. First, note that our analysis of the critical coupling for arbitrary initial spin configurations is based on the dynamics described by Eqs.~(\ref{eq:cl_loc1}-d) where the initial state of the left qubit $\theta_L$ and $\phi_L$ corresponding to the fully populated JC site does not make any significant change and hence is fixed at $\theta_L(0)=\phi_L(0)=0$. Second, since we are looking for a lower bound on the critical ratio $g/J$, we choose $\phi_R(0)=0$ to maximize the coefficient multiplying $g$ in the first term at the RHS of Eq.~(\ref{eq:cl_loc}).
Given these initial conditions, the dynamics of $I_R$ and $\theta_R$ can be combined into a single equation
\begin{align}
\frac{\dot{\theta}^2_R}{2}= 2 g^2 S (\cos\theta_R-\cos\theta_R(0))+ 2 Jg \sqrt{N} (\theta_R-\theta_R(0))
\end{align}
where the initial conditions are $\theta_R(0)\neq 0$ and $\dot{\theta}_R(0)=0$. The second initial condition means that the qubit has no initial kinetic energy, which is a physical assumption.
The above equation is reminiscent of a pendulum moving under gravity (described by the angle $\theta_R$) and subject to a constant torque $\tau=2Jg \sqrt{N}$.
It is easier to understand the localization transition using the pendulum analogy.
The pendulum may do a full rotation (delocalized) or only oscillate with a small amplitude (localized). In the oscillatory mode the motion is bound by the turning points at which $\dot{\theta}_R=0$. So, the existence of turning points can be used to determine the boundary between localized and delocalized dynamics.
The critical value can then be written as $g^{(cl)}_c/2J\sqrt{N}=1/\sin\theta_0$ where $\theta_0$ is the non-zero solution of
\begin{align}
\label{eq:cl_cr}
\sin \theta_0 + \frac{\cos\theta_0-\cos\theta_R(0)}{\theta_0-\theta_R(0)}=0.
\end{align}
Notice that the critical ratio for $\theta_R(0)\geq \pi/2$ is simplified into $g^{(cl)}_c/2J\sqrt{N}=1/\sin\theta_R(0)$ as expected from setting the RHS of Eq.~(\ref{eq:cl_loc}) zero. However, for $\theta_R(0)< \pi/2$, Eq.~(\ref{eq:cl_cr}) gives a lower critical value than the more strict condition $g^{(cl)}_c/2J\sqrt{N}=1/\sin\theta_R(0)$. This means that the spin dynamics must be considered fully along with the cavity field dynamics to yield the correct critical value.
Finally, we confirm our analysis by numerically simulating the full nonlinear classical dynamics (shown as blue circles in Fig.~\ref{fig:cl_phase}) and taking the averaged imbalance as an order parameter to determine the phase boundary between localized and delocalized phases. Note that in numerics we only set the initial condition for the left spin $\theta_L(0)=\phi_L(0)=0$.

Let us now briefly discuss how one can include next order fluctuations in this formalism and why such a program involves some complications. We use $q_i$ ($i=1$ to $8$) to collectively denote the canonical variables of JCDM. The action for small fluctuations around the classical trajectory $q_i(t)=q_i^{(cl)}(t)+r_i(t)$ can be approximated by
\begin{align*}
{\cal S}[q]\approx {\cal S}[q^{(cl)}] + \frac{1}{2} \int dt' \int dt'' r_i(t') G_{ij}(t',t'') r_j(t'')+\cdots
\end{align*}
where the classical solutions satisfy the stationary condition $\delta {\cal S}/\delta q^{(cl)}_i=0$ and the Green's function is defined by
\begin{align*}
G_{ij}(t',t'')=\frac{\delta^2 {\cal S}[q]}{\delta q_i(t') \delta q_j(t'') }{\Big|}_{q_i=q_i^{(cl)}}.
\end{align*}
The difficulty in calculating the Green's function is due to the Berry phase term for spins in Eq.~(\ref{eq:spin_berry}) which is not quadratic and causes $G_{ij}(t',t'')$ to explicitly depend on the time variables $t'$ and $t''$ during a trajectory. Therefore, finding the Green's function for an arbitrary trajectory is generically a difficult task especially since there is no close solution to Eqs.~(\ref{eq:cl_full_dyn}(a)-(d)). This process can however be done numerically.

The coupling to the qubit not only induces a non-linearity that leads to the localization, but also enlarges the phase space available to the system by adding more degrees of freedom.
Unlike the BECDW~\cite{smerzi_pra}, the JCDM is not integrable for non-zero coupling. The study of chaos in the JCDM is beyond the scope of the present paper.

\renewcommand\theequation{B\arabic{equation}}

\section{\label{sec:chaos}Classical chaos and the KAM theorem}
As mentioned in Appendix~\ref{sec:cl}, the non-linearity due to coupling to the qubit may lead to chaos at the classical level. When $g=0$, our system is (Liouville) integrable~\cite{Josebook}, periodic or quasi-periodic motion then occurs on invariant tori with angles as variables and different tori labeled by conserved action
\begin{equation}
  I_i = \frac{1}{2 \pi} \oint_{\gamma_i} p \cdot dq \ .
\end{equation}
The Kolmogorov-Arnold-Moser (KAM) theorem  states that
as a non-linear perturbation is introduced, the rational (resonant/periodic) tori are destroyed and the quasi-periodic ones are deformed. As the interaction is increased, only sufficiently irrational tori continue to survive, those that do form a Cantor set.
Let us now visualize this behavior in our system.
We use the restricted dynamics to construct Poincar\'{e} sections.  For certain initial conditions, the dynamics is restricted to a four dimensional subspace. 
In this construction a 4d phase space gives rise to a 3d fixed energy surface.
To construct the Poincar\'{e} section, one follows the dynamics and records the state of the system when one of the degrees of freedom, called periodic variable, reaches a certain value. This specifies a point in the 2d plane corresponding to the other
degrees of freedom. The section is then filled in by sampling initial conditions on the fixed energy
surface and following the dynamics for each initial state for a long time.

\begin{figure}
\includegraphics[scale=0.15]{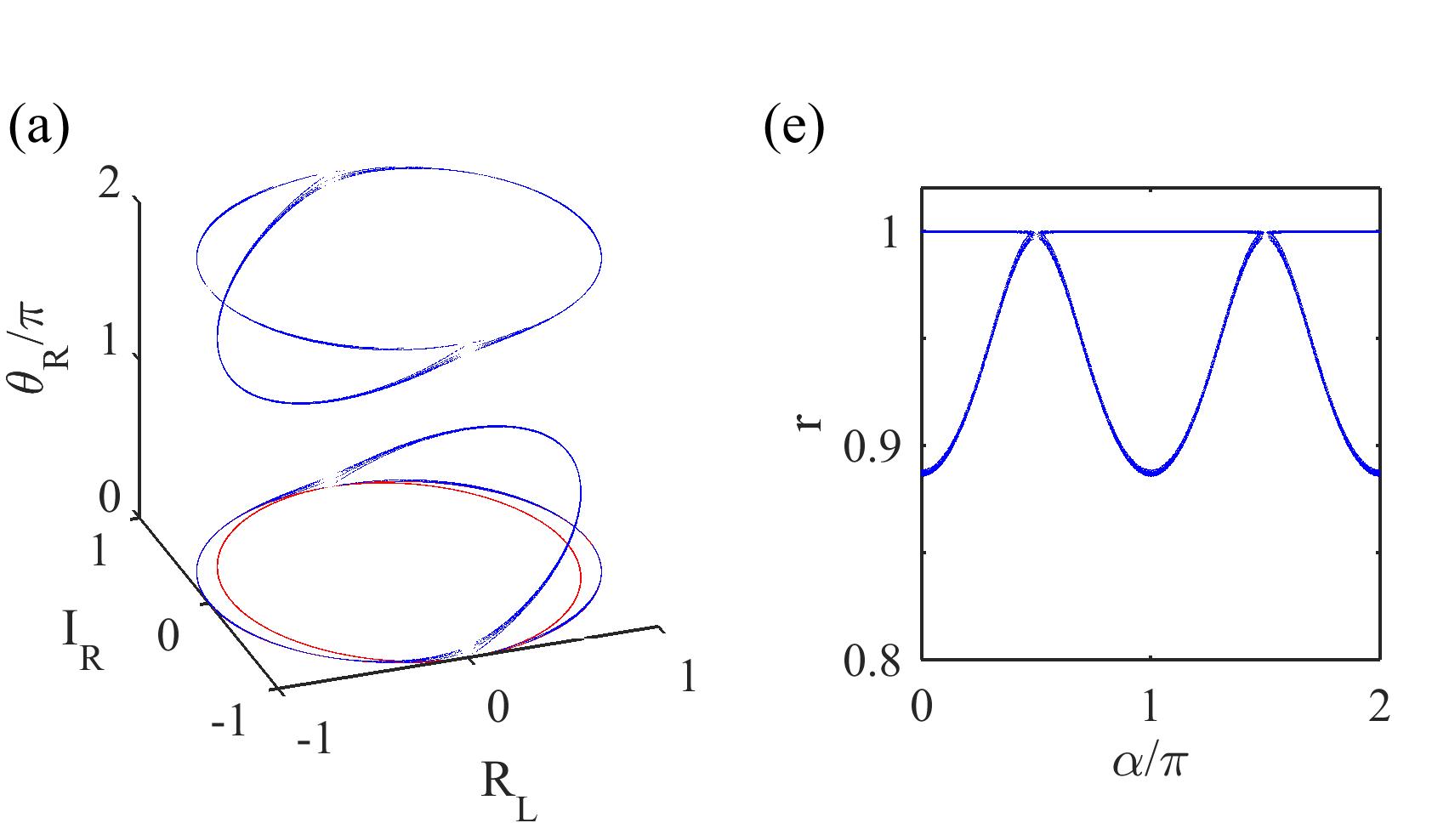}
\includegraphics[scale=0.15]{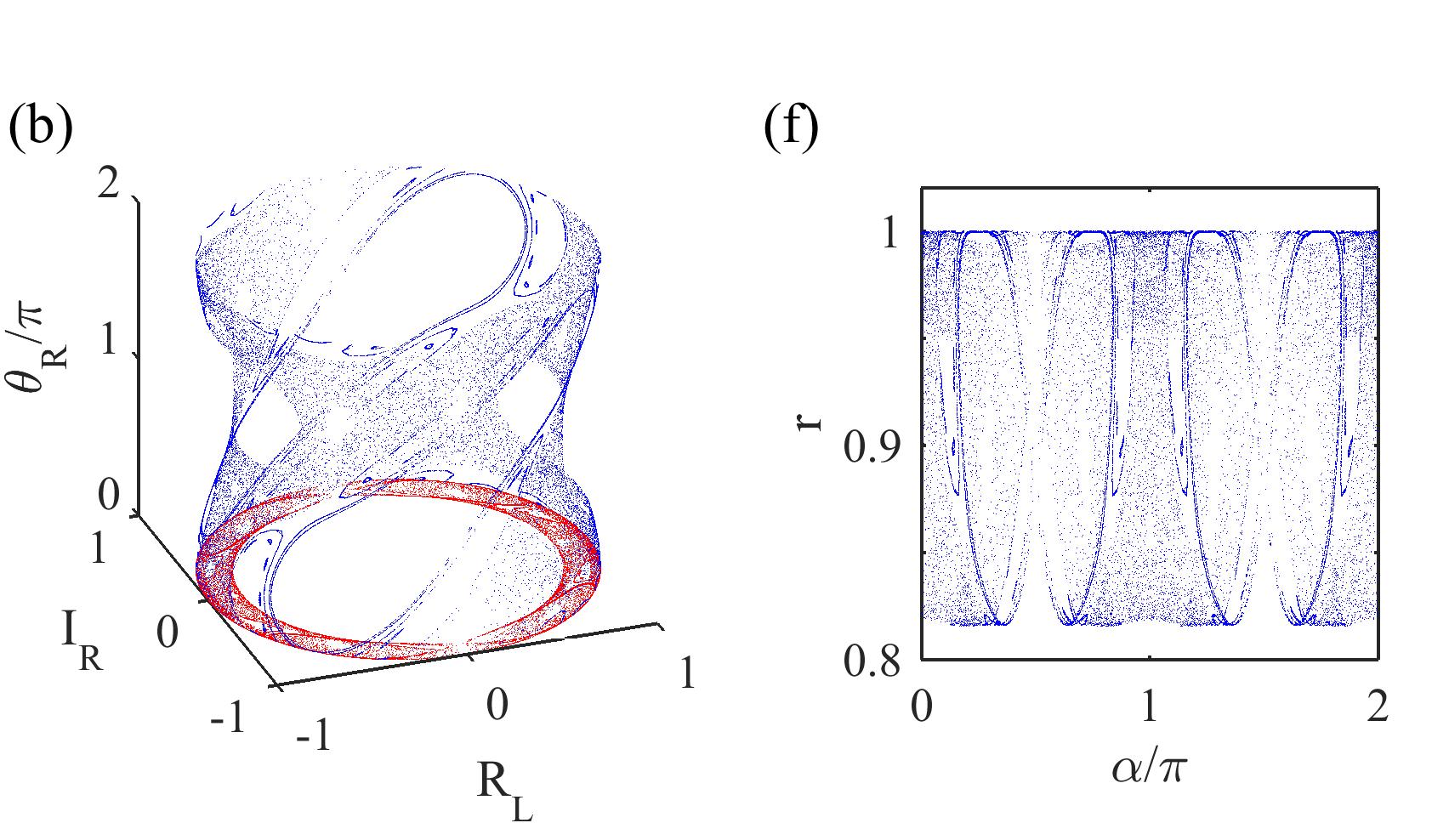}
\includegraphics[scale=0.15]{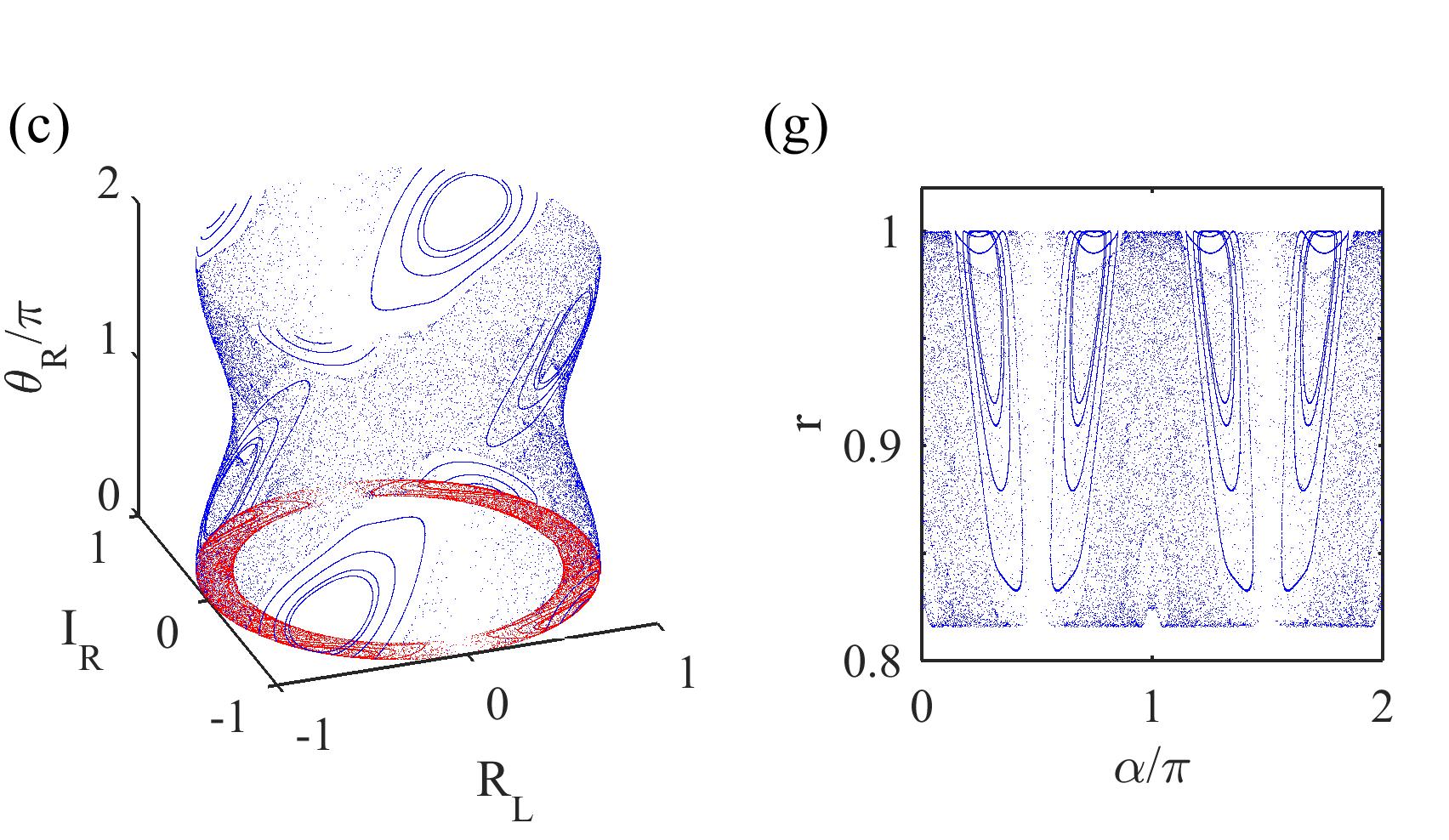}
\includegraphics[scale=0.15]{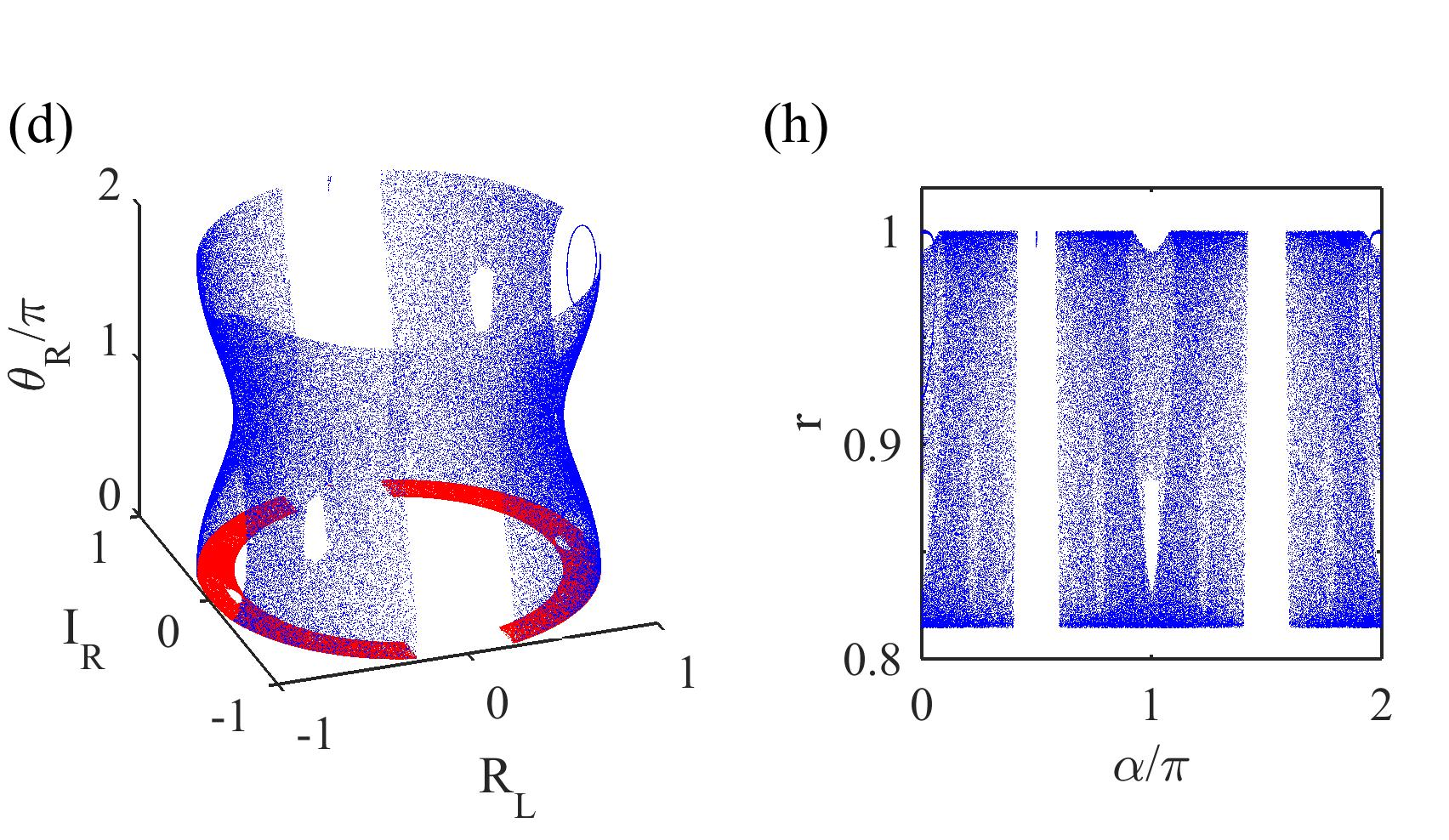}
\caption{\label{fig:classical_chaos} (Color online) Poincar\'e sections for various values of $g/2J\sqrt{N}$. (a)-(d) 3D diagrams showing explicitly one of spin angles, (e)-(h) 2D diagrams (equivalent to the red sections of its left figure) using the polar coordinates $(r,\alpha)=(\sqrt{(R_L^2+I_R^2)/N},\tan^{-1}(I_R/R_L))$. From top to bottom $g/2J\sqrt{N}=0.088,\ 0.311,\ 0.442,\ \text{and}\ 1.41$.}
\end{figure}

An example of such sections is given in Fig.~\ref{fig:classical_chaos} for various values of $g/J$.
Here, the initial conditions are taken as $I_L=R_R=0$, $\phi_L=\pi/2$, and $\phi_R=0$ such that  the dynamics is restricted to a 4d subspace.
This choice leads to a set of four coupled equations, which preserve this choice.
This sub-manifold contains the dynamics corresponding to an initial condition
with perfect imbalance (e.g. $Z \equiv n_L-n_R = N$ for $R_L=\sqrt{N}$ and $I_R=0$ at $t=0$).
The reduced equations of motion are then
\begin{subequations}
\begin{align}
\dot{\theta}_L &=  2g R_L \\
\dot{\theta}_R &=  2g I_R  \\
\dot{R}_L &=-gS \sin\theta_L - J  I_R \\
\dot{ I}_R  &=-gS \sin\theta_R + J R_L
\end{align}
\end{subequations}
together with
\begin{align}
  \dot{R}_R =
  \dot{I}_L =
  \dot{\phi}_L =
  \dot{\phi}_R = 0.
\end{align}
In this four dimensional subspace the tunneling potential in Eq.~(\ref{eq:tunnel_action}) is identically zero, which means
the energy associated with this subspace is independent of the hopping amplitude $J$.
 In Fig.~\ref{fig:classical_chaos}, the periodic variable is chosen to be $\theta_L(t)$ (with $2\pi$ period) and the other two variables are the cavity fields $R_L(t)$ and $I_R(t)$. As we see, for sufficiently small values of $g/J$ as in Fig.~\ref{fig:classical_chaos} (a) and (e) the motion is periodic while it becomes more and more chaotic as the ratio $g/J$ is increased. Nevertheless, there are few surviving quasi-periodic orbits for intermediate interactions.

Interestingly, the restricted equations of motion we study for classical chaos are identical to the classical
equations of motion for electrons moving in a 2d periodic potential subject to an external
magnetic field. This system has been previously studied~\cite{chaos_in_electron} and is known to display a variety of ballistic,
1d and 2d normal diffusive and anomalous diffusive transport, related to the chaotic behavior.

\renewcommand\theequation{C\arabic{equation}}

\section{\label{sec:BEC} Review of the Bose-Einstein condensate double-well problem}
The BECDW system is considerably simpler than the JCDM as there is no spin degrees of freedom. The Schr\"odinger equation has the same form as in Eq.~(\ref{eq:schrodinger}) with two differences that the hopping term is $B(x)=J/2 \sqrt{(1+x+h)(1-x+h)}$ and
the nonlinearity is explicit in the potential energy term $W(x)=\gamma x^2 + \varepsilon x$ where $\gamma$ is the interaction strength and $\varepsilon$ is an \emph{ad hoc} term to break the left-right parity symmetry.
We choose the ansatz $\Psi  (x,t) = \alpha(x,t) e^{\frac{i}{h} S(x,t)}$ with a real-valued phase and amplitude.
The real and imaginary parts of the Schr\"odinger equation up to first order in $h$ are found to be
\begin{align} \label{eq:BEC_HJ}
-\partial_t S=& -2B_0 \cos\left(2\partial_x S\right)+ W(x) - 2h B_1 \cos\left(2\partial_x S\right) \nonumber \\
-\partial_t \alpha =& 4 \alpha B_0 \cos\left(2\partial_x S\right) \partial_x^2 S + 2 (\alpha B'_0 +2 \alpha' B_0) \sin\left(2\partial_x S\right)
\end{align}
where
\begin{align*}
B_0 (x)=& \frac{J}{2} \sqrt{1-x^2}\\
B_1(x)=& \frac{J}{2} \frac{1}{\sqrt{1-x^2}}.
\end{align*}
In order to identify the Hamiltonian and the canonical momentum, we use the Hamilton-Jacobi equations
\begin{align*}
H&= -\partial_t S
&\varphi=\partial_x S
\end{align*}
So, the Hamiltonian can be written as
\begin{align*}
H= -2B_0(x) \cos\left(2\varphi \right)+ W(x) + W_Q (x)
\end{align*}
where all quantum corrections to the classical limit are put into the so-called {\it quantum potential}~\cite{Ballentine},
\begin{align*}
 W_Q(x,\varphi)= - 2h B_1(x) \cos\left(2\varphi\right)+O(h^2)\ .
\end{align*}
We introduce the classical velocity
\begin{align*}
v= \frac{\partial H}{\partial \varphi}= 4B_0(x) \sin2\varphi,
\end{align*}
hence, the second identity in Eq.~(\ref{eq:BEC_HJ}) can be rewritten in terms of the probability density $P=\alpha^2$, as
\begin{align*}
\partial_t P+ \partial_x \left(P v \right)=0 .
\end{align*}
which is the continuity equation.
Now that the Hamiltonian is determined, one can solve for the stationary solutions $\Psi(x,t)= \alpha(x) e^{-\frac{i}{h}\epsilon t} e^{\frac{i}{h} S(x)}$ where $S(x)$ is expanded in powers of $h$,
\begin{align*}
S(x)=S^{(0)}(x) +h S^{(1)}(x) + O(h^2),
\end{align*}
thus, the canonical momentum can be also expanded as
\begin{align*}
\varphi=\partial_x S=\varphi^{(0)} +h \varphi^{(1)} + O(h^2).
\end{align*}
Next, we can solve for phase space trajectories with a fixed energy $\epsilon$ by plugging this expansion into the Hamiltonian
\begin{align*}
\epsilon= -2B_0 \cos\left( 2\varphi^{(0)} +2h \varphi^{(1)} \right) + W(x) + W_Q(x,\varphi^{(0)})
\end{align*}
At zeroth order, we get
\begin{align*}
\varphi^{(0)}(x)= \frac{1}{2} \text{Arccos}\left(\frac{\epsilon-W(x)}{-2B_0(x)}\right)
\end{align*}
and the first order correction would be
\begin{align*}
\varphi^{(1)} &= \frac{B_1(x)}{2B_0(x)} \frac{ \cos\left(2\varphi^{(0)}(x)\right)}{ \sin\left(2\varphi^{(0)}(x)\right)} \nonumber \\
&=- \frac{1}{2(1-x^2)} \frac{\epsilon-W(x)}{\left(J(1-x^2) - (\epsilon-W(x))^2\right)^{1/2}}.
\end{align*}
Using these results, quantization rules can be found accordingly~\cite{Keeling}.

\renewcommand\theequation{D\arabic{equation}}

\section{\label{sec:exact_bc} Exact solutions of the Schr\"odinger equation near boundaries}

Here, we briefly discuss the exact solutions at the boundaries we encounter in JCDM polariton bands.
Except for the critical region which will be explained at the end, the linearized Schr\"{o}dinger equation around the classical turning point $x=z_\alpha+\xi$ up to $O(h^2)$ corrections can be easily shown to be
\begin{align} \label{eq:reg_BC}
\partial^2_\xi \psi_i - \beta_\alpha \partial_\xi \psi_i- c_{i,\alpha} \gamma_\alpha \xi\ \psi_i\approx 0,
\end{align}
where $z_\alpha$ is the turning point that is the solution to $\epsilon=V^\alpha_i(z_\alpha)$, in which $\alpha=l, h$ refers to the stationary points with $\varphi_i =0,\ \pi/2$ respectively, and
\begin{align*}
\beta_\alpha &= \frac{z_\alpha}{1-z_\alpha^2}, \\
\gamma_\alpha&= \frac{\sqrt{1-z_\alpha^2}}{2h^2J}\ |\partial_x V_i^{\alpha}(z_\alpha)|,
\end{align*}
here, a sign coefficient $c_{i,\alpha}$ is introduced to be $\pm1$ when $\xi>0$ is classically forbidden (allowed). The turning point is given by the general expression
\begin{align} \label{eq:z_alpha}
|z_\alpha| =& \frac{1}{J^2}{\Big(} (\epsilon^2+J^2-2g'^2)J^2 -2(g'^2\mp \epsilon J)^2 \nonumber \\
&+ 2 |g'^2 \mp \epsilon J| \sqrt{g'^2+2J^2\mp 2\epsilon J} {\Big)}^{1/2}.
\end{align}
where the $\pm$ signs are both accepted only if there are four turning points on both inner and outer boundaries.
The solution of the linearized differential equation can be written analytically in terms of the Airy functions
\begin{align*}
\psi_i= e^{\beta_\alpha \xi/2} [C_a \text{Ai} (c_{i,\alpha} \gamma_\alpha^{1/3} \tilde{\xi}) + C_b \text{Bi} ( c_{i,\alpha} \gamma_\alpha^{1/3} \tilde{\xi})]
\end{align*}
where $\tilde{\xi}=\xi+c_{i,\alpha}\ \beta_\alpha^2/4\gamma_\alpha$. It is important to note that in deriving the above equation near the turning point at which $\varphi_i \to \pi/2$, the wave-function is oscillating so fast $\psi_i\propto e^{i \pi x/2h}$; hence, we write down the linearized Schr\"odinger equation for the slowly varying part of the actual wave-function by mutliplying the wave-function by $e^{- i \pi x/2h}$ to cancel the oscillating factors.

In the case of upper-upper or lower-lower bands, there is a critical region, near energy $\epsilon_{c,1}=2g'-J$ or $\epsilon_{c,4}=-2g'+J$ respectively,  between the localized and delocalized states where the potential term $V_1^l(x)$ or $V_4^h(x)$ is extremum at the turning point $x=0$. This situation requires a separate analysis as the exact solution is no longer the Airy functions and the quantization condition will be different from the usual Bohr-Sommerfeld quantization.
Near the critical level, energy can be written as $\epsilon=\epsilon_{c,k}+\lambda/N$ where the expanded Schr\"{o}dinger equation near $x=0+\xi$ is simplified into
\begin{align} \label{eq:cr_BC}
\partial_\xi^2\psi_k - \xi\ \partial_\xi \psi_k + \frac{c_k}{2Jh^2} [-\lambda h +\frac{\xi^2}{2}(\frac{g'}{2}-J) ] \psi_k \approx 0
\end{align}
in which the sign coefficient $c_k$ is  $c_1=-1$ for the upper-upper component or $c_4=+1$ for the lower-lower component.
Note that this is not a linear differential equation (as opposed to the regular boundary in Eq.~(\ref{eq:reg_BC})) and the corresponding solutions are parabolic cylinder functions  instead of Airy functions~\cite{Keeling,Whittakerbook},
\begin{align*}
\psi_k =& e^{\xi^2/4}\ [C_a D_{i\chi-1/2} (c_k e^{i\pi/4} \xi \sqrt{\mu/h}) \\
&+ C_b D_{i\chi-1/2} (c_k e^{-i3\pi/4} \xi \sqrt{\mu/h}) ],
\end{align*}
where we neglect the terms of order unity compared to $1/h$ and
\begin{align} \label{eq:cr_par}
&\mu =  \sqrt{\frac{g'}{2J}-1}, &  \chi= \frac{\lambda}{2 \mu J}\ .
\end{align}



\renewcommand\theequation{E\arabic{equation}}

\section{\label{sec:der_ll}Derivation of quantization conditions for the lower-lower band}
Here we explicitly show how to derive the quantization rules for lower-lower polaritons.

{\it (i) Delocalized states, $\epsilon< J-2g'$}:

Here, the particle is confined in a two-sided potential.
Since the classically allowed region is defined between two turning points $|x| <z_l$, the WKB wave-function (Fig.~\ref{fig:wf_ll}(b)) would simply be
\begin{align} \label{eq:wf_ll}
\psi_4 (x) = \begin{cases}
\alpha_4(x) \left( C_+ e^{\frac{i}{h} S_4(x)} +  C_- e^{-\frac{i}{h} S_4(x)}\right), & |x|< z_l \\
\alpha_4(x) C'_- e^{\frac{-1}{h} Q_4(x)}, & x > z_l \\
\alpha_4(x) C'_+ e^{\frac{1}{h} Q_4(x)}. & x <- z_l
\end{cases}
\end{align}
and the other modes are described by
\begin{align*}
\psi_{2}(x)=& \alpha_2 (x) e^{\frac{-1}{h} Q_2(x)}, \\
\psi_{3}(x)=& \alpha_3 (x) e^{\frac{1}{h} Q_3(x)}.
\end{align*}
The above modes will add a correction of order $h$ to the wave-function exponents in Eq.~(\ref{eq:wf_ll}) through Eq.~(\ref{eq:forbid}); however, these terms are exponentially small and will be neglected.
Matching the WKB wave-function with the Airy functions at $x=\pm z_l$ leads to the following connection formulas
\begin{align*}
i \frac{C_-}{C_+}&= e^{\frac{2i}{h}\Delta S_4} ,& C_b=0
\end{align*}
where $\Delta S_4=  \int_0^{z_l} \varphi_4 dx $.
Similarly, one can derive the connection formula for $x=-z_l+\xi$
\begin{align*}
i \frac{C_+}{C_-}= e^{\frac{2i}{h}\Delta S_4}.
\end{align*}
Thus, we get $\frac{4}{h}  \Delta S_4 = (2n+1) \pi $ which means
\begin{align} \label{eq:BS1_ll}
\frac{1}{h}  \int_{-z_l}^{z_l} \varphi_4(x) dx = \left( n+\frac{1}{2}\right) \pi
\end{align}
This is similar to the familiar Bohr-Sommerfeld quantization for a particle confined in a potential well~\cite{Ballentine}.

{\it (ii) Localized states, $J-2g'<\epsilon<-g'\sqrt{2}$}:

The potential profile consists of two isolated classically allowed regions close to $x=\pm1$. Hence, there are two sets of turning points $z_l$ and $z_h$:
One is bouncing off the $\varphi_4=0$ potential, which is similar to the previous case $\epsilon=V_4^l(x=z_l)$,
and the other is bouncing off the $2\varphi_4=\pi$ potential, which is the solution of $\epsilon=V_4^h(x=z_h)$.
Provided that the classically allowed region is $z_h <|x| <z_l$, the WKB wave-function (Fig.~\ref{fig:wf_ll}(c)) would be
\begin{align*}
\psi_4 (x) = \begin{cases}
\alpha_4(x) C'_- e^{\frac{-1}{h} Q_4(x)}, & x > z_l \\
\alpha_4(x) \left( C_{R,+} e^{\frac{i}{h} S_4(x)} +  C_{R,-} e^{-\frac{i}{h} S_4(x)}\right), & z_h< x< z_l \\
\alpha_4(x) \left( C''_+ e^{\frac{1}{h} Q_4(x)} +  C''_- e^{-\frac{1}{h} Q_4(x)}\right), & |x|< z_h \\
\alpha_4(x) \left( C_{L,+} e^{\frac{i}{h} S_4(x)} +  C_{L,-} e^{-\frac{i}{h} S_4(x)}\right), & -z_l< x< -z_h \\
\alpha_4(x) C'_+ e^{\frac{1}{h} Q_4(x)}. & x < -z_l
\end{cases}
\end{align*}
Close to $|x|=z_l$ the connection formula is the same as the previous case
\begin{align}
i \frac{C_{R,-}}{C_{R,+}}&= e^{\frac{2i}{h}\Delta S_4}, \nonumber \\
\label{eq:far_BC}
i \frac{C_{L,+}}{C_{L,-}}&= e^{\frac{2i}{h}\Delta S_4}
\end{align}
where the boundary of the integral is only slightly different as in $\Delta S_4=\frac{1}{2} \int_{y_r}^{z_l} \varphi_4 dx $ and $y_r$ is the solution to the equation $\epsilon=W_{44}(x,\varphi_4)$ corresponding to $\varphi_4=\pi/4$,
\begin{align} \label{eq:y_r}
y_r= \frac{|\epsilon|}{g'} \sqrt{1-(\frac{\epsilon}{2g'})^2}\ .
\end{align}
Near $x=z_h$, $\varphi_4 \to \pi$ and the wave-function is oscillating so fast and $\psi_4\propto e^{i \pi x/2h}$; therefore, as mentioned in Appendix~\ref{sec:exact_bc}, only the slowly varying part of the wave-function will be matched to the Airy functions and there will be an additional term $\pi y_r/2h$ in the quantization rule as a result of this manipulation.

Close to $x=z_h$ and on the forbidden side, the connection formula yields
\begin{align}
\frac{C_a}{C_b}= 2 e^{\frac{1}{h} \Delta Q_4}
\end{align}
where $\Delta Q_4= \int_{-z_h}^{z_h} \phi_4 dx$. The matching condition for the allowed side is simplified into
\begin{align*}
\left|\frac{C_{R,+}e^{\frac{i}{h} \Delta S_4'-i\pi y_r/2h+i\pi/4}+C_{R,-}e^{-\frac{i}{h} \Delta S_4' +i\pi y_r/2h-i\pi/4}}{-C_{R,+}e^{\frac{i}{h} \Delta S_4'-i\pi y_r/2h+i\pi/4}+C_{R,-}e^{-\frac{i}{h} \Delta S_4' +i\pi y_r/2h-i\pi/4}} \right|= \left|\frac{C_b}{C_a}\right|,
\end{align*}
where $\Delta S_4'= \int_{z_h}^{y_r} \tilde{\varphi}_4 dx$, and
\begin{align} \label{eq:phi_neg}
 \tilde{\varphi}_4(x)=\frac{1}{2} \text{Arccos}\left(\frac{\epsilon-W_{44}(x)}{2B_{44}^{(0)}(x')} \right).
\end{align}
It is worth noting that there is no minus sign inside the argument as opposed to Eq.~(\ref{eq:allowed}).
Combining this with Eq.~(\ref{eq:far_BC}), we get
\begin{align*}
\frac{1- \cos 2 \Delta S_4^{\text{(eff)}}}{1+ \cos{2 \Delta S_4^{\text{(eff)}}}} &=\frac{1}{4} e^{-\frac{2}{h} \Delta Q_4}
\end{align*}
where $\Delta S_4^{\text{(eff)}} =\Delta S_4 -\Delta S_4' + \pi y_r/2h =\int_{y_r}^{z_l} \varphi_4 dx -\int_{z_h}^{y_r} \tilde{\varphi}_4 dx + \pi y_r/2h$.
This can be recast in the following quantization condition
\begin{align}  \label{eq:BS2_ll}
\frac{1}{h} \Delta S_4^{\text{(eff)}}(\epsilon) =  n \pi \pm \frac{1}{4} e^{-\frac{1}{h} \Delta Q_4(\epsilon)}\ .
\end{align}
As a result, the tunneling rate is related to the energy splitting in the energy spectrum through
\begin{align*}
\delta \epsilon_{\text{(split)}} = \frac{1}{2} e^{-\frac{1}{h} \Delta Q_4(\epsilon)} \left|\frac{d\Delta S_4^{\text{(eff)}}}{d\epsilon} \right|^{-1}.
\end{align*}

{\it (iii) Critical region: $\epsilon\sim J-2g'$}:

Consider a small deviation from the critical energy $\epsilon=J-2g'+\lambda/N$. The WKB solutions close to the critical point $x=0$  is found to be
\begin{align}
\partial_x S^{(0)} &= \frac{\pi}{2} - \frac{\mu x}{2} ,\nonumber \\
\partial_x S^{(1)} &=\frac{\lambda}{2J\mu\ x}  ,\nonumber \\
\label{eq:cr_S}
\alpha_4 (x) &= \frac{\alpha_{0,4}}{\sqrt{2\mu J x}} e^{x^2/4}.
\end{align}
So, the wave-function becomes
\begin{align}
\psi_4(x) = C_{\pm} \exp\left(\pm \frac{i\pi}{2h}x \mp i\frac{\mu x^2}{4 h}+ (\pm i \chi-1/2) \log (x \sqrt{\mu/h})  \right)
\end{align}
and the quantization formula is derived to be
\begin{align} \label{eq:BS_cr1}
\text{arg} \left[e^{-\frac{2i}{h}\Delta S^{(\text{eff})}_4} \frac{\sqrt{2\pi}}{\Gamma(1/2-i\chi)}- e^{\pi\chi/2} \right]= n\pi + \pi/2
\end{align}
where
\begin{align}
\Delta S^{(\text{eff})}_4= \int_{y_r}^{z_l} \varphi_4 dx - \int_{0}^{y_r} \tilde{\varphi}_4 dx + \frac{\pi y_r}{2},
\end{align}
$\chi$ and $y_r$ are defined in Eqs.~(\ref{eq:cr_par}) and (\ref{eq:y_r}), respectively.



\renewcommand\theequation{F\arabic{equation}}

\section{\label{sec:der_others} Quantization conditions for other polariton bands}
In this appendix, we show the quantization conditions for upper-upper and middle polariton bands.

{\bf Upper-upper band:}
The upper-upper mode is populated for energies $g'\sqrt{2}\leq \epsilon \leq J+2g'$, as illustrated in Fig.~\ref{fig:spec_WKB}.
The situation is very similar to the lower-lower polariton band with the only difference that the localized states appear in the bottom of the band and the delocalized states appear at higher energies.
Likewise, there are three regions: one with delocalized wave-functions centering around $x=0$, one with localized wave-functions in which the probability density is maximized close to the boundary points $x=\pm 1$, and one at the critical region separating the localized and delocalized states. This band is similar to the BECDW with attractive interaction. The derivation of the quantization rules is very similar to Appendix~\ref{sec:der_ll} and we shall only quote the final results.

{\it (i) Delocalized states, $\epsilon> -J+2g'$}:
The particle is confined in a two-sided potential well and the wave-function is large close to the center ($x=0$). It is worth noting that the end points are at $2 \varphi_1=\pi$ and the boundary matching must be done for a slowly varying component of the wave-function.
The quantization condition is found by
\begin{align}  \label{eq:BS1_uu}
\frac{1}{h} \int_{-z_l}^{z_l} \tilde{\varphi}_1(x) dx = \left( n+\frac{1}{2}\right) \pi
\end{align}
where
\begin{align} \label{eq:phi_neg}
 \tilde{\varphi}_1(x)=\frac{1}{2} \text{Arccos}\left(\frac{\epsilon-W_{11}(x)}{2B_{11}^{(0)}(x')} \right).
\end{align}
Notice that there is no minus sign in the denominator as opposed to Eq.~(\ref{eq:allowed}) which has to do with only taking the slowly varying multiple of the wave-function.

{\it (ii) Localized states, $g'\sqrt{2}<\epsilon< -J+2g'$}:
The potential looks like a double-well with two minima at $x=\pm1$. The quantization rule would be
\begin{align}  \label{eq:BS2_uu}
\frac{1}{h} \Delta S_1^{\text{(eff)}}(\epsilon) =  n \pi \pm e^{-\frac{1}{h} \Delta Q_1(\epsilon)}
\end{align}
where
\begin{align*}
\Delta S_1^{\text{(eff)}} =& \int_{y_r}^{z_l}  \tilde{\varphi}_1 dx- \int_{z_h}^{y_r}  \varphi_1 dx + \frac{\pi}{2} y_r , \\
\Delta Q_1 =& \int_{-z_h}^{z_h}  \phi_1 dx ,
\end{align*}
in which $\tilde{\varphi}_1$ is defined in Eq.~(\ref{eq:phi_neg}) and other parameters are introduced as in Appendix~\ref{sec:der_ll}.

{\it (iii) Critical region, $\epsilon \sim -J+2g'$}:
Without much difference from the lower-lower band, here the quantization formula is found to be
\begin{align} \label{eq:BS_cr2}
\text{arg} \left[e^{-\frac{2i}{h}\Delta S^{(\text{eff})}_1} \frac{\sqrt{2\pi}}{\Gamma(1/2+i\chi)}- e^{-\pi\chi/2} \right]= n\pi + \pi/2
\end{align}
where
\begin{align*}
\Delta S^{(\text{eff})}_1= \int_{y_r}^{z_l} \tilde{\varphi}_1 dx - \int_{0}^{y_r} \varphi_1 dx + \frac{\pi y_r}{2},
\end{align*}
$\chi$ and $y_r$ are defined in Eqs.~(\ref{eq:cr_par}) and (\ref{eq:y_r}), respectively.

{\bf Middle bands:}
The middle polariton bands correspond to the energy range $|\epsilon| \leq g'\sqrt{2}$.
The wave-function has two non-zero components: the lower-upper and upper-lower polariton modes. This region in turn can be divided to two subregions:
In the first region, when $J \leq \epsilon \leq g'\sqrt{2}$ or $-g'\sqrt{2} \leq \epsilon \leq -J$, one component is dominant in each half ($x>0$ or $x<0$) since only one component is classically allowed and the other one is forbidden. In the second region, when
$|\epsilon|<J$ both components are non-zero at the same region of $x$. However, in terms of quantization relations the treatments for both cases are identical as the first order correction, in Eq.~(\ref{eq:1st_coupling}), does not couple the second and third components directly and all corrections come in as higher order contributions (at least of order $h^2$). Therefore, it does not really matter whether these two components overlap over a range of $x$ or not.

The upper-lower (second) polariton component is confined between two turning points $z_l\leq x\leq z_h$ where $z_l$ and $z_h$ are solutions to $\epsilon=V_2^l(x)$ and $\epsilon=V_2^h(x)$ given in Eq.~(\ref{eq:z_alpha}). The lower-upper (third) polariton component is then confined in $-z_h\leq x \leq-z_l$ due to symmetry (see Fig.~\ref{fig:spec_WKB}).
So, the WKB wave-function takes the form $\Psi=[0,\psi_2,\psi_3,0]^T$ where
\begin{align*}
\psi_2 (x) = \begin{cases}
\alpha_2(x) \left( C_{2,+} e^{\frac{i}{h} S_2(x)} +  C_{2,-} e^{-\frac{i}{h} S_2(x)}\right), & z_l < x < z_h \\
\alpha_2(x) C'_{2,-} e^{\frac{-1}{h} Q_2(x)}, & x > z_h \\
\alpha_2(x) C'_{2,+} e^{\frac{1}{h} Q_2(x)}, & x < z_l
\end{cases}
\end{align*}
and
\begin{align*}
\psi_3 (x) = \begin{cases}
\alpha_3(x) \left( C_{3,+} e^{\frac{i}{h} S_3(x)} +  C_{3,-} e^{-\frac{i}{h} S_3(x)}\right), & -z_h < x < - z_l \\
\alpha_3(x) C'_{3,-} e^{\frac{-1}{h} Q_3(x)}, & x > -z_l \\
\alpha_3(x) C'_{3,+} e^{\frac{1}{h} Q_3(x)}. & x <- z_h
\end{cases}
\end{align*}
Starting from the connection formulas at the two boundaries for each component, it is straightforward to derive the quantization conditions
\begin{subequations}
\label{eq:BS3}
\begin{align}
\frac{1}{h} \Delta S_2^{\text{(eff)}}(\epsilon) =  n \pi - \frac{\pi}{2} y_r\\
\frac{1}{h} \Delta S_3^{\text{(eff)}}(\epsilon) =  n \pi + \frac{\pi}{2} y_r
\end{align}
\end{subequations}
where
\begin{align*}
\Delta S_2^{\text{(eff)}}&=\int_{y_r}^{z_l} \varphi_2 dx -\int_{z_h}^{y_r} \tilde{\varphi}_2 dx  \\
\Delta S_3^{\text{(eff)}}&=\int_{-y_r}^{-z_l} \varphi_3 dx  -\int_{-y_r}^{-z_h} \tilde{\varphi}_3 dx
\end{align*}
and $y_r$ is defined in Eq.~(\ref{eq:y_r}). It is easy to show that these two relations are indeed identical as the integrands are even functions of $x$.


\renewcommand\theequation{G\arabic{equation}}

\section{\label{sec:valid} Validity of WKB analysis}
The derivation of WKB relies on our original assumption that the total polariton number $N$ is large and a $1/N$ expansion is applicable. A natural question is to what extent this assumption can be justified.
In order to check the validity of the WKB approximation, we use the Husimi-Kano $Q$ representation~\cite{Husimi_orig} to compare our semiclassical picture and the quantum eigenstates (see a comprehensive discussion of the Husimi function in Ref.~\cite{Lee_Husimi}). Similar analyses have been done for the BEC double-well system~\cite{Mahmud_BEC}.

\begin{figure}
\includegraphics[scale=0.8]{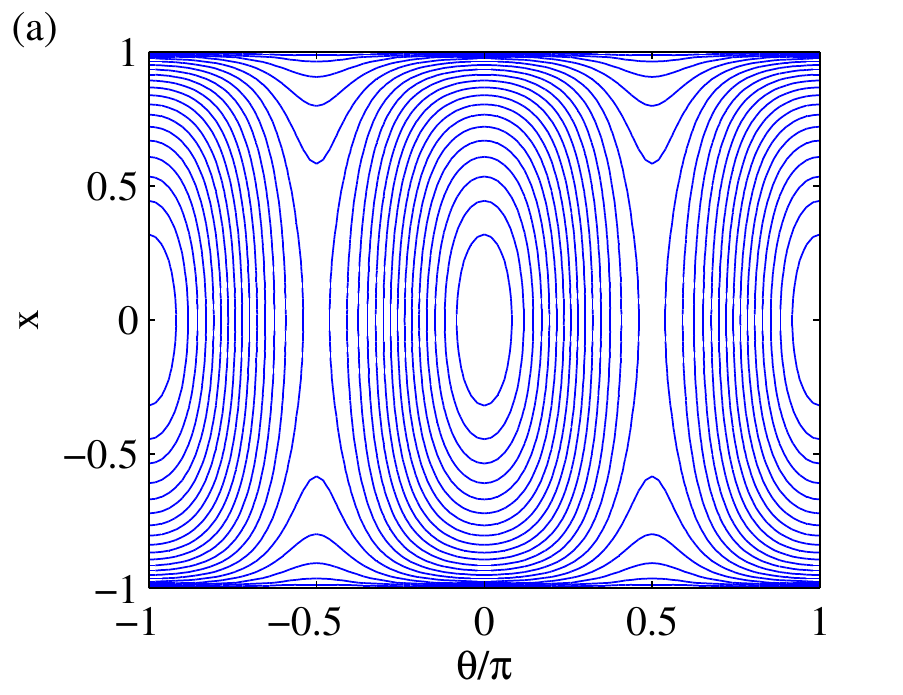}
\includegraphics[scale=0.8]{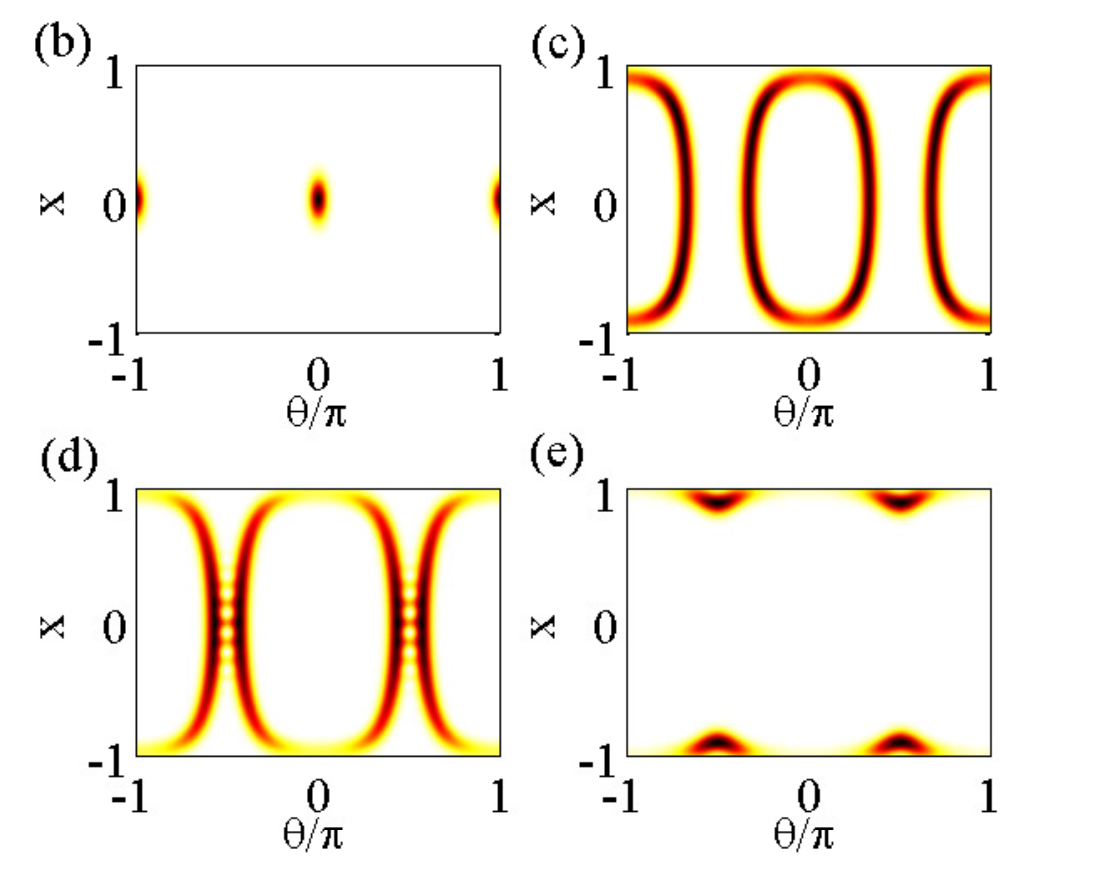}
\caption{\label{fig:Husimi_N100} (Color online) Comparison of the classical phase space and the quantum Husimi distribution in the lower-lower polariton band. (a) classical energy contours. The $Q$ distribution is shown (b) for the ground state, (c) for the mid band (oscillatory) state $52$-th, (d) close to the critical level (separatrix in (a)) $73$-rd, (e) for the localized state $97$-th state. Here $N=100$ and $J/g'=1/3$.}
\end{figure}

\begin{figure*}
\begin{tikzpicture}[scale=0.9]
\node[anchor=center,inner sep=0] at (0.5,-2) {\includegraphics[scale=0.8]{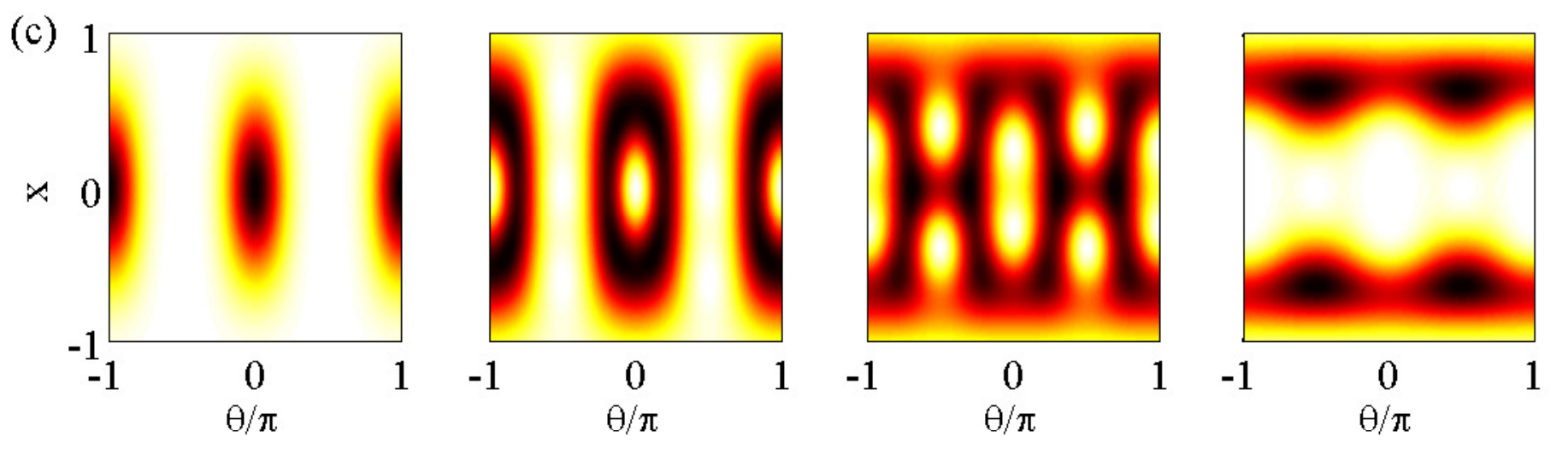}};
\node[anchor=center,inner sep=0] at (0.5,2.8) {\includegraphics[scale=0.8]{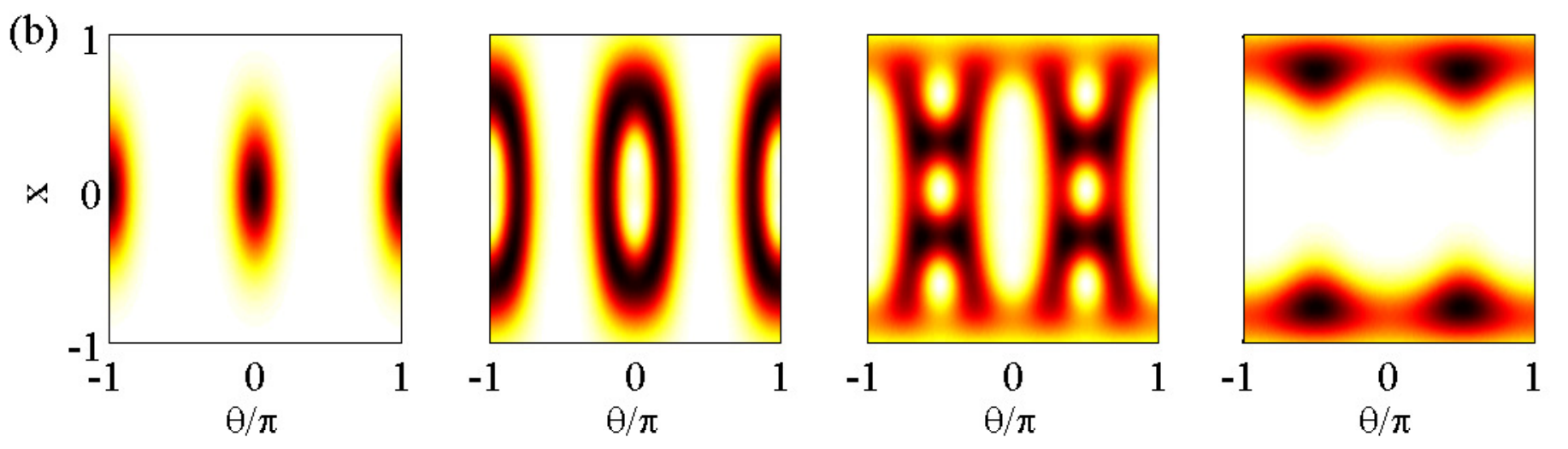}};
\node[anchor=center,inner sep=0] at (0.5,7.5) {\includegraphics[scale=0.8]{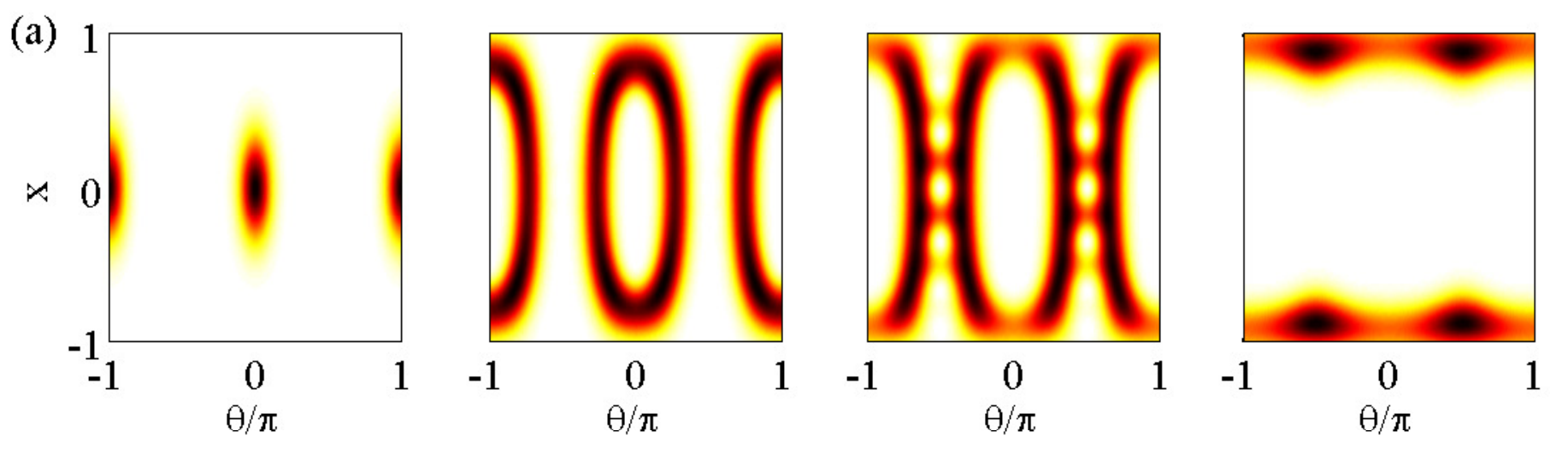}};
\draw[->,>=latex,line width=2pt] (9,4.5)--(9,1.6);
\draw[->,>=latex,line width=2pt] (-1.5,9.8)--(2.8,9.8);
\draw (0.5,10) node {\normalsize going up in spectrum};
\draw (9.2,4.5) node[right,rotate=-90] {\normalsize decreasing N};
\end{tikzpicture}
\caption{\label{fig:Husimi_N_cmp} (Color online) Husimi $Q$ representation for various values of system size $N$ in the lower-lower polariton band: (a) $N=20$ (b) $N=10$, and (c) $N=6$. From left to right, each row shows the ground state,  a mid-band (oscillatory) state, a state close to the critical level, and a localized state. Here $J/g'=1/3$.}
\end{figure*}

As we have seen in Sec.~\ref{sec:fock}, dynamics within $i$-th polariton band can be described by two conjugate variables the ``position'' $x=Z/N$ and the ``momentum'' $\varphi_i$. Let us focus on the lower-lower polariton band and define $\theta=\varphi_4$. The Poisson bracket is promoted to the canonical commutation relation $[x,\theta]=ih$ at the quantum level where the Planck's constant is $h=1/N$. A squeezed coherent-state can be represented in position and momentum,
\begin{align}
\langle \theta' |\theta+i x \rangle &= \frac{1}{(\pi\kappa^2)^{1/4} } \exp \left( -i \frac{x \theta'}{h} - \frac{(\theta'-\theta)^2}{2\kappa^2} \right) \nonumber \\
\langle x' |\theta+i x \rangle &= \frac{1}{(\pi/\kappa^2)^{1/4} } \exp \left( i \frac{x' \theta}{h} - \kappa^2 \frac{(x'-x)^2}{2} \right).
\end{align}
The squeezing parameter $\kappa$ is the key parameter in connecting the classical picture and quantum eigenstates~\cite{Husimi_orig,Lee_Husimi}. It determines the relative resolution in the phase space $(x,\theta)$. The optimum $\kappa$ generally depends on the explicit form of  the Hamiltonian in terms of the conjugate variables of interest.
For a simple harmonic oscillator, this parameter must be chosen equal to the zero-point fluctuations $\kappa=(m\omega/\hbar)^{1/2}$ in $(x,p)$-space and equal to one $\kappa=1$ in ladder operators $(a,a^\dag)$-space (the usual $Q$ function in quantum optics~\cite{Scullybook}). For our case, $\kappa$ must be tuned to an optimum value to obtain the best resolution. In general, we write $\kappa= s \kappa_0 $ where $s>1$ is the tuning parameter which is set to be $s=1$ for the ground state and is slightly increased for higher excited states.
The value $\kappa_0=( N^3 g^2/8J^2)^{1/8}$ comes from the harmonic approximation of the Hamiltonian close to the ground state
\begin{align}
H_4 \approx -2g'-J + \frac{J}{2} \theta^2 + \frac{g'}{4} x^2.
\end{align}
For a pure state $|\psi\rangle$, the Husimi function is defined by
\begin{align}
Q(x,\theta) = | \langle \theta+ix | \psi\rangle |^2
\end{align}

In order to compute the inner product numerically, we use the following identity
\begin{align}
\langle \theta+i x| \psi\rangle= \frac{1}{(\pi /\kappa)^{1/4}} \sum_{x'=-1}^{1} C_4({x}') \exp \left(i \frac{x' \theta}{h} - \kappa^2 \frac{(x'-x)^2}{2} \right)
\end{align}
where $C_4(Z)$ is the fourth-component of the wave-function in the polariton basis.
Note that in these calculations only $C_4(Z)$ is taken into account as the rest of the components are negligible in the lower-lower band.
Figure~\ref{fig:Husimi_N100} shows that the Husimi functions of the Hamiltonian eigenstates match quite well with the classical phase space energy contours for large values of $N$. As we see in Fig.~\ref{fig:Husimi_N100}(b), the ground state is represented by a point-like distribution meaning that it is a minimum uncertainty (Gaussian) wave-packet, consistent with our harmonic approximation above.
The next excited states (Fig.~\ref{fig:Husimi_N100}(c)) are oscillatory states. At the critical level $\epsilon_{c,4}=-2g'+J$, we arrive at the separatrix (bifurcation point) of the classical phase space, which is also manifest in the Husimi function as in Fig.~\ref{fig:Husimi_N100}(d). Ultimately, above the critical level we obtain the localized states which are represented by two seperate branches close to $x=1$ and $-1$ in the quantum picture, Fig.~\ref{fig:Husimi_N100}(e). This is because the Hamiltonian preserves the parity (left-right) symmetry and the localized states in the classical picture are indeed Schr\"{o}dinger cat states in the eigenspectrum of the Hamiltonian. 
\begin{figure}
\includegraphics[scale=0.76]{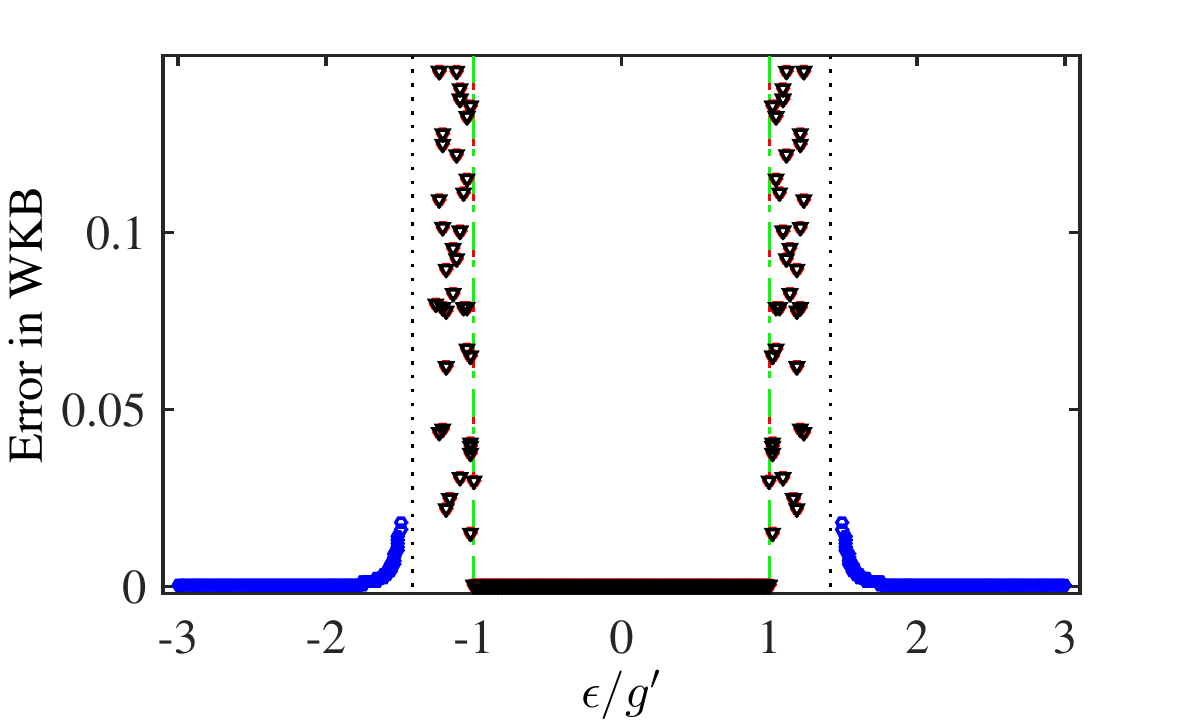}
\caption{\label{fig:quant_dev} (Color online) Error in the Bohr-Sommerfeld quantization rules in the entire spectrum. Different symbols refer to different quantization conditions and the vertical lines show various energy scales; both are explained in the caption of Fig.~\ref{fig:BS_quant}. Here $J/g'=1$ and $N=400$.}
\end{figure}
To see how much the above correspondence survives for smaller system sizes (polariton numbers) we compute the Husimi function for $N=20$, $10$, and $6$ in Figs.~\ref{fig:Husimi_N_cmp}(a)-(c). The classical phase space is independent of $N$ and is shown in Fig.~\ref{fig:Husimi_N100}(a). It is clear that the fluctuations are more pronounced as $N$ becomes smaller; however, even up to such small values as $N=6$ the classical and the quantum phase space pictures seem to be remarkably consistent.
Therefore, the WKB approach would remain a reasonable approximation up to small polariton numbers although its power as an analytical tool is more appreciable in the large system sizes where the exact treatments are exponentially difficult.

Another remark is that we have neglected the interband coupling (quantum potential) as we are interested in the strong coupling regime where the polariton bands are separated from each other. However, in the weak coupling regime the corrections from Eq.~(\ref{eq:WKB_1st}) would be non-negligible as the bands overlap and our quantization rules start to fail. Interestingly, even in the weak coupling regime, the quantization rules remain valid if we stay far enough from the overlapping regions. Such an example for $J/g'$ is illustrated in Fig.~\ref{fig:quant_dev}.

\end{appendix}

\bibliography{JCD_v11.bib}

\providecommand{\noopsort}[1]{}\providecommand{\singleletter}[1]{#1}%
\begin{thebibliography}{61}%
\makeatletter
\providecommand \@ifxundefined [1]{%
 \@ifx{#1\undefined}
}%
\providecommand \@ifnum [1]{%
 \ifnum #1\expandafter \@firstoftwo
 \else \expandafter \@secondoftwo
 \fi
}%
\providecommand \@ifx [1]{%
 \ifx #1\expandafter \@firstoftwo
 \else \expandafter \@secondoftwo
 \fi
}%
\providecommand \natexlab [1]{#1}%
\providecommand \enquote  [1]{``#1''}%
\providecommand \bibnamefont  [1]{#1}%
\providecommand \bibfnamefont [1]{#1}%
\providecommand \citenamefont [1]{#1}%
\providecommand \href@noop [0]{\@secondoftwo}%
\providecommand \href [0]{\begingroup \@sanitize@url \@href}%
\providecommand \@href[1]{\@@startlink{#1}\@@href}%
\providecommand \@@href[1]{\endgroup#1\@@endlink}%
\providecommand \@sanitize@url [0]{\catcode `\\12\catcode `\$12\catcode
  `\&12\catcode `\#12\catcode `\^12\catcode `\_12\catcode `\%12\relax}%
\providecommand \@@startlink[1]{}%
\providecommand \@@endlink[0]{}%
\providecommand \url  [0]{\begingroup\@sanitize@url \@url }%
\providecommand \@url [1]{\endgroup\@href {#1}{\urlprefix }}%
\providecommand \urlprefix  [0]{URL }%
\providecommand \Eprint [0]{\href }%
\providecommand \doibase [0]{http://dx.doi.org/}%
\providecommand \selectlanguage [0]{\@gobble}%
\providecommand \bibinfo  [0]{\@secondoftwo}%
\providecommand \bibfield  [0]{\@secondoftwo}%
\providecommand \translation [1]{[#1]}%
\providecommand \BibitemOpen [0]{}%
\providecommand \bibitemStop [0]{}%
\providecommand \bibitemNoStop [0]{.\EOS\space}%
\providecommand \EOS [0]{\spacefactor3000\relax}%
\providecommand \BibitemShut  [1]{\csname bibitem#1\endcsname}%
\let\auto@bib@innerbib\@empty
\bibitem [{\citenamefont {Greentree}\ \emph {et~al.}(2006)\citenamefont
  {Greentree}, \citenamefont {Tahan}, \citenamefont {Cole},\ and\ \citenamefont
  {Hollenberg}}]{Greentree2006}%
  \BibitemOpen
  \bibfield  {author} {\bibinfo {author} {\bibfnamefont {A.~D.}\ \bibnamefont
  {Greentree}}, \bibinfo {author} {\bibfnamefont {C.}~\bibnamefont {Tahan}},
  \bibinfo {author} {\bibfnamefont {J.~H.}\ \bibnamefont {Cole}}, \ and\
  \bibinfo {author} {\bibfnamefont {L.~C.~L.}\ \bibnamefont {Hollenberg}},\
  }\href {\doibase 10.1038/nphys466} {\bibfield  {journal} {\bibinfo  {journal}
  {Nat Phys}\ }\textbf {\bibinfo {volume} {2}},\ \bibinfo {pages} {856}
  (\bibinfo {year} {2006})}\BibitemShut {NoStop}%
\bibitem [{\citenamefont {Hartmann}\ \emph {et~al.}(2006)\citenamefont
  {Hartmann}, \citenamefont {Brandao},\ and\ \citenamefont
  {Plenio}}]{Hartmann2006}%
  \BibitemOpen
  \bibfield  {author} {\bibinfo {author} {\bibfnamefont {M.~J.}\ \bibnamefont
  {Hartmann}}, \bibinfo {author} {\bibfnamefont {F.~G. S.~L.}\ \bibnamefont
  {Brandao}}, \ and\ \bibinfo {author} {\bibfnamefont {M.~B.}\ \bibnamefont
  {Plenio}},\ }\href {\doibase 10.1038/nphys462} {\bibfield  {journal}
  {\bibinfo  {journal} {Nat Phys}\ }\textbf {\bibinfo {volume} {2}},\ \bibinfo
  {pages} {849} (\bibinfo {year} {2006})}\BibitemShut {NoStop}%
\bibitem [{\citenamefont {Hartmann}\ \emph {et~al.}(2008)\citenamefont
  {Hartmann}, \citenamefont {Brandão},\ and\ \citenamefont
  {Plenio}}]{Hartmann2008}%
  \BibitemOpen
  \bibfield  {author} {\bibinfo {author} {\bibfnamefont {M.}~\bibnamefont
  {Hartmann}}, \bibinfo {author} {\bibfnamefont {F.}~\bibnamefont {Brandão}},
  \ and\ \bibinfo {author} {\bibfnamefont {M.}~\bibnamefont {Plenio}},\ }\href
  {\doibase 10.1002/lpor.200810046} {\bibfield  {journal} {\bibinfo  {journal}
  {Laser and Photonics Reviews}\ }\textbf {\bibinfo {volume} {2}},\ \bibinfo
  {pages} {527} (\bibinfo {year} {2008})}\BibitemShut {NoStop}%
\bibitem [{\citenamefont {Tomadin}\ and\ \citenamefont
  {Fazio}(2010)}]{Tomadin2010}%
  \BibitemOpen
  \bibfield  {author} {\bibinfo {author} {\bibfnamefont {A.}~\bibnamefont
  {Tomadin}}\ and\ \bibinfo {author} {\bibfnamefont {R.}~\bibnamefont
  {Fazio}},\ }\href {\doibase 10.1364/JOSAB.27.00A130} {\bibfield  {journal}
  {\bibinfo  {journal} {J. Opt. Soc. Am. B}\ }\textbf {\bibinfo {volume}
  {27}},\ \bibinfo {pages} {A130} (\bibinfo {year} {2010})}\BibitemShut
  {NoStop}%
\bibitem [{\citenamefont {Houck}\ \emph {et~al.}(2012)\citenamefont {Houck},
  \citenamefont {T\"{u}reci},\ and\ \citenamefont {Koch}}]{Houck2012}%
  \BibitemOpen
  \bibfield  {author} {\bibinfo {author} {\bibfnamefont {A.~A.}\ \bibnamefont
  {Houck}}, \bibinfo {author} {\bibfnamefont {H.~E.}\ \bibnamefont
  {T\"{u}reci}}, \ and\ \bibinfo {author} {\bibfnamefont {J.}~\bibnamefont
  {Koch}},\ }\href {\doibase 10.1038/nphys2251} {\bibfield  {journal} {\bibinfo
   {journal} {Nat Phys}\ }\textbf {\bibinfo {volume} {8}},\ \bibinfo {pages}
  {292} (\bibinfo {year} {2012})}\BibitemShut {NoStop}%
\bibitem [{\citenamefont {Schmidt}\ and\ \citenamefont
  {Koch}(2013)}]{Schmidt2013}%
  \BibitemOpen
  \bibfield  {author} {\bibinfo {author} {\bibfnamefont {S.}~\bibnamefont
  {Schmidt}}\ and\ \bibinfo {author} {\bibfnamefont {J.}~\bibnamefont {Koch}},\
  }\href {\doibase 10.1002/andp.201200261} {\bibfield  {journal} {\bibinfo
  {journal} {Annalen der Physik}\ }\textbf {\bibinfo {volume} {525}},\ \bibinfo
  {pages} {395} (\bibinfo {year} {2013})}\BibitemShut {NoStop}%
\bibitem [{\citenamefont {Angelakis}\ \emph {et~al.}(2011)\citenamefont
  {Angelakis}, \citenamefont {Huo}, \citenamefont {Kyoseva},\ and\
  \citenamefont {Kwek}}]{PhysRevLett.106.153601}%
  \BibitemOpen
  \bibfield  {author} {\bibinfo {author} {\bibfnamefont {D.~G.}\ \bibnamefont
  {Angelakis}}, \bibinfo {author} {\bibfnamefont {M.}~\bibnamefont {Huo}},
  \bibinfo {author} {\bibfnamefont {E.}~\bibnamefont {Kyoseva}}, \ and\
  \bibinfo {author} {\bibfnamefont {L.~C.}\ \bibnamefont {Kwek}},\ }\href
  {\doibase 10.1103/PhysRevLett.106.153601} {\bibfield  {journal} {\bibinfo
  {journal} {Phys. Rev. Lett.}\ }\textbf {\bibinfo {volume} {106}},\ \bibinfo
  {pages} {153601} (\bibinfo {year} {2011})}\BibitemShut {NoStop}%
\bibitem [{\citenamefont {Schir\'o}\ \emph {et~al.}(2012)\citenamefont
  {Schir\'o}, \citenamefont {Bordyuh}, \citenamefont {\"Oztop},\ and\
  \citenamefont {T\"ureci}}]{Schiro2012}%
  \BibitemOpen
  \bibfield  {author} {\bibinfo {author} {\bibfnamefont {M.}~\bibnamefont
  {Schir\'o}}, \bibinfo {author} {\bibfnamefont {M.}~\bibnamefont {Bordyuh}},
  \bibinfo {author} {\bibfnamefont {B.}~\bibnamefont {\"Oztop}}, \ and\
  \bibinfo {author} {\bibfnamefont {H.~E.}\ \bibnamefont {T\"ureci}},\ }\href
  {\doibase 10.1103/PhysRevLett.109.053601} {\bibfield  {journal} {\bibinfo
  {journal} {Phys. Rev. Lett.}\ }\textbf {\bibinfo {volume} {109}},\ \bibinfo
  {pages} {053601} (\bibinfo {year} {2012})}\BibitemShut {NoStop}%
\bibitem [{\citenamefont {Jin}\ \emph {et~al.}(2013)\citenamefont {Jin},
  \citenamefont {Rossini}, \citenamefont {Fazio}, \citenamefont {Leib},\ and\
  \citenamefont {Hartmann}}]{PhysRevLett.110.163605}%
  \BibitemOpen
  \bibfield  {author} {\bibinfo {author} {\bibfnamefont {J.}~\bibnamefont
  {Jin}}, \bibinfo {author} {\bibfnamefont {D.}~\bibnamefont {Rossini}},
  \bibinfo {author} {\bibfnamefont {R.}~\bibnamefont {Fazio}}, \bibinfo
  {author} {\bibfnamefont {M.}~\bibnamefont {Leib}}, \ and\ \bibinfo {author}
  {\bibfnamefont {M.~J.}\ \bibnamefont {Hartmann}},\ }\href {\doibase
  10.1103/PhysRevLett.110.163605} {\bibfield  {journal} {\bibinfo  {journal}
  {Phys. Rev. Lett.}\ }\textbf {\bibinfo {volume} {110}},\ \bibinfo {pages}
  {163605} (\bibinfo {year} {2013})}\BibitemShut {NoStop}%
\bibitem [{\citenamefont {Otterbach}\ \emph {et~al.}(2013)\citenamefont
  {Otterbach}, \citenamefont {Moos}, \citenamefont {Muth},\ and\ \citenamefont
  {Fleischhauer}}]{PhysRevLett.111.113001}%
  \BibitemOpen
  \bibfield  {author} {\bibinfo {author} {\bibfnamefont {J.}~\bibnamefont
  {Otterbach}}, \bibinfo {author} {\bibfnamefont {M.}~\bibnamefont {Moos}},
  \bibinfo {author} {\bibfnamefont {D.}~\bibnamefont {Muth}}, \ and\ \bibinfo
  {author} {\bibfnamefont {M.}~\bibnamefont {Fleischhauer}},\ }\href {\doibase
  10.1103/PhysRevLett.111.113001} {\bibfield  {journal} {\bibinfo  {journal}
  {Phys. Rev. Lett.}\ }\textbf {\bibinfo {volume} {111}},\ \bibinfo {pages}
  {113001} (\bibinfo {year} {2013})}\BibitemShut {NoStop}%
\bibitem [{\citenamefont {Le~Boit\'e}\ \emph {et~al.}(2013)\citenamefont
  {Le~Boit\'e}, \citenamefont {Orso},\ and\ \citenamefont
  {Ciuti}}]{PhysRevLett.110.233601}%
  \BibitemOpen
  \bibfield  {author} {\bibinfo {author} {\bibfnamefont {A.}~\bibnamefont
  {Le~Boit\'e}}, \bibinfo {author} {\bibfnamefont {G.}~\bibnamefont {Orso}}, \
  and\ \bibinfo {author} {\bibfnamefont {C.}~\bibnamefont {Ciuti}},\ }\href
  {\doibase 10.1103/PhysRevLett.110.233601} {\bibfield  {journal} {\bibinfo
  {journal} {Phys. Rev. Lett.}\ }\textbf {\bibinfo {volume} {110}},\ \bibinfo
  {pages} {233601} (\bibinfo {year} {2013})}\BibitemShut {NoStop}%
\bibitem [{\citenamefont {Nissen}\ \emph {et~al.}(2012)\citenamefont {Nissen},
  \citenamefont {Schmidt}, \citenamefont {Biondi}, \citenamefont {Blatter},
  \citenamefont {T\"ureci},\ and\ \citenamefont
  {Keeling}}]{PhysRevLett.108.233603}%
  \BibitemOpen
  \bibfield  {author} {\bibinfo {author} {\bibfnamefont {F.}~\bibnamefont
  {Nissen}}, \bibinfo {author} {\bibfnamefont {S.}~\bibnamefont {Schmidt}},
  \bibinfo {author} {\bibfnamefont {M.}~\bibnamefont {Biondi}}, \bibinfo
  {author} {\bibfnamefont {G.}~\bibnamefont {Blatter}}, \bibinfo {author}
  {\bibfnamefont {H.~E.}\ \bibnamefont {T\"ureci}}, \ and\ \bibinfo {author}
  {\bibfnamefont {J.}~\bibnamefont {Keeling}},\ }\href {\doibase
  10.1103/PhysRevLett.108.233603} {\bibfield  {journal} {\bibinfo  {journal}
  {Phys. Rev. Lett.}\ }\textbf {\bibinfo {volume} {108}},\ \bibinfo {pages}
  {233603} (\bibinfo {year} {2012})}\BibitemShut {NoStop}%
\bibitem [{\citenamefont {Carusotto}\ and\ \citenamefont
  {Ciuti}(2013)}]{RevModPhys.85.299}%
  \BibitemOpen
  \bibfield  {author} {\bibinfo {author} {\bibfnamefont {I.}~\bibnamefont
  {Carusotto}}\ and\ \bibinfo {author} {\bibfnamefont {C.}~\bibnamefont
  {Ciuti}},\ }\href {\doibase 10.1103/RevModPhys.85.299} {\bibfield  {journal}
  {\bibinfo  {journal} {Rev. Mod. Phys.}\ }\textbf {\bibinfo {volume} {85}},\
  \bibinfo {pages} {299} (\bibinfo {year} {2013})}\BibitemShut {NoStop}%
\bibitem [{\citenamefont {Le~Hur}\ \emph {et~al.}(2015)\citenamefont {Le~Hur},
  \citenamefont {Henriet}, \citenamefont {Petrescu}, \citenamefont {Plekhanov},
  \citenamefont {Roux},\ and\ \citenamefont {Schir\'{o}}}]{LeHur_Marco}%
  \BibitemOpen
  \bibfield  {author} {\bibinfo {author} {\bibfnamefont {K.}~\bibnamefont
  {Le~Hur}}, \bibinfo {author} {\bibfnamefont {L.}~\bibnamefont {Henriet}},
  \bibinfo {author} {\bibfnamefont {A.}~\bibnamefont {Petrescu}}, \bibinfo
  {author} {\bibfnamefont {K.}~\bibnamefont {Plekhanov}}, \bibinfo {author}
  {\bibfnamefont {G.}~\bibnamefont {Roux}}, \ and\ \bibinfo {author}
  {\bibfnamefont {M.}~\bibnamefont {Schir\'{o}}},\ }\href
  {http://arxiv.org/abs/1505.00167} {\bibfield  {journal} {\bibinfo  {journal}
  {arXiv:1505.00167}\ } (\bibinfo {year} {2015})}\BibitemShut {NoStop}%
\bibitem [{\citenamefont {Koch}\ \emph {et~al.}(2010)\citenamefont {Koch},
  \citenamefont {Houck}, \citenamefont {Hur},\ and\ \citenamefont
  {Girvin}}]{PhysRevA.82.043811}%
  \BibitemOpen
  \bibfield  {author} {\bibinfo {author} {\bibfnamefont {J.}~\bibnamefont
  {Koch}}, \bibinfo {author} {\bibfnamefont {A.~A.}\ \bibnamefont {Houck}},
  \bibinfo {author} {\bibfnamefont {K.~L.}\ \bibnamefont {Hur}}, \ and\
  \bibinfo {author} {\bibfnamefont {S.~M.}\ \bibnamefont {Girvin}},\ }\href
  {\doibase 10.1103/PhysRevA.82.043811} {\bibfield  {journal} {\bibinfo
  {journal} {Phys. Rev. A}\ }\textbf {\bibinfo {volume} {82}},\ \bibinfo
  {pages} {043811} (\bibinfo {year} {2010})}\BibitemShut {NoStop}%
\bibitem [{\citenamefont {Petrescu}\ \emph {et~al.}(2012)\citenamefont
  {Petrescu}, \citenamefont {Houck},\ and\ \citenamefont
  {Le~Hur}}]{PhysRevA.86.053804}%
  \BibitemOpen
  \bibfield  {author} {\bibinfo {author} {\bibfnamefont {A.}~\bibnamefont
  {Petrescu}}, \bibinfo {author} {\bibfnamefont {A.~A.}\ \bibnamefont {Houck}},
  \ and\ \bibinfo {author} {\bibfnamefont {K.}~\bibnamefont {Le~Hur}},\ }\href
  {\doibase 10.1103/PhysRevA.86.053804} {\bibfield  {journal} {\bibinfo
  {journal} {Phys. Rev. A}\ }\textbf {\bibinfo {volume} {86}},\ \bibinfo
  {pages} {053804} (\bibinfo {year} {2012})}\BibitemShut {NoStop}%
\bibitem [{\citenamefont {Cho}\ \emph {et~al.}(2008)\citenamefont {Cho},
  \citenamefont {Angelakis},\ and\ \citenamefont
  {Bose}}]{PhysRevLett.101.246809}%
  \BibitemOpen
  \bibfield  {author} {\bibinfo {author} {\bibfnamefont {J.}~\bibnamefont
  {Cho}}, \bibinfo {author} {\bibfnamefont {D.~G.}\ \bibnamefont {Angelakis}},
  \ and\ \bibinfo {author} {\bibfnamefont {S.}~\bibnamefont {Bose}},\ }\href
  {\doibase 10.1103/PhysRevLett.101.246809} {\bibfield  {journal} {\bibinfo
  {journal} {Phys. Rev. Lett.}\ }\textbf {\bibinfo {volume} {101}},\ \bibinfo
  {pages} {246809} (\bibinfo {year} {2008})}\BibitemShut {NoStop}%
\bibitem [{\citenamefont {Hafezi}\ \emph
  {et~al.}(2013{\natexlab{a}})\citenamefont {Hafezi}, \citenamefont {Mittal},
  \citenamefont {Fan}, \citenamefont {Migdall},\ and\ \citenamefont
  {Taylor}}]{Hafezi2013}%
  \BibitemOpen
  \bibfield  {author} {\bibinfo {author} {\bibfnamefont {M.}~\bibnamefont
  {Hafezi}}, \bibinfo {author} {\bibfnamefont {S.}~\bibnamefont {Mittal}},
  \bibinfo {author} {\bibfnamefont {J.}~\bibnamefont {Fan}}, \bibinfo {author}
  {\bibfnamefont {A.}~\bibnamefont {Migdall}}, \ and\ \bibinfo {author}
  {\bibfnamefont {J.~M.}\ \bibnamefont {Taylor}},\ }\href {\doibase
  10.1038/nphoton.2013.274} {\bibfield  {journal} {\bibinfo  {journal} {Nat
  Photon}\ }\textbf {\bibinfo {volume} {7}},\ \bibinfo {pages} {1001} (\bibinfo
  {year} {2013}{\natexlab{a}})}\BibitemShut {NoStop}%
\bibitem [{\citenamefont {Umucalilar}\ and\ \citenamefont
  {Carusotto}(2011)}]{PhysRevA.84.043804}%
  \BibitemOpen
  \bibfield  {author} {\bibinfo {author} {\bibfnamefont {R.~O.}\ \bibnamefont
  {Umucalilar}}\ and\ \bibinfo {author} {\bibfnamefont {I.}~\bibnamefont
  {Carusotto}},\ }\href {\doibase 10.1103/PhysRevA.84.043804} {\bibfield
  {journal} {\bibinfo  {journal} {Phys. Rev. A}\ }\textbf {\bibinfo {volume}
  {84}},\ \bibinfo {pages} {043804} (\bibinfo {year} {2011})}\BibitemShut
  {NoStop}%
\bibitem [{\citenamefont {Hafezi}\ \emph
  {et~al.}(2013{\natexlab{b}})\citenamefont {Hafezi}, \citenamefont {Lukin},\
  and\ \citenamefont {Taylor}}]{Hafezi_FQH}%
  \BibitemOpen
  \bibfield  {author} {\bibinfo {author} {\bibfnamefont {M.}~\bibnamefont
  {Hafezi}}, \bibinfo {author} {\bibfnamefont {M.~D.}\ \bibnamefont {Lukin}}, \
  and\ \bibinfo {author} {\bibfnamefont {J.~M.}\ \bibnamefont {Taylor}},\
  }\href {http://stacks.iop.org/1367-2630/15/i=6/a=063001} {\bibfield
  {journal} {\bibinfo  {journal} {New J. Phys.}\ }\textbf {\bibinfo {volume}
  {15}},\ \bibinfo {pages} {063001} (\bibinfo {year}
  {2013}{\natexlab{b}})}\BibitemShut {NoStop}%
\bibitem [{\citenamefont {Brune}\ \emph {et~al.}(1996)\citenamefont {Brune},
  \citenamefont {Schmidt-Kaler}, \citenamefont {Maali}, \citenamefont {Dreyer},
  \citenamefont {Hagley}, \citenamefont {Raimond},\ and\ \citenamefont
  {Haroche}}]{Haroche1996}%
  \BibitemOpen
  \bibfield  {author} {\bibinfo {author} {\bibfnamefont {M.}~\bibnamefont
  {Brune}}, \bibinfo {author} {\bibfnamefont {F.}~\bibnamefont
  {Schmidt-Kaler}}, \bibinfo {author} {\bibfnamefont {A.}~\bibnamefont
  {Maali}}, \bibinfo {author} {\bibfnamefont {J.}~\bibnamefont {Dreyer}},
  \bibinfo {author} {\bibfnamefont {E.}~\bibnamefont {Hagley}}, \bibinfo
  {author} {\bibfnamefont {J.~M.}\ \bibnamefont {Raimond}}, \ and\ \bibinfo
  {author} {\bibfnamefont {S.}~\bibnamefont {Haroche}},\ }\href {\doibase
  10.1103/PhysRevLett.76.1800} {\bibfield  {journal} {\bibinfo  {journal}
  {Phys. Rev. Lett.}\ }\textbf {\bibinfo {volume} {76}},\ \bibinfo {pages}
  {1800} (\bibinfo {year} {1996})}\BibitemShut {NoStop}%
\bibitem [{\citenamefont {Birnbaum}\ \emph {et~al.}(2005)\citenamefont
  {Birnbaum}, \citenamefont {Boca}, \citenamefont {Miller}, \citenamefont
  {Boozer}, \citenamefont {Northup},\ and\ \citenamefont
  {Kimble}}]{Kimble2005}%
  \BibitemOpen
  \bibfield  {author} {\bibinfo {author} {\bibfnamefont {K.~M.}\ \bibnamefont
  {Birnbaum}}, \bibinfo {author} {\bibfnamefont {A.}~\bibnamefont {Boca}},
  \bibinfo {author} {\bibfnamefont {R.}~\bibnamefont {Miller}}, \bibinfo
  {author} {\bibfnamefont {A.~D.}\ \bibnamefont {Boozer}}, \bibinfo {author}
  {\bibfnamefont {T.~E.}\ \bibnamefont {Northup}}, \ and\ \bibinfo {author}
  {\bibfnamefont {H.~J.}\ \bibnamefont {Kimble}},\ }\href {\doibase
  10.1038/nature03804} {\bibfield  {journal} {\bibinfo  {journal} {Nature}\
  }\textbf {\bibinfo {volume} {436}},\ \bibinfo {pages} {87} (\bibinfo {year}
  {2005})}\BibitemShut {NoStop}%
\bibitem [{\citenamefont {Reithmaier}\ \emph {et~al.}(2004)\citenamefont
  {Reithmaier}, \citenamefont {Sek}, \citenamefont {Loffler}, \citenamefont
  {Hofmann}, \citenamefont {Kuhn}, \citenamefont {Reitzenstein}, \citenamefont
  {Keldysh}, \citenamefont {Kulakovskii}, \citenamefont {Reinecke},\ and\
  \citenamefont {Forchel}}]{Reithmaier2004}%
  \BibitemOpen
  \bibfield  {author} {\bibinfo {author} {\bibfnamefont {J.~P.}\ \bibnamefont
  {Reithmaier}}, \bibinfo {author} {\bibfnamefont {G.}~\bibnamefont {Sek}},
  \bibinfo {author} {\bibfnamefont {A.}~\bibnamefont {Loffler}}, \bibinfo
  {author} {\bibfnamefont {C.}~\bibnamefont {Hofmann}}, \bibinfo {author}
  {\bibfnamefont {S.}~\bibnamefont {Kuhn}}, \bibinfo {author} {\bibfnamefont
  {S.}~\bibnamefont {Reitzenstein}}, \bibinfo {author} {\bibfnamefont {L.~V.}\
  \bibnamefont {Keldysh}}, \bibinfo {author} {\bibfnamefont {V.~D.}\
  \bibnamefont {Kulakovskii}}, \bibinfo {author} {\bibfnamefont {T.~L.}\
  \bibnamefont {Reinecke}}, \ and\ \bibinfo {author} {\bibfnamefont
  {A.}~\bibnamefont {Forchel}},\ }\href {\doibase 10.1038/nature02969}
  {\bibfield  {journal} {\bibinfo  {journal} {Nature}\ }\textbf {\bibinfo
  {volume} {432}},\ \bibinfo {pages} {197} (\bibinfo {year}
  {2004})}\BibitemShut {NoStop}%
\bibitem [{\citenamefont {Yoshie}\ \emph {et~al.}(2004)\citenamefont {Yoshie},
  \citenamefont {Scherer}, \citenamefont {Hendrickson}, \citenamefont
  {Khitrova}, \citenamefont {Gibbs}, \citenamefont {Rupper}, \citenamefont
  {Ell}, \citenamefont {Shchekin},\ and\ \citenamefont {Deppe}}]{Yoshie2004}%
  \BibitemOpen
  \bibfield  {author} {\bibinfo {author} {\bibfnamefont {T.}~\bibnamefont
  {Yoshie}}, \bibinfo {author} {\bibfnamefont {A.}~\bibnamefont {Scherer}},
  \bibinfo {author} {\bibfnamefont {J.}~\bibnamefont {Hendrickson}}, \bibinfo
  {author} {\bibfnamefont {G.}~\bibnamefont {Khitrova}}, \bibinfo {author}
  {\bibfnamefont {H.~M.}\ \bibnamefont {Gibbs}}, \bibinfo {author}
  {\bibfnamefont {G.}~\bibnamefont {Rupper}}, \bibinfo {author} {\bibfnamefont
  {C.}~\bibnamefont {Ell}}, \bibinfo {author} {\bibfnamefont {O.~B.}\
  \bibnamefont {Shchekin}}, \ and\ \bibinfo {author} {\bibfnamefont {D.~G.}\
  \bibnamefont {Deppe}},\ }\href {\doibase 10.1038/nature03119} {\bibfield
  {journal} {\bibinfo  {journal} {Nature}\ }\textbf {\bibinfo {volume} {432}},\
  \bibinfo {pages} {200} (\bibinfo {year} {2004})}\BibitemShut {NoStop}%
\bibitem [{\citenamefont {Peter}\ \emph {et~al.}(2005)\citenamefont {Peter},
  \citenamefont {Senellart}, \citenamefont {Martrou}, \citenamefont
  {Lema\^{i}tre}, \citenamefont {Hours}, \citenamefont {G\'erard},\ and\
  \citenamefont {Bloch}}]{Bloch2005}%
  \BibitemOpen
  \bibfield  {author} {\bibinfo {author} {\bibfnamefont {E.}~\bibnamefont
  {Peter}}, \bibinfo {author} {\bibfnamefont {P.}~\bibnamefont {Senellart}},
  \bibinfo {author} {\bibfnamefont {D.}~\bibnamefont {Martrou}}, \bibinfo
  {author} {\bibfnamefont {A.}~\bibnamefont {Lema\^{i}tre}}, \bibinfo {author}
  {\bibfnamefont {J.}~\bibnamefont {Hours}}, \bibinfo {author} {\bibfnamefont
  {J.~M.}\ \bibnamefont {G\'erard}}, \ and\ \bibinfo {author} {\bibfnamefont
  {J.}~\bibnamefont {Bloch}},\ }\href {\doibase 10.1103/PhysRevLett.95.067401}
  {\bibfield  {journal} {\bibinfo  {journal} {Phys. Rev. Lett.}\ }\textbf
  {\bibinfo {volume} {95}},\ \bibinfo {pages} {067401} (\bibinfo {year}
  {2005})}\BibitemShut {NoStop}%
\bibitem [{\citenamefont {Wallraff}\ \emph {et~al.}(2004)\citenamefont
  {Wallraff}, \citenamefont {Schuster}, \citenamefont {Blais}, \citenamefont
  {Frunzio}, \citenamefont {Huang}, \citenamefont {Majer}, \citenamefont
  {Kumar}, \citenamefont {Girvin},\ and\ \citenamefont
  {Schoelkopf}}]{Wallraff2004}%
  \BibitemOpen
  \bibfield  {author} {\bibinfo {author} {\bibfnamefont {A.}~\bibnamefont
  {Wallraff}}, \bibinfo {author} {\bibfnamefont {D.~I.}\ \bibnamefont
  {Schuster}}, \bibinfo {author} {\bibfnamefont {A.}~\bibnamefont {Blais}},
  \bibinfo {author} {\bibfnamefont {L.}~\bibnamefont {Frunzio}}, \bibinfo
  {author} {\bibfnamefont {R.-S.}\ \bibnamefont {Huang}}, \bibinfo {author}
  {\bibfnamefont {J.}~\bibnamefont {Majer}}, \bibinfo {author} {\bibfnamefont
  {S.}~\bibnamefont {Kumar}}, \bibinfo {author} {\bibfnamefont {S.~M.}\
  \bibnamefont {Girvin}}, \ and\ \bibinfo {author} {\bibfnamefont {R.~J.}\
  \bibnamefont {Schoelkopf}},\ }\href {\doibase 10.1038/nature02851} {\bibfield
   {journal} {\bibinfo  {journal} {Nature}\ }\textbf {\bibinfo {volume}
  {431}},\ \bibinfo {pages} {162} (\bibinfo {year} {2004})}\BibitemShut
  {NoStop}%
\bibitem [{\citenamefont {Blais}\ \emph {et~al.}(2004)\citenamefont {Blais},
  \citenamefont {Huang}, \citenamefont {Wallraff}, \citenamefont {Girvin},\
  and\ \citenamefont {Schoelkopf}}]{Blais2004}%
  \BibitemOpen
  \bibfield  {author} {\bibinfo {author} {\bibfnamefont {A.}~\bibnamefont
  {Blais}}, \bibinfo {author} {\bibfnamefont {R.-S.}\ \bibnamefont {Huang}},
  \bibinfo {author} {\bibfnamefont {A.}~\bibnamefont {Wallraff}}, \bibinfo
  {author} {\bibfnamefont {S.~M.}\ \bibnamefont {Girvin}}, \ and\ \bibinfo
  {author} {\bibfnamefont {R.~J.}\ \bibnamefont {Schoelkopf}},\ }\href
  {\doibase 10.1103/PhysRevA.69.062320} {\bibfield  {journal} {\bibinfo
  {journal} {Phys. Rev. A}\ }\textbf {\bibinfo {volume} {69}},\ \bibinfo
  {pages} {062320} (\bibinfo {year} {2004})}\BibitemShut {NoStop}%
\bibitem [{\citenamefont {Schmidt}\ \emph {et~al.}(2010)\citenamefont
  {Schmidt}, \citenamefont {Gerace}, \citenamefont {Houck}, \citenamefont
  {Blatter},\ and\ \citenamefont {T\"ureci}}]{Tureci}%
  \BibitemOpen
  \bibfield  {author} {\bibinfo {author} {\bibfnamefont {S.}~\bibnamefont
  {Schmidt}}, \bibinfo {author} {\bibfnamefont {D.}~\bibnamefont {Gerace}},
  \bibinfo {author} {\bibfnamefont {A.~A.}\ \bibnamefont {Houck}}, \bibinfo
  {author} {\bibfnamefont {G.}~\bibnamefont {Blatter}}, \ and\ \bibinfo
  {author} {\bibfnamefont {H.~E.}\ \bibnamefont {T\"ureci}},\ }\href {\doibase
  10.1103/PhysRevB.82.100507} {\bibfield  {journal} {\bibinfo  {journal} {Phys.
  Rev. B}\ }\textbf {\bibinfo {volume} {82}},\ \bibinfo {pages} {100507}
  (\bibinfo {year} {2010})}\BibitemShut {NoStop}%
\bibitem [{\citenamefont {Milburn}\ \emph {et~al.}(1997)\citenamefont
  {Milburn}, \citenamefont {Corney}, \citenamefont {Wright},\ and\
  \citenamefont {Walls}}]{Milburn}%
  \BibitemOpen
  \bibfield  {author} {\bibinfo {author} {\bibfnamefont {G.~J.}\ \bibnamefont
  {Milburn}}, \bibinfo {author} {\bibfnamefont {J.}~\bibnamefont {Corney}},
  \bibinfo {author} {\bibfnamefont {E.~M.}\ \bibnamefont {Wright}}, \ and\
  \bibinfo {author} {\bibfnamefont {D.~F.}\ \bibnamefont {Walls}},\ }\href
  {\doibase 10.1103/PhysRevA.55.4318} {\bibfield  {journal} {\bibinfo
  {journal} {Phys. Rev. A}\ }\textbf {\bibinfo {volume} {55}},\ \bibinfo
  {pages} {4318} (\bibinfo {year} {1997})}\BibitemShut {NoStop}%
\bibitem [{\citenamefont {Smerzi}\ \emph {et~al.}(1997)\citenamefont {Smerzi},
  \citenamefont {Fantoni}, \citenamefont {Giovanazzi},\ and\ \citenamefont
  {Shenoy}}]{smerzi_prl}%
  \BibitemOpen
  \bibfield  {author} {\bibinfo {author} {\bibfnamefont {A.}~\bibnamefont
  {Smerzi}}, \bibinfo {author} {\bibfnamefont {S.}~\bibnamefont {Fantoni}},
  \bibinfo {author} {\bibfnamefont {S.}~\bibnamefont {Giovanazzi}}, \ and\
  \bibinfo {author} {\bibfnamefont {S.~R.}\ \bibnamefont {Shenoy}},\ }\href
  {\doibase 10.1103/PhysRevLett.79.4950} {\bibfield  {journal} {\bibinfo
  {journal} {Phys. Rev. Lett.}\ }\textbf {\bibinfo {volume} {79}},\ \bibinfo
  {pages} {4950} (\bibinfo {year} {1997})}\BibitemShut {NoStop}%
\bibitem [{\citenamefont {Albiez}\ \emph {et~al.}(2005)\citenamefont {Albiez},
  \citenamefont {Gati}, \citenamefont {F\"olling}, \citenamefont {Hunsmann},
  \citenamefont {Cristiani},\ and\ \citenamefont {Oberthaler}}]{BEC_exp1}%
  \BibitemOpen
  \bibfield  {author} {\bibinfo {author} {\bibfnamefont {M.}~\bibnamefont
  {Albiez}}, \bibinfo {author} {\bibfnamefont {R.}~\bibnamefont {Gati}},
  \bibinfo {author} {\bibfnamefont {J.}~\bibnamefont {F\"olling}}, \bibinfo
  {author} {\bibfnamefont {S.}~\bibnamefont {Hunsmann}}, \bibinfo {author}
  {\bibfnamefont {M.}~\bibnamefont {Cristiani}}, \ and\ \bibinfo {author}
  {\bibfnamefont {M.~K.}\ \bibnamefont {Oberthaler}},\ }\href {\doibase
  10.1103/PhysRevLett.95.010402} {\bibfield  {journal} {\bibinfo  {journal}
  {Phys. Rev. Lett.}\ }\textbf {\bibinfo {volume} {95}},\ \bibinfo {pages}
  {010402} (\bibinfo {year} {2005})}\BibitemShut {NoStop}%
\bibitem [{\citenamefont {Levy}\ \emph {et~al.}(2007)\citenamefont {Levy},
  \citenamefont {Lahoud}, \citenamefont {Shomroni},\ and\ \citenamefont
  {Steinhauer}}]{BEC_exp2}%
  \BibitemOpen
  \bibfield  {author} {\bibinfo {author} {\bibfnamefont {S.}~\bibnamefont
  {Levy}}, \bibinfo {author} {\bibfnamefont {E.}~\bibnamefont {Lahoud}},
  \bibinfo {author} {\bibfnamefont {I.}~\bibnamefont {Shomroni}}, \ and\
  \bibinfo {author} {\bibfnamefont {J.}~\bibnamefont {Steinhauer}},\ }\href
  {\doibase 10.1038/nature06186} {\bibfield  {journal} {\bibinfo  {journal}
  {Nature}\ }\textbf {\bibinfo {volume} {449}},\ \bibinfo {pages} {579}
  (\bibinfo {year} {2007})}\BibitemShut {NoStop}%
\bibitem [{\citenamefont {Esteve}\ \emph {et~al.}(2008)\citenamefont {Esteve},
  \citenamefont {Gross}, \citenamefont {Weller}, \citenamefont {Giovanazzi},\
  and\ \citenamefont {Oberthaler}}]{BEC_exp3}%
  \BibitemOpen
  \bibfield  {author} {\bibinfo {author} {\bibfnamefont {J.}~\bibnamefont
  {Esteve}}, \bibinfo {author} {\bibfnamefont {C.}~\bibnamefont {Gross}},
  \bibinfo {author} {\bibfnamefont {A.}~\bibnamefont {Weller}}, \bibinfo
  {author} {\bibfnamefont {S.}~\bibnamefont {Giovanazzi}}, \ and\ \bibinfo
  {author} {\bibfnamefont {M.~K.}\ \bibnamefont {Oberthaler}},\ }\href
  {\doibase 10.1038/nature07332} {\bibfield  {journal} {\bibinfo  {journal}
  {Nature}\ }\textbf {\bibinfo {volume} {455}},\ \bibinfo {pages} {1216}
  (\bibinfo {year} {2008})}\BibitemShut {NoStop}%
\bibitem [{\citenamefont {Abbarchi}\ \emph {et~al.}(2013)\citenamefont
  {Abbarchi}, \citenamefont {Amo}, \citenamefont {Sala}, \citenamefont
  {Solnyshkov}, \citenamefont {Flayac}, \citenamefont {Ferrier}, \citenamefont
  {Sagnes}, \citenamefont {Galopin}, \citenamefont {Lemaitre}, \citenamefont
  {Malpuech},\ and\ \citenamefont {Bloch}}]{Bloch2013}%
  \BibitemOpen
  \bibfield  {author} {\bibinfo {author} {\bibfnamefont {M.}~\bibnamefont
  {Abbarchi}}, \bibinfo {author} {\bibfnamefont {A.}~\bibnamefont {Amo}},
  \bibinfo {author} {\bibfnamefont {V.~G.}\ \bibnamefont {Sala}}, \bibinfo
  {author} {\bibfnamefont {D.~D.}\ \bibnamefont {Solnyshkov}}, \bibinfo
  {author} {\bibfnamefont {H.}~\bibnamefont {Flayac}}, \bibinfo {author}
  {\bibfnamefont {L.}~\bibnamefont {Ferrier}}, \bibinfo {author} {\bibfnamefont
  {I.}~\bibnamefont {Sagnes}}, \bibinfo {author} {\bibfnamefont
  {E.}~\bibnamefont {Galopin}}, \bibinfo {author} {\bibfnamefont
  {A.}~\bibnamefont {Lemaitre}}, \bibinfo {author} {\bibfnamefont
  {G.}~\bibnamefont {Malpuech}}, \ and\ \bibinfo {author} {\bibfnamefont
  {J.}~\bibnamefont {Bloch}},\ }\href {\doibase 10.1038/nphys2609} {\bibfield
  {journal} {\bibinfo  {journal} {Nat Phys}\ }\textbf {\bibinfo {volume} {9}},\
  \bibinfo {pages} {275} (\bibinfo {year} {2013})}\BibitemShut {NoStop}%
\bibitem [{\citenamefont {Raftery}\ \emph {et~al.}(2014)\citenamefont
  {Raftery}, \citenamefont {Sadri}, \citenamefont {Schmidt}, \citenamefont
  {T\"ureci},\ and\ \citenamefont {Houck}}]{James}%
  \BibitemOpen
  \bibfield  {author} {\bibinfo {author} {\bibfnamefont {J.}~\bibnamefont
  {Raftery}}, \bibinfo {author} {\bibfnamefont {D.}~\bibnamefont {Sadri}},
  \bibinfo {author} {\bibfnamefont {S.}~\bibnamefont {Schmidt}}, \bibinfo
  {author} {\bibfnamefont {H.~E.}\ \bibnamefont {T\"ureci}}, \ and\ \bibinfo
  {author} {\bibfnamefont {A.~A.}\ \bibnamefont {Houck}},\ }\href {\doibase
  10.1103/PhysRevX.4.031043} {\bibfield  {journal} {\bibinfo  {journal} {Phys.
  Rev. X}\ }\textbf {\bibinfo {volume} {4}},\ \bibinfo {pages} {031043}
  (\bibinfo {year} {2014})}\BibitemShut {NoStop}%
\bibitem [{\citenamefont {Braun}(1993)}]{Braun_RMP}%
  \BibitemOpen
  \bibfield  {author} {\bibinfo {author} {\bibfnamefont {P.~A.}\ \bibnamefont
  {Braun}},\ }\href {\doibase 10.1103/RevModPhys.65.115} {\bibfield  {journal}
  {\bibinfo  {journal} {Rev. Mod. Phys.}\ }\textbf {\bibinfo {volume} {65}},\
  \bibinfo {pages} {115} (\bibinfo {year} {1993})}\BibitemShut {NoStop}%
\bibitem [{\citenamefont {Altshuler}\ \emph {et~al.}(1997)\citenamefont
  {Altshuler}, \citenamefont {Gefen}, \citenamefont {Kamenev},\ and\
  \citenamefont {Levitov}}]{Altshuler1997}%
  \BibitemOpen
  \bibfield  {author} {\bibinfo {author} {\bibfnamefont {B.~L.}\ \bibnamefont
  {Altshuler}}, \bibinfo {author} {\bibfnamefont {Y.}~\bibnamefont {Gefen}},
  \bibinfo {author} {\bibfnamefont {A.}~\bibnamefont {Kamenev}}, \ and\
  \bibinfo {author} {\bibfnamefont {L.~S.}\ \bibnamefont {Levitov}},\ }\href
  {\doibase 10.1103/PhysRevLett.78.2803} {\bibfield  {journal} {\bibinfo
  {journal} {Phys. Rev. Lett.}\ }\textbf {\bibinfo {volume} {78}},\ \bibinfo
  {pages} {2803} (\bibinfo {year} {1997})}\BibitemShut {NoStop}%
\bibitem [{\citenamefont {J\"a\"askel\"ainen}\ and\ \citenamefont
  {Meystre}(2005)}]{PhysRevA.71.043603}%
  \BibitemOpen
  \bibfield  {author} {\bibinfo {author} {\bibfnamefont {M.}~\bibnamefont
  {J\"a\"askel\"ainen}}\ and\ \bibinfo {author} {\bibfnamefont
  {P.}~\bibnamefont {Meystre}},\ }\href {\doibase 10.1103/PhysRevA.71.043603}
  {\bibfield  {journal} {\bibinfo  {journal} {Phys. Rev. A}\ }\textbf {\bibinfo
  {volume} {71}},\ \bibinfo {pages} {043603} (\bibinfo {year}
  {2005})}\BibitemShut {NoStop}%
\bibitem [{\citenamefont {Juli\'a-D\'{i}az}\ \emph {et~al.}(2010)\citenamefont
  {Juli\'a-D\'{i}az}, \citenamefont {Dagnino}, \citenamefont {Lewenstein},
  \citenamefont {Martorell},\ and\ \citenamefont {Polls}}]{JuliaDiaz}%
  \BibitemOpen
  \bibfield  {author} {\bibinfo {author} {\bibfnamefont {B.}~\bibnamefont
  {Juli\'a-D\'{i}az}}, \bibinfo {author} {\bibfnamefont {D.}~\bibnamefont
  {Dagnino}}, \bibinfo {author} {\bibfnamefont {M.}~\bibnamefont {Lewenstein}},
  \bibinfo {author} {\bibfnamefont {J.}~\bibnamefont {Martorell}}, \ and\
  \bibinfo {author} {\bibfnamefont {A.}~\bibnamefont {Polls}},\ }\href
  {\doibase 10.1103/PhysRevA.81.023615} {\bibfield  {journal} {\bibinfo
  {journal} {Phys. Rev. A}\ }\textbf {\bibinfo {volume} {81}},\ \bibinfo
  {pages} {023615} (\bibinfo {year} {2010})}\BibitemShut {NoStop}%
\bibitem [{\citenamefont {Fu}\ and\ \citenamefont
  {Liu}(2006)}]{FuLiu_dynamics}%
  \BibitemOpen
  \bibfield  {author} {\bibinfo {author} {\bibfnamefont {L.}~\bibnamefont
  {Fu}}\ and\ \bibinfo {author} {\bibfnamefont {J.}~\bibnamefont {Liu}},\
  }\href {\doibase 10.1103/PhysRevA.74.063614} {\bibfield  {journal} {\bibinfo
  {journal} {Phys. Rev. A}\ }\textbf {\bibinfo {volume} {74}},\ \bibinfo
  {pages} {063614} (\bibinfo {year} {2006})}\BibitemShut {NoStop}%
\bibitem [{\citenamefont {Graefe}\ and\ \citenamefont
  {Korsch}(2007)}]{PhysRevA.76.032116}%
  \BibitemOpen
  \bibfield  {author} {\bibinfo {author} {\bibfnamefont {E.~M.}\ \bibnamefont
  {Graefe}}\ and\ \bibinfo {author} {\bibfnamefont {H.~J.}\ \bibnamefont
  {Korsch}},\ }\href {\doibase 10.1103/PhysRevA.76.032116} {\bibfield
  {journal} {\bibinfo  {journal} {Phys. Rev. A}\ }\textbf {\bibinfo {volume}
  {76}},\ \bibinfo {pages} {032116} (\bibinfo {year} {2007})}\BibitemShut
  {NoStop}%
\bibitem [{\citenamefont {Shchesnovich}\ and\ \citenamefont
  {Trippenbach}(2008)}]{PhysRevA.78.023611}%
  \BibitemOpen
  \bibfield  {author} {\bibinfo {author} {\bibfnamefont {V.~S.}\ \bibnamefont
  {Shchesnovich}}\ and\ \bibinfo {author} {\bibfnamefont {M.}~\bibnamefont
  {Trippenbach}},\ }\href {\doibase 10.1103/PhysRevA.78.023611} {\bibfield
  {journal} {\bibinfo  {journal} {Phys. Rev. A}\ }\textbf {\bibinfo {volume}
  {78}},\ \bibinfo {pages} {023611} (\bibinfo {year} {2008})}\BibitemShut
  {NoStop}%
\bibitem [{\citenamefont {Nissen}\ and\ \citenamefont
  {Keeling}(2010)}]{Keeling}%
  \BibitemOpen
  \bibfield  {author} {\bibinfo {author} {\bibfnamefont {F.}~\bibnamefont
  {Nissen}}\ and\ \bibinfo {author} {\bibfnamefont {J.}~\bibnamefont
  {Keeling}},\ }\href {\doibase 10.1103/PhysRevA.81.063628} {\bibfield
  {journal} {\bibinfo  {journal} {Phys. Rev. A}\ }\textbf {\bibinfo {volume}
  {81}},\ \bibinfo {pages} {063628} (\bibinfo {year} {2010})}\BibitemShut
  {NoStop}%
\bibitem [{\citenamefont {Ballentine}(1998)}]{Ballentine}%
  \BibitemOpen
  \bibfield  {author} {\bibinfo {author} {\bibfnamefont {L.~E.}\ \bibnamefont
  {Ballentine}},\ }\href@noop {} {\emph {\bibinfo {title} {Qunatum Mechanics A
  Modern Development}}},\ \bibinfo {edition} {1st}\ ed.\ (\bibinfo  {publisher}
  {World Scientific Publishing Co., Singapore},\ \bibinfo {year}
  {1998})\BibitemShut {NoStop}%
\bibitem [{\citenamefont {Brody}(1973)}]{Brody_orig}%
  \BibitemOpen
  \bibfield  {author} {\bibinfo {author} {\bibfnamefont {T.}~\bibnamefont
  {Brody}},\ }\href {\doibase 10.1007/BF02727859} {\bibfield  {journal}
  {\bibinfo  {journal} {Lettere Al Nuovo Cimento Series 2}\ }\textbf {\bibinfo
  {volume} {7}},\ \bibinfo {pages} {482} (\bibinfo {year} {1973})}\BibitemShut
  {NoStop}%
\bibitem [{\citenamefont {Manfredi}(1984)}]{Manfredi}%
  \BibitemOpen
  \bibfield  {author} {\bibinfo {author} {\bibfnamefont {V.}~\bibnamefont
  {Manfredi}},\ }\href {\doibase 10.1007/BF02747113} {\bibfield  {journal}
  {\bibinfo  {journal} {Lettere al Nuovo Cimento}\ }\textbf {\bibinfo {volume}
  {40}},\ \bibinfo {pages} {135} (\bibinfo {year} {1984})}\BibitemShut
  {NoStop}%
\bibitem [{\citenamefont {Tikhonenkov}\ \emph {et~al.}(2013)\citenamefont
  {Tikhonenkov}, \citenamefont {Vardi}, \citenamefont {Anglin},\ and\
  \citenamefont {Cohen}}]{Brody_trimer}%
  \BibitemOpen
  \bibfield  {author} {\bibinfo {author} {\bibfnamefont {I.}~\bibnamefont
  {Tikhonenkov}}, \bibinfo {author} {\bibfnamefont {A.}~\bibnamefont {Vardi}},
  \bibinfo {author} {\bibfnamefont {J.~R.}\ \bibnamefont {Anglin}}, \ and\
  \bibinfo {author} {\bibfnamefont {D.}~\bibnamefont {Cohen}},\ }\href
  {\doibase 10.1103/PhysRevLett.110.050401} {\bibfield  {journal} {\bibinfo
  {journal} {Phys. Rev. Lett.}\ }\textbf {\bibinfo {volume} {110}},\ \bibinfo
  {pages} {050401} (\bibinfo {year} {2013})}\BibitemShut {NoStop}%
\bibitem [{\citenamefont {Sciolla}\ and\ \citenamefont
  {Biroli}(2013)}]{Biroli}%
  \BibitemOpen
  \bibfield  {author} {\bibinfo {author} {\bibfnamefont {B.}~\bibnamefont
  {Sciolla}}\ and\ \bibinfo {author} {\bibfnamefont {G.}~\bibnamefont
  {Biroli}},\ }\href {\doibase 10.1103/PhysRevB.88.201110} {\bibfield
  {journal} {\bibinfo  {journal} {Phys. Rev. B}\ }\textbf {\bibinfo {volume}
  {88}},\ \bibinfo {pages} {201110} (\bibinfo {year} {2013})}\BibitemShut
  {NoStop}%
\bibitem [{\citenamefont {Kadanoff}\ and\ \citenamefont
  {Baym}(1962)}]{Kadanoffbook}%
  \BibitemOpen
  \bibfield  {author} {\bibinfo {author} {\bibfnamefont {L.~P.}\ \bibnamefont
  {Kadanoff}}\ and\ \bibinfo {author} {\bibfnamefont {G.}~\bibnamefont
  {Baym}},\ }\href@noop {} {\emph {\bibinfo {title} {Quantum Statistical
  Mechanics}}}\ (\bibinfo  {publisher} {W. A. Benjamin, Inc., New York},\
  \bibinfo {year} {1962})\BibitemShut {NoStop}%
\bibitem [{\citenamefont {Kamenev}(2011)}]{Kamenevbook}%
  \BibitemOpen
  \bibfield  {author} {\bibinfo {author} {\bibfnamefont {A.}~\bibnamefont
  {Kamenev}},\ }\href@noop {} {\emph {\bibinfo {title} {Field Theory of
  Non-equilibrium Systems}}}\ (\bibinfo  {publisher} {Cambridge University
  Press, Cambridge},\ \bibinfo {year} {2011})\BibitemShut {NoStop}%
\bibitem [{\citenamefont {Mandt}\ \emph {et~al.}(2015)\citenamefont {Mandt},
  \citenamefont {Sadri}, \citenamefont {Houck},\ and\ \citenamefont
  {T\"ureci}}]{Mandt}%
  \BibitemOpen
  \bibfield  {author} {\bibinfo {author} {\bibfnamefont {S.}~\bibnamefont
  {Mandt}}, \bibinfo {author} {\bibfnamefont {D.}~\bibnamefont {Sadri}},
  \bibinfo {author} {\bibfnamefont {A.~A.}\ \bibnamefont {Houck}}, \ and\
  \bibinfo {author} {\bibfnamefont {H.~E.}\ \bibnamefont {T\"ureci}},\ }\href
  {http://stacks.iop.org/1367-2630/17/i=5/a=053018} {\bibfield  {journal}
  {\bibinfo  {journal} {New J. Phys.}\ }\textbf {\bibinfo {volume} {17}},\
  \bibinfo {pages} {053018} (\bibinfo {year} {2015})}\BibitemShut {NoStop}%
\bibitem [{\citenamefont {Fradkin}(2013)}]{fradkinbook}%
  \BibitemOpen
  \bibfield  {author} {\bibinfo {author} {\bibfnamefont {E.}~\bibnamefont
  {Fradkin}},\ }\href@noop {} {\emph {\bibinfo {title} {Field Theories of
  Condensed Matter Physics}}},\ \bibinfo {edition} {2nd}\ ed.\ (\bibinfo
  {publisher} {Cambridge University Press, Cambridge},\ \bibinfo {year}
  {2013})\BibitemShut {NoStop}%
\bibitem [{\citenamefont {Altland}\ and\ \citenamefont
  {Simons}(2010)}]{Altlandbook}%
  \BibitemOpen
  \bibfield  {author} {\bibinfo {author} {\bibfnamefont {A.}~\bibnamefont
  {Altland}}\ and\ \bibinfo {author} {\bibfnamefont {B.}~\bibnamefont
  {Simons}},\ }\href@noop {} {\emph {\bibinfo {title} {Condensed Matter Field
  Theory}}},\ \bibinfo {edition} {2nd}\ ed.\ (\bibinfo  {publisher} {Cambridge
  University Press, Cambridge},\ \bibinfo {year} {2010})\BibitemShut {NoStop}%
\bibitem [{\citenamefont {Raghavan}\ \emph {et~al.}(1999)\citenamefont
  {Raghavan}, \citenamefont {Smerzi}, \citenamefont {Fantoni},\ and\
  \citenamefont {Shenoy}}]{smerzi_pra}%
  \BibitemOpen
  \bibfield  {author} {\bibinfo {author} {\bibfnamefont {S.}~\bibnamefont
  {Raghavan}}, \bibinfo {author} {\bibfnamefont {A.}~\bibnamefont {Smerzi}},
  \bibinfo {author} {\bibfnamefont {S.}~\bibnamefont {Fantoni}}, \ and\
  \bibinfo {author} {\bibfnamefont {S.~R.}\ \bibnamefont {Shenoy}},\ }\href
  {\doibase 10.1103/PhysRevA.59.620} {\bibfield  {journal} {\bibinfo  {journal}
  {Phys. Rev. A}\ }\textbf {\bibinfo {volume} {59}},\ \bibinfo {pages} {620}
  (\bibinfo {year} {1999})}\BibitemShut {NoStop}%
\bibitem [{\citenamefont {Jos\'e}\ and\ \citenamefont
  {Saletan}(1998)}]{Josebook}%
  \BibitemOpen
  \bibfield  {author} {\bibinfo {author} {\bibfnamefont {J.~V.}\ \bibnamefont
  {Jos\'e}}\ and\ \bibinfo {author} {\bibfnamefont {E.~J.}\ \bibnamefont
  {Saletan}},\ }\href@noop {} {\emph {\bibinfo {title} {Classical dynamics a
  contemporary approach}}}\ (\bibinfo  {publisher} {Cambridge University Press,
  Cambridge},\ \bibinfo {year} {1998})\BibitemShut {NoStop}%
\bibitem [{\citenamefont {Geisel}\ \emph {et~al.}(1990)\citenamefont {Geisel},
  \citenamefont {Wagenhuber}, \citenamefont {Niebauer},\ and\ \citenamefont
  {Obermair}}]{chaos_in_electron}%
  \BibitemOpen
  \bibfield  {author} {\bibinfo {author} {\bibfnamefont {T.}~\bibnamefont
  {Geisel}}, \bibinfo {author} {\bibfnamefont {J.}~\bibnamefont {Wagenhuber}},
  \bibinfo {author} {\bibfnamefont {P.}~\bibnamefont {Niebauer}}, \ and\
  \bibinfo {author} {\bibfnamefont {G.}~\bibnamefont {Obermair}},\ }\href
  {\doibase 10.1103/PhysRevLett.64.1581} {\bibfield  {journal} {\bibinfo
  {journal} {Phys. Rev. Lett.}\ }\textbf {\bibinfo {volume} {64}},\ \bibinfo
  {pages} {1581} (\bibinfo {year} {1990})}\BibitemShut {NoStop}%
\bibitem [{\citenamefont {Whittaker}\ and\ \citenamefont
  {Watson}(1927)}]{Whittakerbook}%
  \BibitemOpen
  \bibfield  {author} {\bibinfo {author} {\bibfnamefont {E.}~\bibnamefont
  {Whittaker}}\ and\ \bibinfo {author} {\bibfnamefont {G.}~\bibnamefont
  {Watson}},\ }\href@noop {} {\emph {\bibinfo {title} {A Course of Modern
  Analysis}}}\ (\bibinfo  {publisher} {Cambridge University Press, Cambridge},\
  \bibinfo {year} {1927})\BibitemShut {NoStop}%
\bibitem [{\citenamefont {Husimi}(1940)}]{Husimi_orig}%
  \BibitemOpen
  \bibfield  {author} {\bibinfo {author} {\bibfnamefont {K.}~\bibnamefont
  {Husimi}},\ }\href@noop {} {\bibfield  {journal} {\bibinfo  {journal}
  {Proceedings of the Physico-Mathematical Society of Japan. 3rd Series}\
  }\textbf {\bibinfo {volume} {22}},\ \bibinfo {pages} {264} (\bibinfo {year}
  {1940})}\BibitemShut {NoStop}%
\bibitem [{\citenamefont {Lee}(1995)}]{Lee_Husimi}%
  \BibitemOpen
  \bibfield  {author} {\bibinfo {author} {\bibfnamefont {H.-W.}\ \bibnamefont
  {Lee}},\ }\href {\doibase 10.1016/0370-1573(95)00007-4} {\bibfield  {journal}
  {\bibinfo  {journal} {Physics Reports}\ }\textbf {\bibinfo {volume} {259}},\
  \bibinfo {pages} {147 } (\bibinfo {year} {1995})}\BibitemShut {NoStop}%
\bibitem [{\citenamefont {Mahmud}\ \emph {et~al.}(2005)\citenamefont {Mahmud},
  \citenamefont {Perry},\ and\ \citenamefont {Reinhardt}}]{Mahmud_BEC}%
  \BibitemOpen
  \bibfield  {author} {\bibinfo {author} {\bibfnamefont {K.~W.}\ \bibnamefont
  {Mahmud}}, \bibinfo {author} {\bibfnamefont {H.}~\bibnamefont {Perry}}, \
  and\ \bibinfo {author} {\bibfnamefont {W.~P.}\ \bibnamefont {Reinhardt}},\
  }\href {\doibase 10.1103/PhysRevA.71.023615} {\bibfield  {journal} {\bibinfo
  {journal} {Phys. Rev. A}\ }\textbf {\bibinfo {volume} {71}},\ \bibinfo
  {pages} {023615} (\bibinfo {year} {2005})}\BibitemShut {NoStop}%
\bibitem [{\citenamefont {Scully}\ and\ \citenamefont
  {Zubairy}(1997)}]{Scullybook}%
  \BibitemOpen
  \bibfield  {author} {\bibinfo {author} {\bibfnamefont {M.~O.}\ \bibnamefont
  {Scully}}\ and\ \bibinfo {author} {\bibfnamefont {M.~S.}\ \bibnamefont
  {Zubairy}},\ }\href@noop {} {\emph {\bibinfo {title} {Quantum Optics}}}\
  (\bibinfo  {publisher} {Cambridge university press, Cambridge},\ \bibinfo
  {year} {1997})\BibitemShut {NoStop}%
\end{thebibliography}%
\end{document}